\documentclass[aps,pre,twocolumn,showpacs,floatfix]{revtex4}
\usepackage{amsmath,amssymb,graphicx,times,bm}

\usepackage[latin1]{inputenc}

\newcommand{\be}{\begin{equation}}
\newcommand{\ee}{\end{equation}}
\newcommand{\bea}{\begin{eqnarray}}
\newcommand{\eea}{\end{eqnarray}}
\newcommand{\bi}{\begin{itemize}}
\newcommand{\ei}{\end{itemize}}
\newcommand{\pavg}{\Pi_M(t)}
\newcommand{\kommentar}[1]{}

\begin{document}

%\begin{frontmatter}

\title{Continuous-Time Quantum Walks: \\ 
Models for Coherent Transport on Complex Networks}

\author{Oliver M{\"u}lken and Alexander Blumen}
%\email{blumen@physik.uni-freiburg.de}

\affiliation{
%\address{
Theoretical Polymer Physics, University of Freiburg,
Hermann-Herder-Stra{\ss}e 3, 79104 Freiburg, Germany}

\begin{abstract}
% Text of abstract

This paper reviews recent advances in continuous-time quantum walks (CTQW)
and their application to transport in various systems. The introduction
gives a brief survey of the historical background of CTQW.  After a short
outline of the theoretical ideas behind CTQW and of its relation to
classical continuous-time random walks (CTRW) in Sec.~2, implications for
the efficiency of the transport are presented in Sec.~3. The fourth
section gives an overview of different types of networks on which CTQW
have been studied so far. Extensions of CTQW to systems with long-range
interactions and with static disorder are discussed in section V. Systems
with traps, i.e., systems in which the walker's probability to remain
inside the system is not conserved, are presented in section IV. Relations
to similar approaches to the transport are studied in section VII. The
paper closes with an outlook on possible future directions.

\end{abstract}

\date{\today}

%\begin{keyword}
%Quantum Transport, Complex Networks, Coherent and Incoherent Dynamics,
%Excitons% keywords here, in the form: keyword \sep keyword

%\end{keyword}

%\end{frontmatter}

\maketitle

\tableofcontents

% Text

\section{Introduction} 

Transport of mass, charge or energy is the basis of many physical,
chemical or biological processes. Such transfer mechanisms and their
efficiency depend on the underlying structure of the system, range from
polymer physics to solid state physics to biological physics to even
quantum computation, see Refs.\
\cite{Bouchaud1990,albert2002,dorogovtsev2002,kempe2003,vanKampen,weiss}
for reviews.  The underlying structures could be, for example, simple
crystals, as in solid state physics \cite{Ziman}, more complex molecular
aggregates like polymers \cite{Kenkre}, or general network structures
\cite{albert2002}. 

In quantum mechanics, the potential a particle is moving in specifies the
Hamiltonian of the system, which determines the time evolution. For
instance, the dynamics of electrons in a simple crystal is described by
the Bloch ansatz \cite{Ziman,Ashcroft, Kittel}, which mirrors their
behavior in metals quite accurately. Over the years these models have been
refined and extended to address different phenomena, such as the dynamics
of atoms in optical lattices and the Anderson localization in systems with
energetic disorder \cite{anderson1958}.  In quantum chemistry,
H{\"u}ckel's molecular-orbital theory allows to define a Hamiltonian for
more complex structures, such as molecules \cite{McQuarrie}. This is
also related to transport in polymers, where the connectivity of the
polymer plays a fundamental role in its dynamical and relaxational
properties \cite{Doi-Edwards}. There, (classical) transport processes can
be modeled by continuous-time random walk (CTRW) approaches
\cite{Kenkre,weiss,vanKampen}. An important example is the motion of
Frenkel excitons, whose high-temperature dynamics is often governed by a
master equation with an appropriate (classical) transfer operator which
determines the temporal evolution of the system
\cite{scher1973,scher1975}. CTRW have proven to be a very useful and
general tool in describing incoherent transport in various settings,
ranging, for instance, from percolation on fractals \cite{alexander1982}
to electronic energy transfer in glassy systems
\cite{klafter1984,even1984}. Also, anomalous diffusional behavior has been
successfully modeled by CTRW using suitable waiting time distributions
\cite{klafter1987,zumofen1989}.

\subsection{Quantum Walks}

The tight-binding approximation used in sold state physics as well as in
H{\"u}ckel's theory is equivalent to the so-called quantum walks, which
model purely coherent quantum dynamics of excitations on networks
\cite{farhi1998,aharonov1993,kempe2003,mb2005a,bose2003}. What matters is
that the constituting elements (spins, atoms, molecules, etc.) are of the
same type, in the simplest cases acting as two-level systems.  Now, the
extension of classical random walks ideas to the quantum domain is not
unique and allows variants, of which two types capture most of the
interest: discrete-time quantum (random) walks (introduced by Aharonov
{\it et al.}, using an additional internal ``coin'' degree of freedom
\cite{aharonov1993}), and continuous-time quantum walks (CTQW) (introduced
by Farhi and Gutmann, where the connection to CTRW uses the analogy
between the quantum mechanical Hamiltonian and the classical transfer
matrix \cite{farhi1998}). Recently, Strauch has shown how these two
versions are related \cite{strauch2006}. Thus, CTQW and CTRW act as the
two extreme cases of purely coherent and purely incoherent transport,
respectively.

In quantum information theory, quantum walks are used extensively as
algorithmic tools for quantum computation \cite{Nielsen}. Here, the most
prominent examples are Shor's algorithm \cite{shor1994} and Grover's
algorithm \cite{grover1997}. The latter is a quantum algorithm for finding
an item in an (unsorted) database of qubits.  Such an algorithm can also
be related to CTQW \cite{childs2004,abm2010b}.  The idea of using
computers built on quantum mechanical principles instead of classical
computers goes back to Feynman, who already formulated an early version of
quantum walks in 1982 \cite{feynman1982}. From a conceptual point of view,
quantum walks are closely related to quantum cellular automata, see, e.g.,
\cite{meyer1996}. Using CTQW, the transfer of information in complex
systems was discussed by Christandl {\it et al.} in the context of perfect
state transfer \cite{christandl2004}.

CTQW are also in close relation to the so-called quantum graphs (QG), see,
e.g., \cite{kottos1997, schanz2000, kottos2000, kottos2003}. However,
unlike CTQW, QG take into account detailed properties of each bond of the
graph explicitly; in QG, bonds may be directed and have distinct lengths.
The connections between discrete Laplacians on discrete QG and periodic
orbits were recently investigated by Smilansky \cite{smilansky2007}.

\subsection{Experimental implementations of CTQW}

Simple theoretical models have always been very useful for our
understanding of physics. In quantum mechanics, next to the harmonic
oscillator, the particle in a box provides much insight into the quantum
world (e.g.\ \cite{Sakurai}). The problem of a quantum mechanical particle
moving in an infinite box has been reexamined in \cite{kinzel1995,
fgrossmann1997, berry1996}: Remarkably, even this simple system shows
complex but regular spacetime probability structures (so-called quantum
carpets).  In solid state physics and quantum information theory, one of
the most simple systems is associated with a particle moving in a regular
periodic potential. 

In recent years, arrangements close to ideal theoretical models have been
tailored using, e.g., ultra-cold atoms in optical lattices, see
\cite{bloch2005} and references therein.  For both quantum walk variants,
experimental implementations have been proposed, based on microwave
cavities \cite{sanders2003}, on Rydberg atoms \cite{cote2006} or on ground
state atoms \cite{duer2002} in optical lattices or in optical cavities
\cite{knight2003}, or using the orbital angular momentum of photons
\cite{zhang2007}.  Other experimental proposals connected to CTQW are
based on waveguide arrays \cite{hagai2008} or on structured clouds of
Rydberg atoms \cite{mbagrw2007}. 

\subsection{Transport in complex systems}

The range of systems where CTQW can be used to study transport is clearly
not restricted to simple (symmetric) models. In particular, CTQW have been
proven to be very useful in describing the dynamics of excitations in
various complex systems, as will be shown throughout this review.

Since CTQW can be related to the tight-binding approximation in solid
state physics, CTQW are also related to chains of coupled spin 1/2
particles. If all but one spins are prepared in the same state, say, spin
down, and a single spin in the spin up state, one can map this onto a
chain of coupled two-level systems with a single initial excitation at the
node which is identified with the spin up particle
\cite{bose2003,burgarth_phd,bose2007}. The essential dynamics is then
equivalent to a CTQW on a linear network.

However, also the dynamics in topological disordered systems can be
modeled by CTQW. Take for instance a gas of highly excited ultra-cold
Rydberg atoms. At ultra-low temperatures the configuration of the atoms
can be thought of as being frozen on the time-scale on which the following
transport process is taking place
\cite{anderson1998,mourachko1998,westermann2006}: The dynamics is started
by exciting one (or several) of the Rydberg atoms in a higher state which
is resonantly coupled to a lower Rydberg state \cite{cote2006,mbagrw2007}.
Since essentially only two different Rydberg states are involved, one can
map this again onto a network of coupled two-level systems.

Also biological systems share some properties of simple network models.
Recent experiments on the light-harvesting complexes of algae have renewed
the interest in studying the dynamics of excitations in such systems
\cite{engel2007,collini2010}. Here, an initial excitation (a Frenkel
exciton) is created by absorbing a solar photon. The exciton is then
transfered along a network of (bacterio-) chlorophylls (BChl), the
chromophores, to the reaction center, where the excitation energy is
converted into chemical energy. Now, the network of the BChls can be
considered to be static and stable with defined couplings between the
BChls, at all (even ambient) temperatures of interest. Moreover, the
Frenkel exciton can be viewed as a quasi particle moving along the
network.  Therefore, also here the transport can be modeled by CTQW, at
least at (very) low temperatures. The incoherent exciton transport in
dendrimers can be efficiently modelled by random walks, see, for instance,
\cite{heijs2004,vlaming2005,blumen2005,shortreed1997}. However, for
certain systems there is also experimental evidence for coherent
interchromophore transport processes \cite{varnavski2002,varnavski2002b}.
Recent investigations of biological light-harvesting systems range from
the Fenna-Matthews-Olsen complex \cite{engel2007} to marine algae
\cite{collini2010}. There have also been recent theoretical efforts to
understand the coherent features of exciton dynamics found in recent
experiments, see, e.g., \cite{fleming2004,cheng2006, mohseni2008,
caruso2009, olaya-castro2008}.

Clearly, such biological systems cannot be thought of as being isolated
from their environment. Therefore, in these cases the purely coherent
description by CTQW is of limited value. However, it is also possible to
extend the CTQW approach to take the coupling to the environment into
account, which leads to models related to quantum master equations
\cite{Breuer-Petruccione}, see also Sec.~\ref{sec-qme}.

\section{Definitions and terminology of CTQW}

\subsection{Connectivity of network}

Networks involved in the CTQW and CTRW dynamics are characterized by the
form in which their sites are connected.  Starting point of a network is a
collection of $N$ nodes, which are then joined by bonds, the connectivity
matrix $\bm A$ mirroring the way in which the bonding occurs.

The $N\times N$ connectivity matrix $\bm A$ has the elements $A_{kj}$,
where
\be
A_{ {kj}} = \begin{cases} f_j & \mbox{for $k=j$} \\
                -1 & \mbox{if $k$ and $j$ connected} \ A_{kj} = A_{jk} \\ 0 &
                \mbox{else}, \end{cases}
\ee
where $f_j$ is the number of bonds emanating from
node $j$. This matrix has interesting and useful properties: (a) $\bm
A$ is real and symmetric, (b) all its eigenvalues $\lambda_n$ are real and
$\lambda_n\geq0$, and (c) $\bm A$ has a single smallest eigenvalue which is
$\lambda_1=0$.

One can associate with every node of the network a basis vector in an
$N$-dimensional vector space. Now, these basis vectors form a complete
orthonormal basis, which, for instance, is given by
\be
|1\rangle = {\tiny \begin{pmatrix} 1 \\ 0 \\ \vdots \\0
\end{pmatrix}}, \quad
|2\rangle= {\tiny \begin{pmatrix} 0 \\ 1 \\ \vdots \\0
\end{pmatrix}}, \quad
\dots,
|N\rangle = {\tiny \begin{pmatrix} 0 \\ \vdots \\0 \\ 1
\end{pmatrix}}
\ee

Take as an example a ring-like network of $N$ nodes, with only nearest
neighbor connections and $|N+1\rangle \equiv |1\rangle$, then the matrix
$\bm A$ reads: 
\be
A_{ij} = \begin{pmatrix} 
2 & -1 & 0 & \dots & & -1\\ 
-1 & 2 & -1 & 0 & \dots & 0\\ 
0 &\ddots & \ddots & \ddots & \ddots & \vdots\\
\vdots \\
 &  & &-1 & 2 & -1\\
-1 & 0 & \dots & & -1 & 2
\end{pmatrix}
\label{a-ring}
\ee
Writing $\bm A$ in a quantum mechanical fashion using the projection operators
$|l\rangle\langle k|$, leads to
\be
{\bm A} = \sum_l 2 |l\rangle\langle l| -
|l-1\rangle\langle l| - |l+1\rangle\langle l|
\ee

Having now specified the network through the connectivity matrix $\bm A$,
one is interested in the dynamics over the network. First, one
distinguishes between purely classical, i.e., incoherent dynamics and
purely quantum-mechanical, i.e., coherent dynamics. The main question one
wishes to answer is: What is the probability to be at node $k$ after time
$t$ when starting at node $j$?  While here the focus is mainly on
localized initial conditions at a single node $j$, in general, the initial
state can be any distribution involving all the nodes. 

\subsection{Transfer matrix for CTRW}

We start by considering classical transport.  Let the initial node be $j$,
such that the initial state of the system is $| j \rangle$, and denote the
transition probability to go in time $t$ from node $j$ to node $k$ by
$p_{k,j}(t)$.  Therefore, the initial condition is $\langle k | j \rangle
\equiv p_{k,j}(0) = \delta_{k,j}$, where $\delta_{k,j} = 1$ for $k=j$ and
$0$ otherwise. The state after time $t$ is $| j;t \rangle$, such that the
overlap with node $k$ reads $\langle k | j;t\rangle \equiv p_{k,j}(t)$.

The dynamics resulting in the state $|j;t\rangle$ follows from the
transition rates per unit time between two nodes. Those transition rates
are the elements of the so-called transfer matrix $\bm T$, $T_{kj} \equiv
\langle k | {\bm T} | j\rangle$, which is therefore related to the spatial
gain or loss. One can now write the temporal change of the probability
after time $\epsilon\ll1$ as
\be
p_{k,j}(\epsilon) = p_{k,j}(0) + \epsilon
T_{kj} = \delta_{k,j} + \epsilon
T_{kj}.
\ee
Assuming a Markovian process, the following master equation can be shown
to hold \cite{vanKampen, weiss}:
\be
\frac{d}{dt} p_{k,j}(t) =  \sum_l
T_{kl} p_{l,j}(t).
\label{mastereq}
\ee
This equation defines the CTRW, which (depending on the specific form of
the transfer matrix) have been applied to various problems in physics,
chemistry, biology, and also social sciences.

In the simplest case, where the rates for all bonds are equal, say,
$\gamma$, the transfer matrix is related to the connectivity matrix
through $\bm T= - \gamma \bm A$.  Then the master equation describes a
diffusive motion over the network.

The formal solution of Eq.(\ref{mastereq}) is $p_{ {k,j}}(t) = \langle   k
| e^{{\bm T} t} | j \rangle = \langle   k | e^{-\gamma{\bm A} t} | j
\rangle$.  Denoting the eigenstates of $\bm A$ by ${{|q_n\rangle}}$, one
has
\be
p_{k,j}(t) = \langle k | e^{-\gamma{\bm A}t} | j\rangle = \sum_n
e^{-\lambda_n \gamma t} \langle k |
q_n \rangle \langle q_n | j\rangle.
\label{meq-solution}
\ee
Since the eigenvalues are positive ($\lambda_n>0$ for $n>1$ and
$\lambda_1=0$), the long-time limit follows directly: For $t\gg1$ in the
sum of Eq.(\ref{meq-solution}) all exponential terms but one decay
quickly to zero. The only term which survives is the one for
$\lambda_1=0$, with the corresponding eigenstate $|q_1 \rangle =
\frac{1}{N} \sum_l |l\rangle$.  Therefore, the long-time limit of all
transition probabilities is $\lim_{t\to\infty} p_{k,j}(t) = 1/N$, which is
independent of the connectivity of the network. This means that every CTRW
whose transfer matrix follows directly from the connectivity matrix will
eventually decay at long times to the equipartition value $1/N$, a fact
which is sometimes referred to as ``ground state dominance''.

\subsection{Hamiltonian for CTQW}

Turning now to the quantum-mechanical dynamics, one faces the fact that
the transport has to be formulated in Hilbert space. For this one assumes
that the states $|j\rangle$ representing the nodes span the whole
accessible Hilbert space. As before, it is assumed that the states are
orthonormal and complete, i.e., $\langle k|j\rangle = \delta_{k,j}$,
$\sum_j |j\rangle \langle j| = \bm 1$. The dynamics is then governed by a
specific Hamiltonian $\bm H$, such that Schr\"odinger's equation for the
transition {\sl amplitudes} $\alpha_{k,j}(t) \equiv \langle k | j;t
\rangle$ reads 
\be
{\frac{d}{d t} \alpha_{k,j}(t) = -i \sum_l H_{k,l} \
\alpha_{l,j}(t)}.
\label{sgl}
\ee
Similar to the CTRW, the formal solution for the transition amplitudes is
given by
\be
\alpha_{k,j}(t) = \langle k | e^{-i \bm H t} | j \rangle,
\ee
where $e^{-i \bm H t}$ is the quantum-mechanical time-evolution operator.
The transition probabilities follow as usual as $\pi_{k,j}(t) \equiv
|\alpha_{k,j}(t)|^2$. 

Comparing Eq.(\ref{sgl}) to Eq.(\ref{mastereq}) one immediatly notices the
very similar structure of the two equations, except for the imaginary unit $i$
appearing in Eq.(\ref{sgl}). However, while Eq.(\ref{mastereq}) is an
equation for the transition probabilities $p_{k,j}(t)$, Eq.(\ref{sgl}) is
an equation for the transition amplitudes $\alpha_{k,j}(t)$. 

One can push the similarities further by identifying the
quantum-mechanical Hamiltonian $\bm H$ with the classical transfer matrix
$\bm T$, i.e., $\bm H \equiv -\bm T$.  This approach, pioneered by Farhi
and Gutmann in Ref.~\cite{farhi1998}, allows to compare the two extremes
of transport on the same topology.  On the one hand there is the purely
incoherent CTRW, while on the other hand, there is Schr\"odinger's equation
which now defines CTQW. Assuming again that all transition rates between
different connected nodes are the same, the Hamiltonian can be directly
related to the connectivity, $\bm H = \gamma \bm A$. Therefore, the
underlying topological network is determined both for CTRW and for CTQW by
the connectivity matrix $\bm A$. This allows to study the role of the
connectivity in parallel for CTRW and for CTQW.

Let us donote the eigenvalues of $\bm H$ by  ${E_n}$ and the eigenstates
of $\bm H$ by ${|\Psi_n\rangle}$. Evidently,  the eigenvalues and
eigenstates of $\bm H$ and $\bm T$ are practically the same; nonetheless,
we will keep the distinction between $\bm H$ and $\bm T$ due to later
purposes. Now, the quantum-mechanical transition probabilities read 
\be
\pi_{k,j}(t) = \Big|\sum_n e^{-i E_n t} \langle k |
\Psi_n \rangle \langle \Psi_n | j \rangle \Big|^2.
\label{pi_allg}
\ee
Unlike the situation for CTRW, for CTQW there is no unique long-time limit
of $\pi_{k,j}(t)$ due to the unitary time evolution. In order to compare
to the long-time behavior of the CTRW one uses the long time average
\cite{aharonov2001,mvb2005a}
\bea
&& 
\chi_{k,j} \equiv \lim_{T\to\infty} \frac{1}{T} \int_0^T dt \
\pi_{k,j}(t)
\nonumber \\
&& 
= \sum_{n,m}
\delta_{E_n,E_m}
\langle k |
\Psi_n \rangle \langle \Psi_n |
j \rangle \langle j | \Psi_m
\rangle \langle \Psi_m | k \rangle, \qquad
\label{limprob}
\eea
where $\delta_{E_n,E_m} = 1$ if $E_n=E_m$ and $0$ otherwise. The long-time
average $\chi_{k,j}$ still depends on the initial and final nodes $j$ and
$k$. 

\subsection{Example: A discrete ring}

As an example which illustrates the differences and similarities between
CTQW and CTRW we consider a discrete ring of $N$ nodes, whose connectivity
matrix is given above in Eq.(\ref{a-ring}). This one-dimensional structure
with the periodic boundary conditions $|N+1\rangle = |1\rangle$ allows for
a full analytical solution. When realizing that the matrix $\bm A$ is
nothing but a tight-binding matrix for a particle moving in a regular
one-dimensional crystal, one readily obtains the eigenvalues and
eigenstates from a Bloch ansatz \cite{Kittel,Ziman}. Namely, the Bloch
states are linear combinations of the localized states $|j\rangle$ and are
given by
\be
|\Psi_\theta\rangle = \frac{1}{\sqrt N}\sum_{j=1}^N e^{-i \theta j}
|j\rangle.
\label{blochef}
\ee
Now the energy is obtained as
\be
E_\theta = 2 - 2 \cos\theta.
\ee
For small $\theta$ the energy is given by $E_\theta \approx \theta^2$
which resembles the energy spectrum of a free particle.

By inverting Eq.(\ref{blochef}) one may describe the state $|j\rangle$
localized at node $j$ as a Wannier state \cite{Kittel,Ziman}
\be
|j\rangle
= \frac{1}{\sqrt N} \sum_\theta e^{i\theta j} | \Psi_\theta \rangle.
\ee
Since the states $|j\rangle$ span the whole accessible Hilbert space, one
has $\langle k | j \rangle = \delta_{kj}$ and, via Eq.(\ref{blochef}),
also $\langle \Psi_{\theta'} | \Phi_\theta \rangle =
\delta_{\theta'\theta}$.  Then the transition amplitude reads
\cite{mb2005b}
\bea
\alpha_{kj}(t) &=&
\frac{1}{N} \sum_{\theta,\theta'} \langle \Psi_{\theta'} | e^{-i\theta k}
e^{-i{\bm H}t} e^{i\theta' j} | \Psi_\theta \rangle 
\nonumber \\
&=& 
\frac{1}{N} \sum_{\theta} e^{-iE_\theta t} e^{-i\theta(k-j)}.
\label{ampl_bloch}
\eea
In an analogous way one obtains the classical transition probabilities
\cite{mb2005b}
\be
p_{k,j}(t) = \frac{1}{N} \sum_{\theta} e^{-\lambda_\theta t} e^{-i\theta(k-j)},
\ee
because the Bloch states are also the eigenstates $|q_n\rangle$.

The periodic boundary condition for a $1d$ lattice of size $N$ requires
that $\theta = 2 n \pi /N$ with $n\in]0,N]$. Now Eq.(\ref{ampl_bloch}) is
given by
\be
\alpha_{jk}(t) = \frac{e^{-i2t}}{N} \sum_{n} e^{i2t\cos(2n\pi/N)}
e^{-i2\pi n(k-j)/N}.
\label{ampl_bloch2}
\ee
For small $\theta$, this is directly related to the results
obtained for a quantum particle in a box \cite{kinzel1995, fgrossmann1997,
berry1996}, because then $E_n \sim n^2$.  Indeed, some features found for
the particle in a box can also be found in the case of a CTQW on a
discrete ring. For the particle in a box the initial condition is restored
after some revival time. In fact the probability to find the particle at a
certain position in the box is a periodic function.

Analogously but not completely similarily, for the CTQW on the ring, the
initial (localized) condition is only partially restored. Only small rings
of sizes $N=1,2,3,4,$ and $6$ lead to a full revival of the initial
condition. All other sizes only lead to partial revivals.  The revival
time $\tau$ is given by $\alpha_{k,j}(\tau) = \alpha_{k,j}(0)$.  Since the
transition amplitudes are given as a sum over all modes $n$, see
Eq.(\ref{ampl_bloch2}), one has for each mode $n$ its revival time
\cite{mb2005b} 
\be
\tau_n=\frac{r\pi}{2} [1 + \cot^2(n\pi/N)],
\label{revtime}
\ee
where $r\in{\mathbb N}$ (without any loss of generality one sets $r=1$).
From Eq.(\ref{revtime}) one finds that $\tau_n > \tau_{n+1}$ for $n\in
]0,N/2]$ and $\tau_n < \tau_{n+1}$ for $n\in ]N/2,N]$.  For certain values
of $n$, $\tau_n$ will be of order unity, e.g.\ for $n=N/2$ $\tau_n =
\pi/2$.  However, for $n << N$, Eq.(\ref{revtime}) yields $\tau_n =
N^2/2\pi n^2 \equiv \tau_0/n^2$, which is analogous to the particle in the
box and where $\tau_0$ is a universal revival time.  

\begin{figure}
\centerline{
\includegraphics[width=0.425\columnwidth]{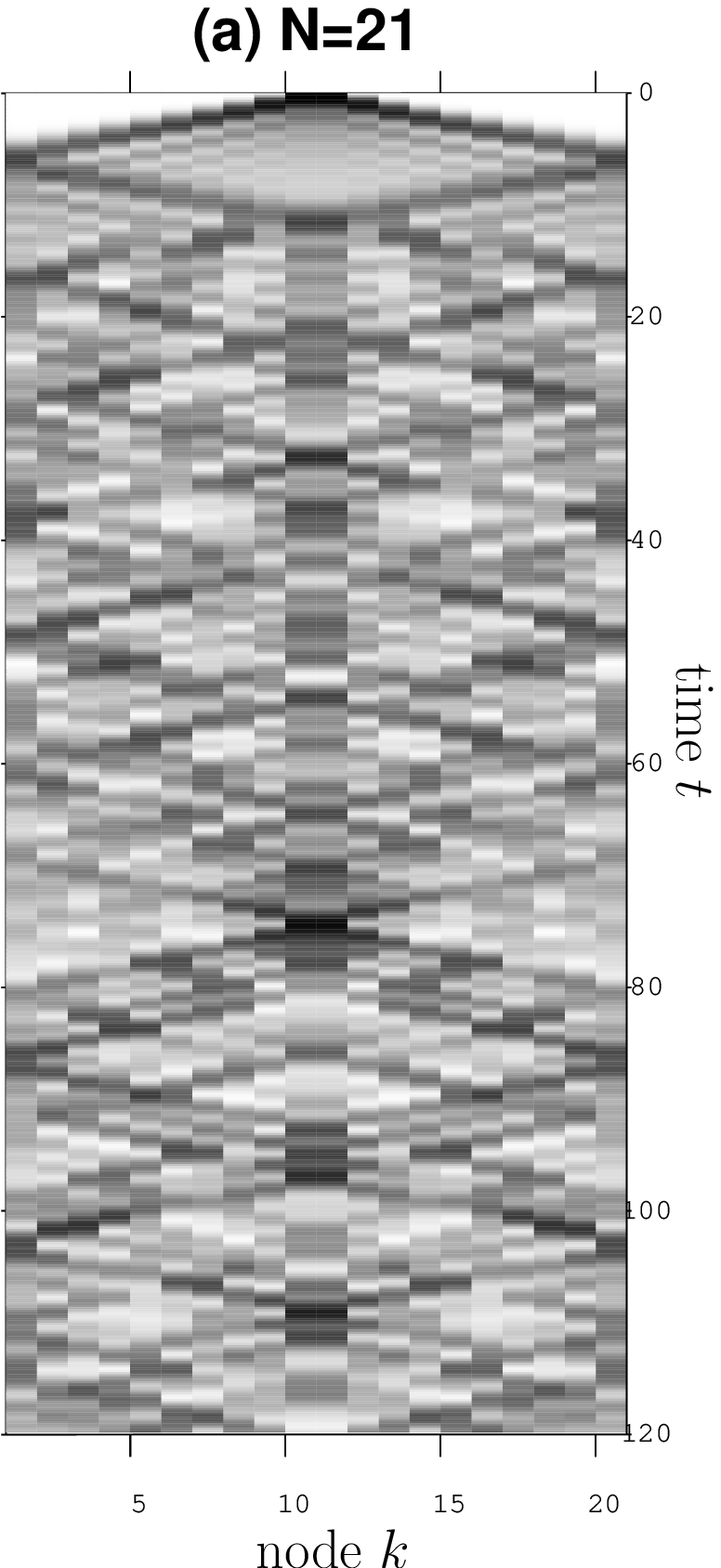}
\includegraphics[width=0.425\columnwidth]{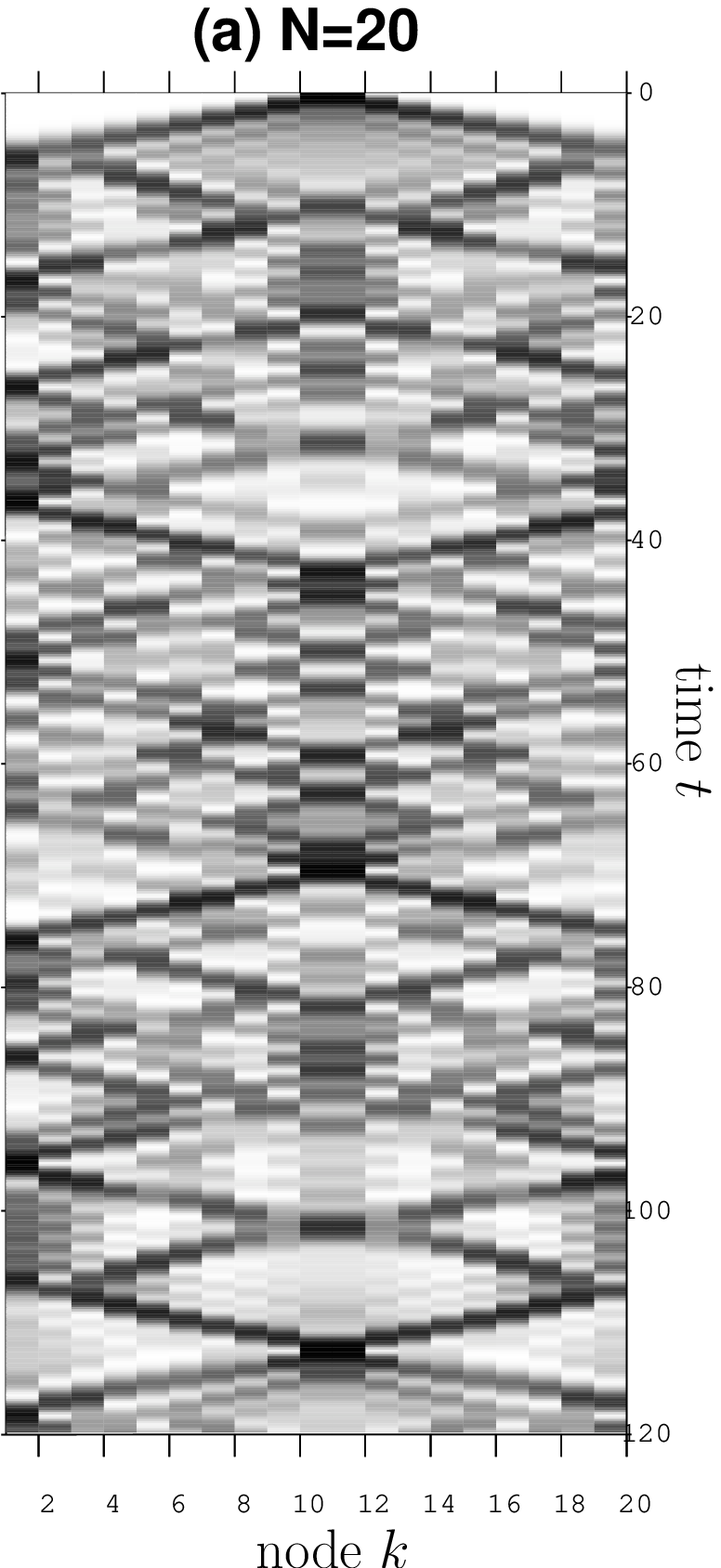}}
\caption{Contour plot of the probability for a CTQW on a circle of length
(a) $N=21$ and (b) $N=20$ over long times $t$. Dark regions denote high
probabilities.  From \cite{mb2005b}.}
\label{1d-circle-contlt}
\end{figure}

Because the revival times $\tau_n$ have large variations in value, one
compares these times to the actual time needed by the CTQW for travelling
through the lattice. As mentioned earlier, interference effects in the
return probability $\pi_{1,1}(t)$ are seen after a time $t\approx N/2$.
The first revival time has to be larger than this, because there cannot be
any revival unless the wave reaches its starting node again. Our
calculations \cite{mb2005b} suggest that the first revival time will be of order
$\tau_0$. From Fig.~\ref{1d-circle-contlt} one sees that the first
(incomplete) revival occurs for $N=20$ at $t\approx 70 > 20^2/2\pi$ and
for $N=21$ at $t\approx 75 > 21^2/2\pi$.

Since the transition probabilities are easily obtained from the Bloch
ansatz, also the long-time averages can be estimated analytically.
Depending on the number of nodes in the network, the long-time averages
are, \cite{mb2005b}, for even $N$:
\be
\chi_{k,j} = \begin{cases}
1/N^2 & k\neq j \\
1/2N & k=j \ \mbox{and } k=j+N/2
\end{cases}
\ee
and for odd $N$:
\be
\chi_{k,j} = \begin{cases}
1/N^2 & k\neq j \\
1/N & k=j 
\end{cases}
\ee
Thus, unlike in the classical CTRW case the long-time average is not
equipartitioned but ``remembers'' the initial condition. Moreover, also
the symmetry arising from the even or odd number $N$ of nodes is reflected
in the $\chi_{k,j}$.

\section{Efficiency of CTQW and CTRW} 
\label{sec-eff}

The performance of CTQW and CTRW depends to a large extent on the
connectivity of the underlying network, i.e., on its topology.  One
focuses on the return probabilities $\pi_{j,j}(t)$ and $p_{j,j}(t)$.  Now,
if these probabilities decay quickly with time, the probabilities
$1-\pi_{j,j}(t)$ and $1-p_{j,j}(t)$, to be at any but the initial node,
increase quickly. This implies a fast transport through the network.  In
order to make a global statement on the performance, one considers the
average return probabilities \cite{mb2006b}
\be
\overline{\pi}(t) \equiv \frac{1}{N} \sum_j \pi_{j,j}(t) \qquad
\mbox{for CTQW}
\ee
and
\be
\overline{p}(t) \equiv \frac{1}{N} \sum_j p_{j,j}(t) \qquad 
\mbox{for CTRW.}
\ee
Again, a quick decay of $\overline{\pi}(t)$ [$\overline{p}(t)$] implies a
fast propagation through the network, while a slow decay implies a slow
propagation.

For CTRW the average return probabilities simplify considerably; one has
namely
\bea
\overline{p}(t)
&=& 
\frac{1}{N}\sum_{n} e^{-\lambda_n t} \langle \Psi_n | \sum_j |j \rangle
\langle j | \Psi_n \rangle 
\nonumber \\
&=& 
\frac{1}{N} \sum_{n} e^{-\lambda_nt},
\label{pavg}
\eea
which only depends on the eigenvalues $\lambda_n$ of the transfer matrix
$\bm T$ but not on its eigenstates. By making use of the Cauchy-Schwarz
inequality one gets a similar expression for CTQW. The average return
probabily $\overline{\pi}(t)$ is related to the average return amplitude
$\overline{\alpha}(t)$ by \cite{mb2006b}
\bea
\overline{\pi}(t) &=& 
\frac{1}{N}\sum_{j}
\Big|\alpha_{j,j}(t)\Big|^2 
\nonumber \\
&\geq&
\Big|\frac{1}{N}\sum_{j} \alpha_{j,j}(t)\Big|^2 \equiv
|\overline{\alpha}(t)|^2.
\label{piavg}
\eea
Similar to $\overline{p}(t)$ for CTRW, $|\overline{\alpha}(t)|^2$ only
depends on the eigenvalues $E_n$ of the Hamiltonian $\bm H$:
\bea
|\overline{\alpha}(t)|^2 &=& 
\Big|\frac{1}{N} \sum_{n} e^{-iE_n t} 
\langle \Psi_n | \sum_{j} | j \rangle \langle j | \Psi_n \rangle \Big|^2
\nonumber \\
&=& 
\Big|\frac{1}{N} \sum_{n} \ e^{-iE_nt}\Big|^2.
\label{pilb}
\eea
Having a quantity which only requires the calculation of the eigenvalues
considerably shortens the computation time, especially for very large
systems. 

Clearly, due to the oscillating terms in Eq.(\ref{pilb}), the lower bound
$|\overline{\alpha}(t)|^2$ will oscillate in most cases. Thus, to compare
to the decay of the classical $\overline{p}(t)$ one uses the envelope of
$|\overline{\alpha}(t)|^2$. Note that when one identifies the Hamiltonian
of the CTQW with the transfer matrix of the CTRW, the eigenvalues of both
are the same, i.e., $E_n = \lambda_n$. Thus, formally the difference in
dynamics is only due to the different functional form of Eq.(\ref{pavg})
and Eq.(\ref{pilb}). Nonetheless, this can lead to drastic effects.

As already discussed, for CTRW the long-time limit of the transition
probabilies reaches the equipartition value $1/N$. In the same way, the
long-time limit of $\overline{p}(t)$ is given by $1/N$. In contrast, for
CTQW neither $\overline{\pi}(t)$ nor $|\overline{\alpha}(t)|^2$ decay to a
given value at long times, but rather oscillate around the corresponnding
long-time average which for $\bar{\pi}(t)$ is given by \cite{m2007a}
\bea \label{eq:chi_bar}
\bar{\chi} &\equiv& \lim_{T \rightarrow \infty} \frac{1}{T} \int_0^T
dt \; \bar{\pi}(t) \nonumber \\
&=& \frac{1}{N} \sum_{n,m}
\delta_{\lambda_n,\lambda_m}|\langle j | \psi_n \rangle|^2 \;
|\langle j | \psi_m \rangle|^2.
\eea
Again, on can obtain a lower bound which does not depend on the
eigenvectors \cite{m2007a}:
\begin{equation} \label{eq:chi_bar_lb}
\bar{\chi} \geq \frac{1}{N^2} \sum_{n,m}
\delta_{\lambda_n, \lambda_m} \equiv \bar{\chi}_{lb}.
\end{equation}

\subsection{Examples}
\label{sec_effex}

\begin{figure}
\centerline{\includegraphics[clip,width=0.4\columnwidth]{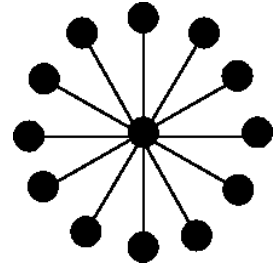}}
\caption{Star with $N-1=12$ arms.  }
\label{star-1}
\end{figure}

The difference in the global efficiencies of CTQW and CTRW is illustrated
by considering two distinct examples: the discrete ring of $N$ nodes and
the star with one core node and $N-1$ nodes attached to it, see
Fig.~\ref{star-1}.

The eigenvalues (and also eigenstates) of the discrete ring have already
been discussed. From these results it is straightforward to
calculate the average return probabilities. Since the ring's eigenstates
are Bloch states, one can easily verify that the lower bound
$|\overline{\alpha}(t)|^2$ is exact, i.e., in this case also
$\overline{\pi}(t)$ does not depend on the eigenstates.  Therefore, the
average return probabilities are given by \cite{mb2006b}
\be
\overline{\pi}(t) =  \Big| \frac{1}{N} \sum_{n=1}^N \exp[-i2t(1-\cos(2\pi
n/N)) \Big|^2
\ee
for CTQW and
\be
\overline{p}(t) = \frac{1}{N} \sum_{n=1}^N \exp[-2t(1-\cos(2\pi n/N))] 
\ee
for CTRW.  In the limit of large $N$, i.e., for continuous $\theta=2\pi
n/N$, one can replace the sums by integrals which leads to
\be
\overline{p}_\gamma(t) = \int dE \ \rho(E) \
\exp(-E t),
\label{pclavginf} 
\ee
and to
\be
\overline{\pi}_\gamma(t) = \Big| \int dE \ \rho(E) \
\exp(-i E t) \Big|^2,
\label{pqmavginf}
\ee
where $\rho(E)$ is the density of states (DOS).  Then one obtains for CTQW
$\overline{\pi}(t)\sim|J_0(2t)|^{2}$, which for  $t\gg1$ can be
approximated by $\overline{\pi}(t)\sim \sin^{2}(2t+\pi/4)/t$
\cite{gradshteyn}. Since the maximum of the ${\rm sine}$ function is $1$,
the envelope of $\overline{\pi}(t)$ decays as $t^{-1}$. For CTRW there is
only a single sum and no additional quadrature, such that a similar
calculation leads to $\overline{p}(t)\sim t^{-1/2}$. This temporal decay
will also be present for finite systems. The time-range over which it will
be visible depends on the value of $N$. Obviously, for small $N$ the
long-time behavior will be reached faster than for larger $N$.

Thus, we just established that for CTQW on a ring the exponent of the
decay power-law is twice as large as the one for CTRW. This fact holds for
a wide class of systems. Namely, all networks whose
density of states $\rho(E) \equiv \frac{1}{N}\sum_{n=1}^N \delta(E-E_n)$
follows a power-law, $\rho(E) \sim (E E_m - E^2)^\nu$, will show this
feature. Here, $E_m$ is the maximal eigenvalue (one assumes the minimum
eigenvalue to be zero). Since the interest is in the large $t$ behavior,
$\overline{p}(t)$ will be mainly determined by the small eigenvalues, such
that for $t\gg1$ one has $\rho(E) \sim E^\nu$. It is straighforward to
show that for CTRW 
\be
\overline{p}(t)\sim t^{-(1+\nu)}.
\label{pl_cl}
\ee
This scaling feature at not too short times is well known, see, e.g.,
\cite{alexander1981}, where $2(1+\nu)\equiv d_s$ is sometimes called the
spectral or fracton dimension.  For CTQW, also $|\overline{\alpha}(t)|^2$
will be in general determined by the small eigenvalues. For $\rho(E) \sim
E^\nu$ one obtains $|\overline{\alpha}(t)|=\overline{p}(t)$. Here, all
quantum mechanical oscillations vanish, because one considers only the
leading term of the DOS.  Therefore, the envelope of the lower bound for
CTQW scales as \cite{mb2006b}
\be
\mbox{env}[|\overline{\alpha}(t)|^2] \sim t^{-2(1+\nu)}.
\label{pl_qm}
\ee

In other cases of interest (e.g., stars, see below) the DOS has highly
degenerate eigenvalues \cite{m2007a}. As an extreme case there may exist a
single eigenvalue $E_l$, whose degeneracy, $D_l$, is of the order ${\cal
O}(N)$, whereas all others are of the order ${\cal O}(1)$ or less.  By
writing
\be
\overline{\alpha}(t) = \frac{1}{N}\Bigg[ D_l \  e^{- iE_lt} +
\sum_{E_n\neq
E_l} \ D_n \ e^{-iE_nt} \Bigg]
\ee
one obtains, up to order ${\cal O}(1/N^2)$ \cite{m2007a}:
\be
\left| \overline{\alpha}(t) \right|^2 \approx \frac{D_l}{N^2} \Bigg\{ D_l
+ \sum_{E_n\neq E_l} D_n \ 2\cos[(E_l - E_n) t] \Bigg\}.
\label{alpha_deg}
\ee
The first term on the right-hand side of eq.~(\ref{alpha_deg}) is of order
${\cal O}(1)$, while the second term is of order ${\cal O}(1/N)$.
Therefore, for few highly degenerate eigenvalues, the lower bound $\left|
\overline{\alpha}(t) \right|^2$ will not show a decay to values which
fluctuate about $1/N$ but will rather fluctuate around $1-1/N$ at all
times. Also, $\overline{\pi}(t)$ will not decay but will fluctuate around
the same value, since $\left| \overline{\alpha}(t) \right|^2$ is a lower
bound.

An example of a system with a single highly degenerate eigenvalue is the star,
having one core node and $N-1$ nodes directly connected to the core but
not to each other, see Fig.~\ref{star-1}. The eigenvalue spectrum has a
very simple structure. There are $3$ distinct eigenvalues, namely $E_1=0$,
$E_2=1$, and $E_3=N$, with degeneracies $D_1=1$, $D_2=N-2$, and
$D_3=1$, respectively.  Therefore, one gets \cite{m2007a}
\begin{eqnarray}
\overline{p}(t) &=& \frac{1}{N} \left[ 1 + (N-2) {\rm e}^{-t} + {\rm
e}^{-(N-2)t}
\right] \\
|\overline{\alpha}(t)|^2 &=& \frac{1}{N^2} \left| 1 + (N-2) {\rm e}^{-{\rm
i}t} +
{\rm e}^{-{\rm i}(N-2)t} \right|^2. \quad
\label{pi_star}
\end{eqnarray}
Obviously, only the term $|(N-2)\exp(-{\rm i}t)|^2/N^2 = (N-2)^2/N^2$ in
Eq.\ (\ref{pi_star}) is of order ${\cal O}(1)$.  All the other terms are
of order ${\cal O}(1/N)$ or ${\cal O}(1/N^2)$ and, therefore, cause only
small oscillations (fluctuating terms) around or negligible shifts
(constant terms) from $(N-2)^2/N^2 \approx 1-1/N$.

\begin{figure}
\centerline{\includegraphics[clip=,width=0.95\columnwidth]{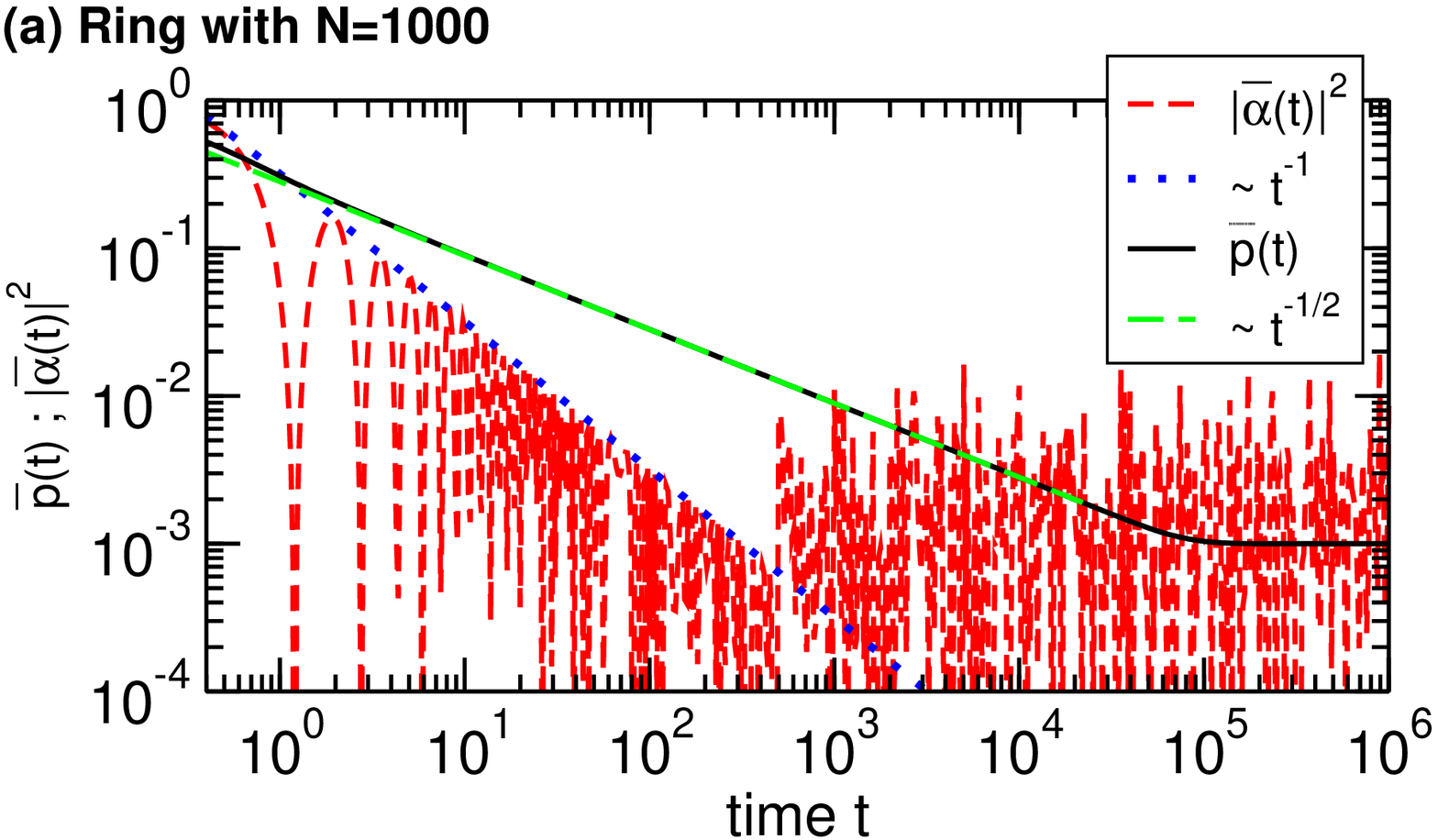}}
\centerline{\includegraphics[clip=,width=0.95\columnwidth]{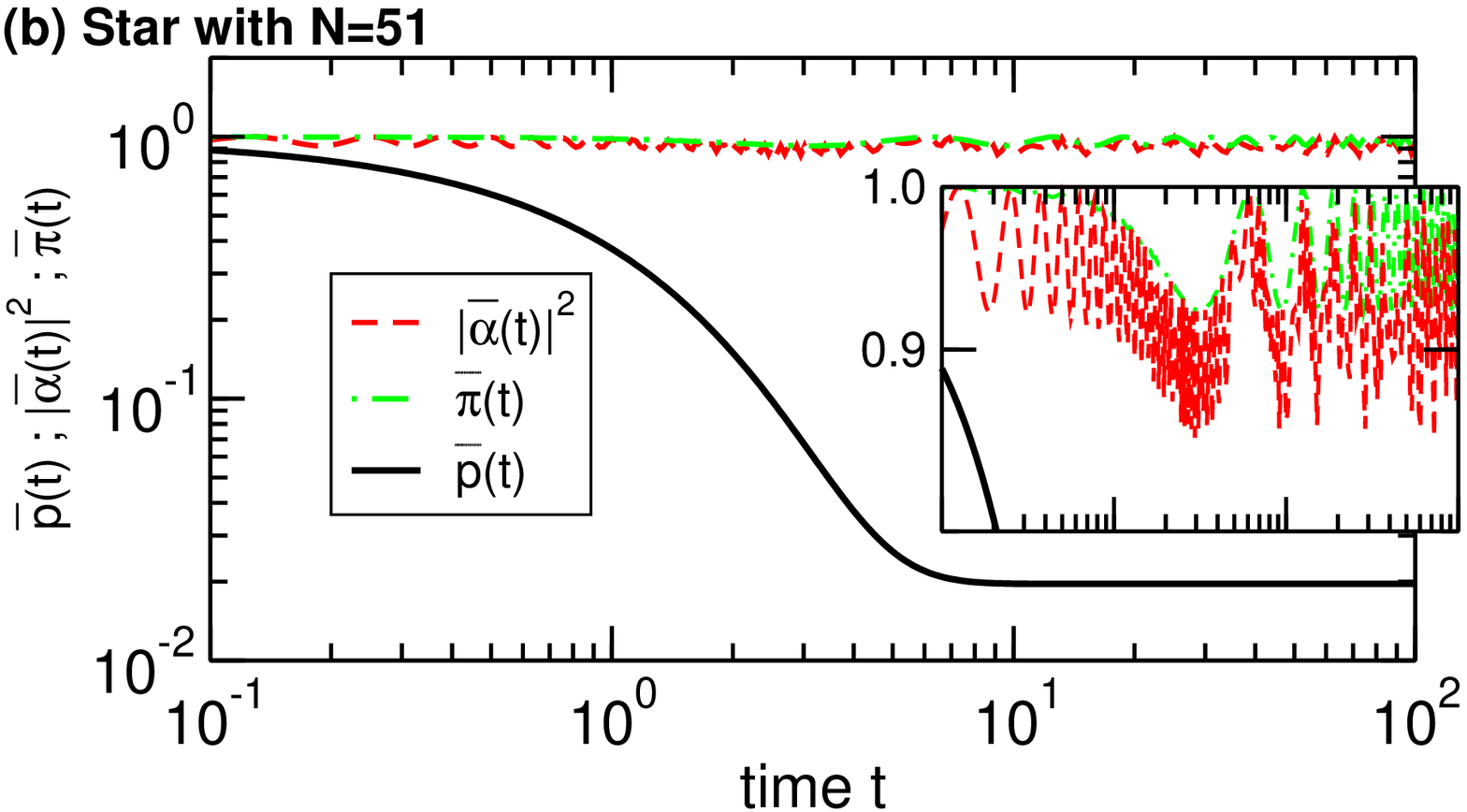}}
\caption{(Colour on-line). Panel (a): $\overline{p}(t)$ and
$|\overline{\alpha}(t)|^2$ with the appropriate scaling $t^{-1/2}$ and
$t^{-1}$, respectively, for a ring of size $N=1000$.  Panel (b):
$\overline{p}(t)$, $\overline{\pi}(t)$, and $|\overline{\alpha}(t)|^2$ for
a star with $N=51$ nodes. The inset shows a close-up of
$\overline{\pi}(t)$ and $|\overline{\alpha}(t)|^2$ in the same time
interval. From \cite{m2007a}.  }
\label{ring-star}
\end{figure}

Figure \ref{ring-star} shows the temporal behavior of $\overline{\pi}(t)$,
$|\overline{\alpha}(t)|^2$, and $\overline{p}(t)$ for a ring (uppper
panel) and for a star (lower panel). The difference, especially for CTQW,
is dramatic. While in both cases the CTRW decay to the equipartition value
$1/N$, this is not so for the CTQW. For the ring
$|\overline{\alpha}(t)|^2$ (thus also $\overline{\pi}(t)$) follows the
$t^{-1}$ decay at intermediate times and oscillates around the long-time
average at long times. For the star, on the other hand, there is no decay
for both $|\overline{\alpha}(t)|^2$ and $\overline{\pi}(t)$.  There are
only some oscillations around the value $1-1/N$.

In the two examples above the efficiencies of the two processes are vastly
different. For the ring there is a considerable increase in efficiency for
CTQW, because for it the probability to return to the origin decays much
faster than for CTRW. The contrary is true for the star. Here, the CTRW
still decay while the CTQW remain close to unity. This implies that the
probability to visit other places than the initial node is - on average -
very low. Note that $\overline{\pi}(t)$, $|\overline{\alpha}(t)|^2$, and
$\overline{p}(t)$ are averaged over all nodes of the network. For the ring
all nodes are equivalent, but the star has the core node which stands out
from the rest. Therefore, the temporal evolution is very different when
the CTQW starts at the core or at one of the peripheral nodes. When
starting at the core, due to the rotational symmetry of the network, the
walk can be mapped into a walk involving only two nodes. When starting at
the periphery, there is no such simple mapping. Since all but one node are
at the periphery, they lead to the poor performance of the CTQW when
compared to the CTRW.

\begin{figure}
\centerline{\includegraphics[clip,width=0.5\columnwidth]{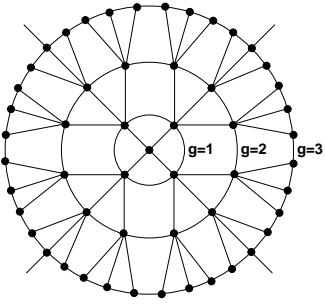}}
\caption{Spidernet graph of generation $g=3$ with $N=53=1+4\sum_{k=1}^g
3^{k-1}$ nodes.  }
\label{spidernet}
\end{figure}

Salimi studied the situation over the semi-regular spidernet graph
\cite{salimi2009c}. This graph is build in a hierarchical, radially
symmetric manner, i.e., one starts from a single core node and builds up
the network generation after generation, as indicated in
Fig.~\ref{spidernet}. The procedure used \cite{mb2006b,m2007a,salimi2009c}
also leads to power-law behaviors for both CTQW and CTRW
\cite{salimi2009c}. When one starts at the core node, one can map the
dynamics onto a line, where all the states corresponding to the nodes
belonging to the same generation are summed up to form a new state,
representing this generation. Salimi showed that in this case the return
to the origin (core node) follows a power-law which for CTRW goes as
$t^{-3/2}$ while for CTQW it goes as $t^{-3}$. Therefore, a behavior
similar to that discussed around Eqs.~(\ref{pl_cl}) and (\ref{pl_qm}) is
also found here, since the exponent of the CTQW decay is twice as large as
the exponent of the CTRW decay.

\section{CTQW on networks}

\subsection{Deterministic networks}
\label{sec-detnet}

\subsubsection{Two dimensional regular networks}

CTQW on regular $2d$ structures carry over many of the properties of their
$1d$ counterparts, since $2d$ regular networks ca be envisaged to be the
direct product of two $2d$ structures, {\it vide infra}. However, some care is
in order, since the symmetries observed in 1d are not always observed in
2d \cite{mvb2005a}.

\begin{figure}
\centerline{\includegraphics[clip,width=0.75\columnwidth]{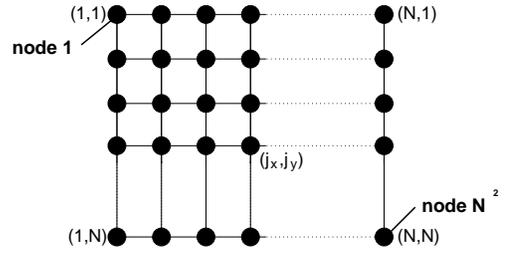}}
\caption{Sketch of a square network arranged as a regular lattice with the
appropriate numbering of the nodes. Note that the actual geometrical
realization can be much more flexible, see text for details. From
\cite{mvb2005a}.  }
\label{lattice}
\end{figure}

Consider now $2d$ regular structures of side length $N$, thus, they
contain $N^2$ nodes giving rise to $N^2$ basis states \cite{mvb2005a}. In
a pair notation one sets $|\boldsymbol j\rangle = |j_x,j_y\rangle$, where
$j_x$ and $j_y$ are integer labels in the two directions, with
$j_x,j_y\in[1,N]$, see Fig.\ref{lattice}. This labeling of the states is
not to be confused with the labeling of the adjacency matrix. Note that
capital bold letters denote matrices, while small bold letters denote the
nodes and the states.

The focus in solid state physics is on systems where Born - von Karman
periodic boundary conditions (PBC) are assumed. Now, for an internal site
of the network (not on an edge or in a corner), the Hamiltonian acting on
a state $|\boldsymbol j\rangle = |j_x,j_y\rangle$ reads
\bea
{\bm H} | j_x,j_y\rangle &=& 2 | j_x,j_y\rangle - | j_x+1,j_y\rangle - |
j_x-1,j_y\rangle \nonumber \\
&+& 2 | j_x,j_y\rangle - | j_x, j_y+1\rangle - | j_x,j_y-1\rangle.
\nonumber \\
\label{honj}
\eea
PBC extend this equation to all the sites of the network by interpreting
every integer $j_x$ and $j_y$ to be taken modulus $N$.  With this
generalization, the time independent SE
\be
{\bm H} | \Psi_{\boldsymbol\theta}\rangle = E_{\boldsymbol\theta}
|\Psi_{\boldsymbol\theta}\rangle
\label{hbloch}
\ee
admits (similar to the 1d case) the following Bloch eigenstates
\be
|\Psi_{\boldsymbol\theta}\rangle = \frac{1}{N} \sum_{j_x,j_y=1}^{N}
e^{-i(\boldsymbol\theta \cdot \boldsymbol j)} |\boldsymbol j\rangle.
\label{blochfct}
\ee
as solutions, where $\boldsymbol\theta\cdot\boldsymbol j$ stands for the
scalar product with $\boldsymbol\theta = (\theta_x,\theta_y)$. The usual
Bloch relation can be obtained by projecting
$|\Psi_{\boldsymbol\theta}\rangle$ on the state $|\boldsymbol j\rangle$
such that $\Psi_{\boldsymbol\theta}(\boldsymbol j) \equiv \langle
\boldsymbol j | \Psi_{\boldsymbol\theta}\rangle = e^{-i(\boldsymbol\theta
\cdot \boldsymbol j)} / N$, thus $\Psi_{\boldsymbol\theta}(j_x+1,j_y+1) =
e^{-i(\theta_x + \theta_y)} \Psi_{\boldsymbol\theta}(j_x,j_y)$. The PBC
restrict the allowed values of $\boldsymbol\theta$.  In the present case
\cite{mvb2005a} (side length $N$), the PBC require that
$\Psi_{\boldsymbol\theta}(N+1,j_y) = \Psi_{\boldsymbol\theta}(1,j_y)$ and
$\Psi_{\boldsymbol\theta}(j_x,N+1) = \Psi_{\boldsymbol\theta}(j_x,1)$. It
follows that one must have $\theta_x = 2n\pi/N$ and $\theta_y=2l\pi/N$,
where $n$ and $l$ are integers and $n,l\in[1,N]$.  It is now a simple
matter to verify that the $|\Psi_{\boldsymbol\theta}\rangle$ also obey
$\langle\Psi_{\boldsymbol\theta}|\Psi_{\boldsymbol\theta'}\rangle =
\delta_{\boldsymbol{\theta,\theta'}}$ and $\sum_{\boldsymbol\theta}
|\Psi_{\boldsymbol\theta}\rangle\langle\Psi_{\boldsymbol\theta}| =
\boldsymbol 1$. 

Furthermore, from Eqs.(\ref{hbloch}) and (\ref{blochfct}) the energy is
obtained as
\be
E_{\boldsymbol\theta} = 4 - 2\cos\theta_x - 2\cos\theta_y = E_{\theta_x} +
E_{\theta_y},
\label{bloch_ev}
\ee
with $E_{\theta_x} = 2 -2\cos\theta_x$ and $E_{\theta_y} = 2 -
2\cos\theta_y$. Under PBC the two-dimensional eigenvalue problem separates
into two one-dimensional problems.

The transition amplitude at time $t$ from state $|\boldsymbol j\rangle$ to
state $|\boldsymbol k\rangle$ is now, using Eq.(\ref{blochfct}) twice,
\cite{mvb2005a}:
\bea
\alpha_{\boldsymbol{k,j}}(t)
&=&
\frac{1}{N^2} \sum_{\boldsymbol{\theta,\theta'}}
\langle\Psi_{\boldsymbol\theta'} | e^{-i(\boldsymbol\theta' \cdot
\boldsymbol k)} e^{-i{\bm
H}t} e^{i(\boldsymbol \theta \cdot \boldsymbol j)} |\Psi_{\boldsymbol
\theta}\rangle
\nonumber \\
&=&
\frac{1}{N^2} \sum_{\boldsymbol{\theta}} e^{-i E_{\boldsymbol\theta}t}
e^{-i
\boldsymbol \theta \cdot (\boldsymbol k - \boldsymbol j)}
\label{transamplbloch}
\eea
In the limit $N\to\infty$, the sums in Eq.(\ref{transamplbloch}) may be
changed to integrals; by making use of Eq.(\ref{bloch_ev}) one obtains
\bea
\lim_{N\to\infty} \alpha_{\boldsymbol{k,j}}(t)
&=&
\frac{e^{-i4t}}{4\pi^2}
\int\limits_{-\pi}^{\pi} d\theta_x \ e^{-i\theta_x(k_x-j_x)}
e^{i2t\cos\theta_x}
\nonumber \\
&& \times
\int\limits_{-\pi}^{\pi} d\theta_y \ e^{-i\theta_y(k_y-j_y)}
e^{i2t\cos\theta_y}, \nonumber \\
\eea
such that
\be
\lim_{N\to\infty} \alpha_{\boldsymbol{k,j}}(t) = 
i^{k_x-j_x} i^{k_y-j_y} e^{-i4t} J_{k_x-j_x}(2t) J_{k_y-j_y}(2t) 
\ee
where $J_n(x)$ is the Bessel function of the first kind \cite{Ito}. Thus,
on a network topologically equivalent to a square lattice with PBC the
transition amplitude between the nodes $\boldsymbol j$ and $\boldsymbol k$
is given by
\be
\lim_{N\to\infty} \pi_{\boldsymbol{k,j}}(t) = [J_{k_x-j_x}(2t)
J_{k_y-j_y}(2t)]^2.
\ee

Now, for finite networks the long-time average $\chi_{k,j}$ gives more
insight into the dynamics for different network sizes $N$. In the 2d case
one has \cite{mvb2005a}
\begin{subequations}
\begin{align}
\chi_{\boldsymbol{k,j}} &=%& 
\lim_{T\to\infty}\frac{1}{T} \int\limits_0^T dt \ \left|
\sum_{n}\langle \boldsymbol k | e^{-i{\bm H}t} |
\boldsymbol q_n \rangle \langle \boldsymbol q_n |
\boldsymbol j \rangle \right|^2 
\nonumber \\
&=
\sum_{n,m} \langle \boldsymbol k |
\boldsymbol q_n \rangle \langle \boldsymbol q_n |
\boldsymbol j \rangle \langle \boldsymbol j | \boldsymbol q_m
\rangle \langle \boldsymbol q_m | \boldsymbol k \rangle
\nonumber \\
&
\times 
\left( \lim_{T\to\infty}\frac{1}{T} \int\limits_0^T dt \
e^{-i(\lambda_n - \lambda_m)\gamma t} \right)
\label{limprob_ev_int}
\\
&=
\sum_{n,m} 
\delta_{\lambda_n,\lambda_m}
\langle \boldsymbol k |
\boldsymbol q_n \rangle \langle \boldsymbol q_n |
\boldsymbol j \rangle \langle \boldsymbol j | \boldsymbol q_m
\rangle \langle \boldsymbol q_m | \boldsymbol k \rangle.
\label{limprob_ev}
\end{align}
\end{subequations}
One notes that the integral in Eq.(\ref{limprob_ev_int}) equals $1$ if
$\lambda_n = \lambda_m$ and $0$ otherwise, i.e., it equals
$\delta_{\lambda_n,\lambda_m}$.  Given that some eigenvalues of $\bm H$
are degenerate, the sum in Eq.(\ref{limprob_ev}) can contain terms
belonging to different eigenstates $|\boldsymbol q_n\rangle$ and
$|\boldsymbol q_m\rangle$. Equation (\ref{limprob_ev}) provides a
numerically very efficient way of computing the $\chi_{\boldsymbol{k,j}}$.
Remarkably, one finds that the $\chi_{\boldsymbol{k,j}}$ depend in an
unexpected way on the exact value of the size $N$ of the finite network
under study.

\begin{figure}
\centerline{\includegraphics[clip,width=\columnwidth]{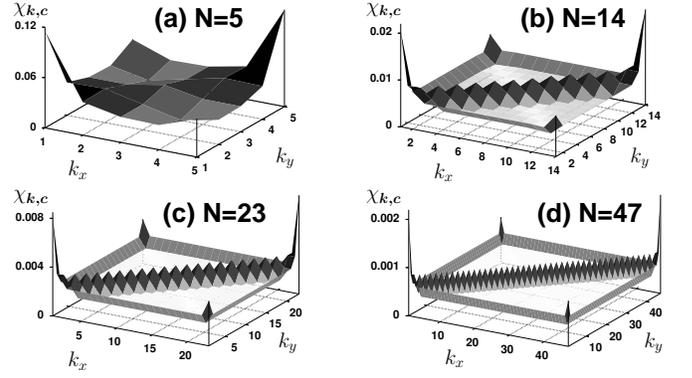}}
\caption{LPs $\chi_{\boldsymbol{k,c}}$ to be at node $\boldsymbol k$ when
starting at the corner node $\boldsymbol c = (1,1)$ for networks of sizes
(a) $N=5$, (b) $N=14$, (c) $N=23$, and (d) $N=47$. From \cite{mvb2005a}.
}
\label{limprobsym}
\end{figure}

When starting at a corner node $\boldsymbol c=(1,1)$, one often finds that
the LPs for the starting node and its ``mirror'' node
$\boldsymbol{oc}=(N,N)$ are equal.  Figure~\ref{limprobsym} shows the
$\chi_{\boldsymbol{k,c}}$ obtained by going from the corner node
$\boldsymbol c=(1,1)$ to the other nodes for networks of sizes $N=5$,
$N=14$, $N=23$, and $N=47$.

\begin{figure}
\centerline{\includegraphics[clip,width=\columnwidth]{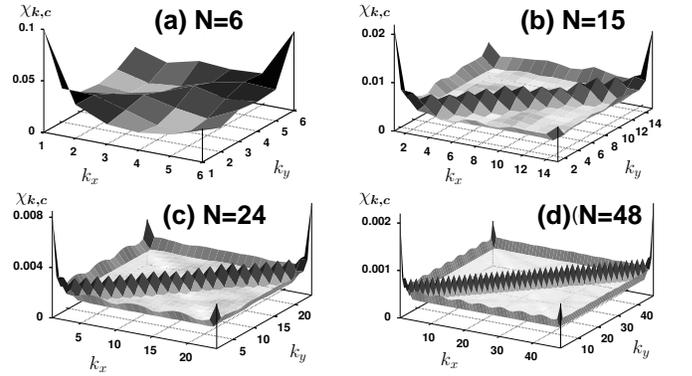}}
\caption{LPs $\chi_{\boldsymbol{k,c}}$ to be at node $\boldsymbol k$ when
starting at the corner node $\boldsymbol c = (1,1)$ for networks of sizes
(a) $N=6$, (b) $N=15$, (c) $N=24$, and (d) $N=48$.  One may note the
asymmetries by comparing to Fig.\ref{limprobsym}. From \cite{mvb2005a}.  }
\label{limprobantisym}
\end{figure}

However, for some particular network sizes the distributions of the LPs
turn out to be asymmetric. For instance, for a network of size $N=15$ the
LP $\chi_{\boldsymbol{oc,c}}$ for the CTQW starting at node $\boldsymbol
c$ to be at the opposite corner node $\boldsymbol{oc}$ is less than the LP
$\chi_{\boldsymbol{c,c}}$ to be at the initial node. The same is true for
the nodes along the edges of the network.  Figure \ref{limprobantisym}
shows that such asymmetries occur for networks of sizes $N=6$, $N=15$,
$N=24$, and $N=48$ (the asymmetries are best seen by looking at
$\chi_{\boldsymbol{c,c}}$ and $\chi_{\boldsymbol{oc,c}}$).  The smallest
network where asymmetries in the distribution of the LPs are detected has
$N=6$. The next ones are found for $N=12, 15, 18, 21, 24, 30, 36, \cdots$.

\begin{figure}
\centerline{\includegraphics[clip,width=0.95\columnwidth]{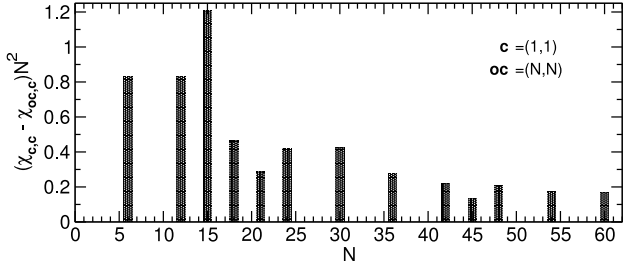}}
\caption{Differences between the LPs for CTQW that start at $\boldsymbol
c=(1,1)$ to be at $\boldsymbol c$, $\chi_{\boldsymbol c,\boldsymbol c}$,
or to be at its ``mirror'' node $\boldsymbol{oc}=(N,N)$,
$\chi_{\boldsymbol{oc},\boldsymbol c}$, as a function of the network size
$N$, for $1\leq N \leq60$. From \cite{mvb2005a}.  }
\label{limprobdiff}
\end{figure}

An asymmetric LP distribution is particularly evident in the difference
between $\chi_{\boldsymbol{c,c}}$ and $\chi_{\boldsymbol{oc,c}}$.  Thus,
as an overview Fig.\ref{limprobdiff} presents as a function of $N$ a plot
of the $(\chi_{\boldsymbol{c,c}} - \chi_{\boldsymbol{oc,c}}) N^2$ values
obtained. Note that all $N$ values in Fig.\ref{limprobdiff} for which
$(\chi_{\boldsymbol{c,c}} - \chi_{\boldsymbol{oc,c}}) \neq 0$ are
divisible by $3$.  However, the converse is not true, one finds symmetric
LP distributions for the networks with $N=3, 9, 27, 33, 39, \cdots$. 

The above effect is also observed for non-square networks, where the
number of nodes in the two directions are not equal \cite{vmb2006a}. While
for periodic boundary conditions in both directions there is no asymmetry,
there are observable asymmetries both for periodic and for open boundary
conditions in only one direction as well as for open boundary conditions
in both directions.

\begin{figure}[t]
\centerline{\includegraphics[clip=,angle=0,width=0.75\columnwidth]{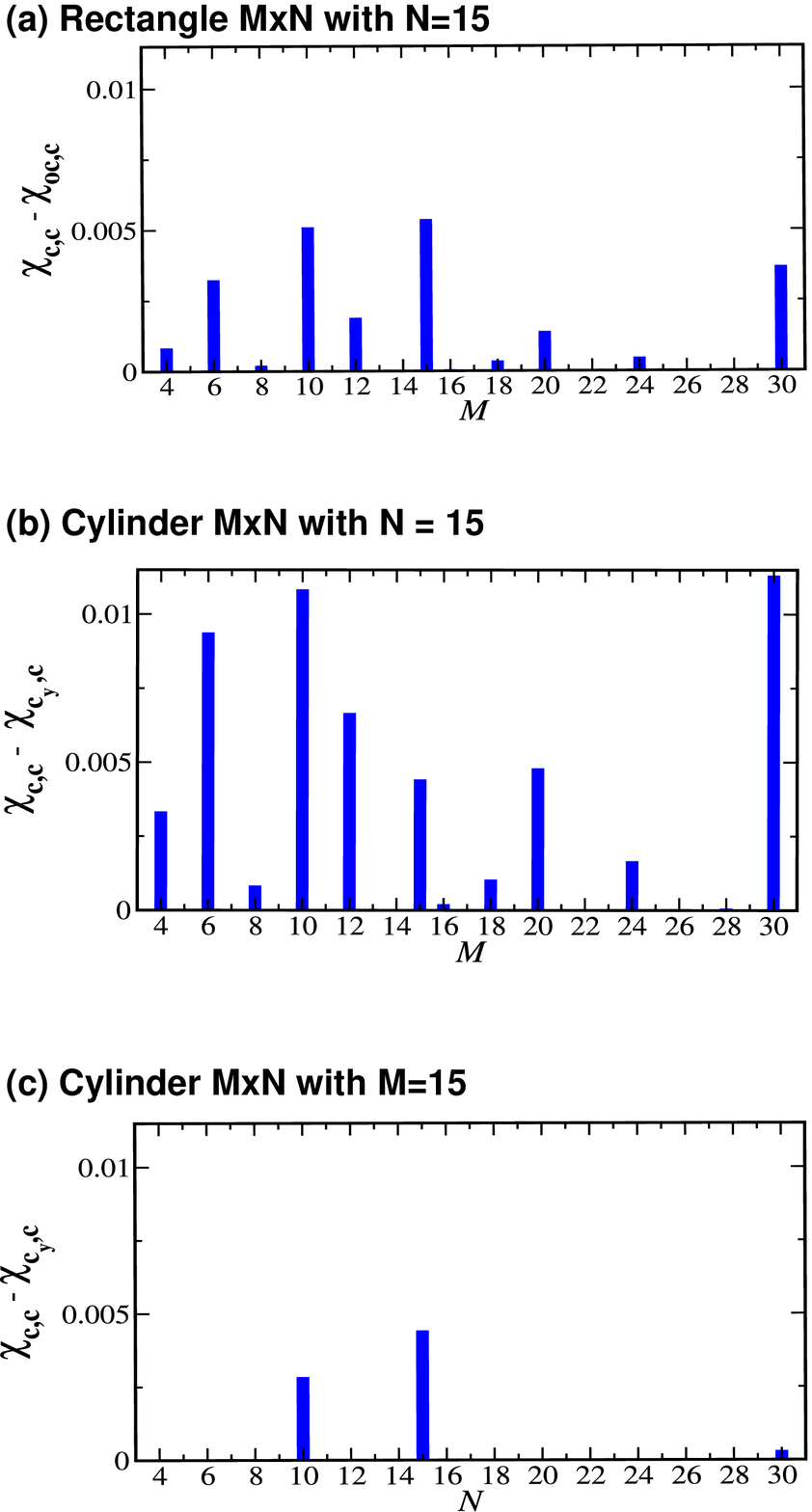}}
\caption{(a) Rectangles $M\times N$ with $N=15$ and varying $M$:
Differences between the LPs for CTQW that start at $\boldsymbol{c}=(1,1)$
to be at $\boldsymbol{c}$, $\chi_{\boldsymbol{c,c}}$, and to be at
$\boldsymbol{oc}=(M,N)$, $\chi_{\boldsymbol{oc,c}}$. (b) Cylinders
$M\times N$ with $N=15$  and varying $M$: Differences between the LPs for
CTQW that start at $\boldsymbol{c}=(1,1)$ to be at $\boldsymbol{c}$, and
to be at the node $\boldsymbol{c_y}=(1,N)$. (c) Cylinders $M\times N$ with
$M=15$  and varying $N$: Differences between the LPs for CTQW that start
at $\boldsymbol{c}=(1,1)$ to be at $\boldsymbol{c}$, and to be at the node
$\boldsymbol{c_y}=(1,N)$. From  \cite{vmb2006a}.  }
\label{chicoc}
\end{figure}

One now fixes $N$ to be $N=15$ and varies $M$, taking $4\leq M\leq 30$.
Figure \ref{chicoc}(a) shows for rectangles the difference between the LPs
on the initial corner $\chi_{\boldsymbol{c,c}}$ and on the corner
$\chi_{\boldsymbol{oc,c}}$. In the analysed range, $4\leq M\leq 30$, the
value $\chi_{\boldsymbol{c,c}}-\chi_{\boldsymbol{oc,c}}$ displays varying
patterns. For $4 \leq M\leq13$ one can associate
$\chi_{\boldsymbol{c,c}}-\chi_{\boldsymbol{oc,c}}=0$ to odd values of $M$
and $\chi_{\boldsymbol{c,c}}-\chi_{\boldsymbol{oc,c}}\neq0$ to even values
of $M$; for larger $M$ the situation becomes more complex. Figure
\ref{chicoc}(b) displays the situation for cylinders with $N=15$ and
$4\leq M\leq 30$. Here the plot shows
$\chi_{\boldsymbol{c,c}}-\chi_{\boldsymbol{c_y,c}}$, with $c_y\equiv
(1,N)$. Figure \ref{chicoc}(c) shows the situation for cylinders with
$M=15$ and $4\leq N\leq 30$. This last case looks quite regular, with
non-vanishing values only for $N=10, 15,$ and $30$. Hence,
Figs.~\ref{chicoc}(b) and \ref{chicoc}(c) ($N<M$) show for cylinders that
changes in radius lead to more asymmetric situations than changes in
length.

An indication of the origin of the asymmetries is obtained by briefly
reviewing some of the details of the calculations of the LPs. Since the
Hamiltonian of the problem separates in the two directions, it is easy to
show that the transition probabilities $\pi_{{\boldsymbol k},{\boldsymbol
j}}(t)$ can be written as the product of the two separate probabilities
for each direction \cite{vmb2006a}, i.e.
\bea
\pi_{{\boldsymbol k},{\boldsymbol j}}(t) &=& \pi_{k_x, j_x}(t) \ \pi_{k_y,
j_y}(t)
\nonumber \\
&=& |\alpha_{k_x, j_x}(t)|^2 \ |\alpha_{k_y, j_y}(t)|^2,
\eea
with $\alpha_{k_x, j_x}(t) = \sum_{\theta_x} \exp(-i\lambda_{\theta_x}t)
\langle k_x | \Psi_{\theta_x} \rangle \langle \Psi_{\theta_x} | j_x
\rangle$ , and similarly for the $y$-direction.\\ Now, according to
Eq.~(\ref{limprob_ev_int}), the LPs are given by \cite{vmb2006a}
\begin{eqnarray}
\chi_{{\boldsymbol k},{\boldsymbol j}} &=& \lim_{T\to\infty}\frac{1}{T}
\int_0^T dt \
\pi_{k_x,
j_x}(t) \ \pi_{k_y, j_y}(t) \nonumber \\
&=& \sum_{\theta_x,\theta'_x,\theta_y, \theta'_y} F_{\boldsymbol{k,j}}
\lim_{T\to\infty} \frac{1}{T} \int_0^T dt 
\nonumber \\
&& \times 
\exp\left[-it(\lambda_{\theta_x}
- \lambda_{\theta'_x} + \lambda_{\theta_y} -
  \lambda_{\theta'_y})\right], \nonumber \\
\label{xi_asymm}
\end{eqnarray}
where $F_{\boldsymbol{k,j}}$ is a time independent function, which depends
on the eigenstates associated with $\theta_x$, $\theta'_x$, $\theta_y$,
and $\theta'_y$. Because of the limit in the time integral in
Eq.~(\ref{xi_asymm}), there are only contributions to $\chi_{{\boldsymbol
k},{\boldsymbol j}}$ if a value $(\lambda_{\theta_x} -
\lambda_{\theta'_x})$ for the $x$-direction has a counterpart
$-(\lambda_{\theta_y} - \lambda_{\theta'_y})$ in the $y$-direction.

A careful analysis of the differences $(\lambda_{\theta_x} -
\lambda_{\theta'_x})$ indicates where the asymmetries stem from. For
finite chains one obtains $(\lambda_{\theta_x} - \lambda_{\theta'_x}) =
2\cos \theta'_x - 2\cos \theta_x$. For simplicity one considers now finite
$N\times N$ networks with OBCs, see \cite{mvb2005a}, because then the
eigenvalues are the same in both directions. It turns out that for
$\theta_x\neq\theta'_x$ the value $(\lambda_{\theta_x} -
\lambda_{\theta'_x})$ appears only once or twice for all symmetric cases.
However, for the asymmetric cases, some of the $(\lambda_{\theta_x} -
\lambda_{\theta'_x})$ values (again for $\theta_x\neq\theta'_x$) appear
more than twice. Therefore, there are more contributions to
$\chi_{{\boldsymbol k},{\boldsymbol j}}$ in the asymmetric cases than in
the symmetric cases. 

\subsubsection{Star-like networks}

Networks which are not regular but, yet, have symmetry properties which
can be exploited, are star-like networks. As already mentioned in
Sec.~\ref{sec_effex} and also discussed in \cite{mb2006b,m2007a}, these
networks consist of a central core node (with label $1$) to which each of
the remaining $N-1$ nodes is attached by an individual bond, see
Fig.~\ref{star-1}. The Hamiltonian has then the following structure 
\be
\bm H = (N-1) |1\rangle\langle 1| + \sum_{j=2}^N \Big( |j\rangle\langle j|
- |1\rangle\langle j| - |j\rangle \langle 1| \Big).
\ee
Using the Gram-Schmidt orthonormalization procedure, Xu obtained the
eigenstates of $\bm H$ \cite{xu2009a}
\be
|\Psi_n\rangle = 
\begin{cases}
\displaystyle \frac{1}{\sqrt{n+1}}\Big(
\sqrt{n}
|n+2\rangle - \frac{1}{\sqrt{n}}
\sum_{m=2}^{n+1} |m\rangle \Big) \\ \hfill \mbox{for } n< N-1 \\
\displaystyle
\frac{1}{\sqrt{N}} \sum_{m=1}^{N} |m\rangle \\ \hfill \mbox{for }
n=N-1 \\
\displaystyle
\frac{1}{\sqrt{N-1}}  \Big(\frac{1}{\sqrt{N}}\sum_{m=1}^{N}  |m\rangle -
\sqrt{N} |1\rangle \Big) \\ \hfill \mbox{for }
n=N
\end{cases}
\ee

This yields analytic expressions for the transition probabilities, e.g., 
\be
\pi_{1,1}(t) = \frac{N^2-2N+2}{N^2} + \frac{2(N-1)}{N^2} \cos(Nt).
\ee
One easily verifies that for $\cos(Nt) =1$, i.e., $t=2\pi r/N$ (where $r$
is an integer), there is a perfect revival. Moreover, for $t=(2r+1)\pi/N$
all the probabilitiy is distributed over all but the core node. This is
not true if one starts the CTQW at any of the other $(N-1)$ nodes, then
one has 
\bea
\pi_{2,2}(t) &=& \Big[ \big(N^4-4N^3+5N^2-2N+2\big) 
\nonumber \\
&+& \big(2N^3 - 6N^2 +4N\big) \cos(t) 
\nonumber \\
&+& \big(2N^2-4N\big) \cos((N-1)t) 
\nonumber \\
&+& (2N-2) \cos(Nt) \Big] \frac{1}{N^2(N-1)^2}, \nonumber \\
\eea
where, without loss of generality, the node $2$ was chosen as initial
node.  One sees, especially for large $N$, that the probabilities are
mainly localized on the initial node.  For all initial conditions, the
classical transition probabilities $p_{k,j}(t)$ approach the equipartition
value $1/N$. 

Having determined the probabilities to return or to still be at the origin
allows to calculate the average return probability $\overline{\pi}(t)$,
see Sec.~\ref{sec-eff}. One has thus
\be
\overline{\pi}(t) = \frac{1}{N} \sum_{k=1}^N \pi_{k,k}(t) = \frac{1}{N}\Big[\pi_{1,1}(t) + (N-1)
\pi_{2,2}(t)\Big],
\ee
which leads to the results given in Eq.~(\ref{pi_star}).

The strong dependence on the initial conditions also carries over to the
long-time averages $\chi_{k,j}$. One obtains namely
\be
\begin{split}
\chi_{1,1} &= (N^2-2N+2)/N^2, \\
\chi_{2,1} &= 2/N^2, \\
\chi_{2,2} &= (N^4-4N^3+5N^2-2N+2)/\big[N^2(N-1)^2\big], \\
\chi_{3,2} &= 2(N^2-N+1)/\big[N^2(N-1)^2\big]. 
\end{split}
\ee
Thus, for large $N$ the probability will be concentrated at the initial node.

\begin{figure}
\centerline{\includegraphics[clip,width=0.95\columnwidth]{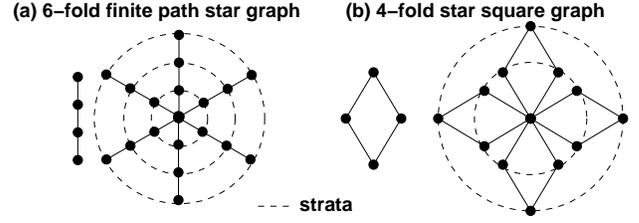}}
\caption{(a) Star with $6$ arms composed of finite segments of length
$4$. (b) Star with $4$ arms composed of circles of length $4$. The
dashed lines are guides to the eye, connecting all nodes belonging to given
strata, see text.  }
\label{stargraph}
\end{figure}

In a slightly more general setup, Salimi considered CTQW on networks
(called {\sl star graphs} in \cite{salimi2009}) with are built from
several sub-networks such that all sub-networks share a single node, see
Fig.~\ref{stargraph}. The adjacency matrix is then a direct product of the
separate adjacency matrices with additonal entries for the newly created
bonds.

By considering so-called strata, i.e., sets formed by all the nodes at the
same chemical distance from the central node, see the dashed lines in
Fig.~\ref{stargraph}, Salimi calculated the transition probabilities to go
from the central node to the different strata. For a star with $N$ arms
each of which having $2$ nodes, see Fig.~2 in \cite{salimi2009}, he
obtains for the transition probabilities
\bea
\pi_{1,1}(t) &=& \left| \frac{1+N\cos(\sqrt{N+1}t)}{N+1}\right|^2, \\
\pi_{2,1}(t) &=& \left|
\frac{\sqrt{N}\sin(\sqrt{N+1}t)}{\sqrt{N+1}}\right|^2, \\
\mbox{and } \pi_{3,1}(t) &=& \left|
\frac{N(1-\cos(\sqrt{N+1}t)}{(N+1)\sqrt{N}}\right|^2 .
\eea 

In the case of a star made out of rings of length $4$, such that all rings
share a single node, there are again 2 strata and the central node, see
Fig.~\ref{stargraph}(b) and \cite{salimi2009}.  The transition
probabilities are given by
\bea
\pi_{1,1}(t) &=& \left| \frac{1+N\cos(\sqrt{2(N+1)}t)}{N+1}\right|^2, \\
\pi_{2,1}(t) &=& \left|
\frac{\sqrt{N}\sin(\sqrt{2(N+1)}t)}{\sqrt{N+1}}\right|^2, \\
\mbox{and } \pi_{3,1}(t) &=& \left|
\frac{N(1-2\cos(\sqrt{N+1}t)}{2(N+1)\sqrt{N}}\right|^2 ,
\eea 
results very similar to the above.

For strata more distant from the core, one can still compute the
transition probabilities for different $N$. Evidently, the $N=1$ case
corresponds to a semi-infinite line, while the $N=2$ case is equivalent to
the infinite line, i.e., the transition probabilities are given by Bessel
functions.

Letting $N$ go to infinity, Salimi showed that in all cases considered
above the transition probabilities reduce to
\bea
\pi_{1,1}(t) &=& \cos^2 t,\\
\pi_{2,1}(t) &=& \sin^2 t,\\
\mbox{and } \pi_{3,1}(t) &=& 0 ,
\eea
which is equivalent to the result for a dimer, i.e., a complete graph
consisting of two nodes.

\subsubsection{Complete graph}

The complete graph (for finite $N$), where all nodes are mutually
connected with each other, shares some properties with the star graph. The
Hamiltonian now reads
\be
\bm H = (N-1)\sum_{j=1}^N |j\rangle\langle j| - \sum_{k\neq j}
|j\rangle\langle k|.
\ee
The graph has two distinct eigenvalues $E_N=0$ and $E_n=N$ for
$n=1,\dots,N-1$.  Xu showed that, similar to the star
\cite{mb2006b,m2007a,xu2009a}, one can calculate the eigenstates using the
Gram-Schmidt orthonormalization procedure \cite{xu2009a}. For the
transition probabilities he then obtains
\be
\pi_{k,j}(t) = 
\begin{cases}
\displaystyle
\frac{N^2-2N+2}{N^2} + \frac{2(N-1)}{N^2} \cos(Nt) \\ \hfill \mbox{for } k=j \\
\displaystyle
\frac{2}{N^2} - \frac{2}{N^2} \cos(Nt) \\ \hfill \mbox{for } k\neq j.
\end{cases}
\ee
Therefore, the transition probabilities have exactly the same form as the ones
for the star graph, when the excitation starts at the core.

\subsubsection{Dendrimers}

\begin{figure}[b]
\centerline{\includegraphics[clip=,width=0.95\columnwidth]{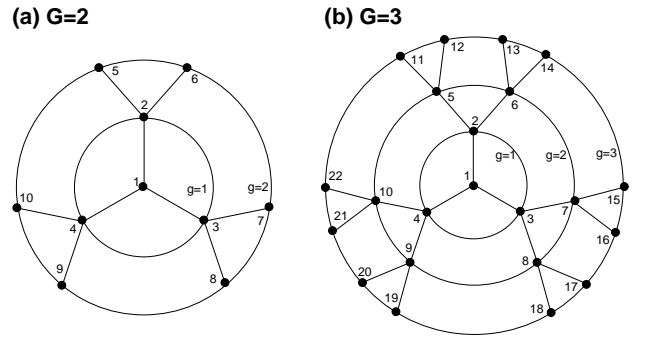}}
\caption{ Dendrimers of functionality $f=3$ and generation $G=2$ (left)
and $G=3$ (right). From \cite{mbb2006a}.  }
\label{topology}
\end{figure}

Star graphs of length $1$ can also be viewed as being dendrimers
(Cayley-trees) of first generation. The stucture of dendrimers is
exemplified in Fig.~\ref{topology} for dendrimers of generations $G=2$ and
$G=3$, with functionality $f=3$, see also \cite{mbb2006a}. In general, the
functionality $f$ gives the number of bonds emanating from each node; the
generation $G$ refers to all the nodes whose shortest distance (in bond
units) from the central node is not larger than $G$. Note that the number
of nodes belonging to the $g$-th generation (where $G\geq g\geq1$) is
$3\cdot2^{g-1}$ and that it grows exponentially with $g$. Moreover, the
total number of nodes in the dendrimer of generation $G$ is
$N=3\cdot2^G-2$.

The connectivity matrix of these dendrimers has a very simple structure.
One has $A_{ii}=3$ for all the nodes in the first $G-1$ generations and
$A_{ii}=1$ for the nodes in generation $G$. The bonds are represented by
the off-diagonal matrix elements $A_{ij}$. Here, every node in generation
$g\ge1$ is connected to two consecutively numbered nodes in generation
$g+1$ and to one node in generation $g-1$.

The eigenmodes of such dendrimers were studied in \cite{cai1997}. There,
for the dendrimers of generations $G=1$ and $G=2$, the eigenvalues and
eigenvectors of ${\bm A}$ were explicitly calculated.  The eigenvectors
determine the eigenmodes of the dendrimer, see e.g.\ Fig.~3 of
\cite{cai1997}. It was further shown that there are $G+1$ nondegenerate
eigenvalues, one of which is always $\lambda_0=0$. 

When an excitation starts at the central node $1$, the dynamics of this
excitation over the dendrimer can be mapped onto a line. Remarkably, for
the $G=2$ dendrimer the transition probabilities are fully periodic when
the coherent excitation starts from the central node $1$ (the same holds
for the $G=1$ dendrimer, too). Note that due to rotational symmetry, the
transition probabilities from the central node to nodes belonging to the
same generation are equal.  Because of this one only has to list three
different transition probabilities. It  follows that there is a perfect
revival of the initial state, which resembles results obained for
continuous \cite{kinzel1995,fgrossmann1997} and discrete quantum carpets
\cite{iwanow2005,mb2005b}. 

If the initial excitation is placed at one of the nodes of the outermost
generation $g=G$ of the dendrimer, the picture changes. Classically, the
propagation through the dendrimer gets to be much slower than in the
previous case, see e.g.\ \cite{barhaim1997,rana2003,heijs2004}.
Nevertheless, eventually the excitation will classically propagate through
the whole graph and in the long time limit the probability will be
equipartitioned among all nodes. Quantum mechanically this effect is even
more dramatic. The main fraction of $\pi_{k,j}(t)$ stays in a small region
closely connected by bonds to the initial node $j$, and the transfer to
other sites is highly unlikely. Also at long times the limiting
probability for the excitation to reach the other branches of the
dendrimer stays very low.

\begin{figure}[t]
\centerline{\includegraphics[clip=,width=0.9\columnwidth]{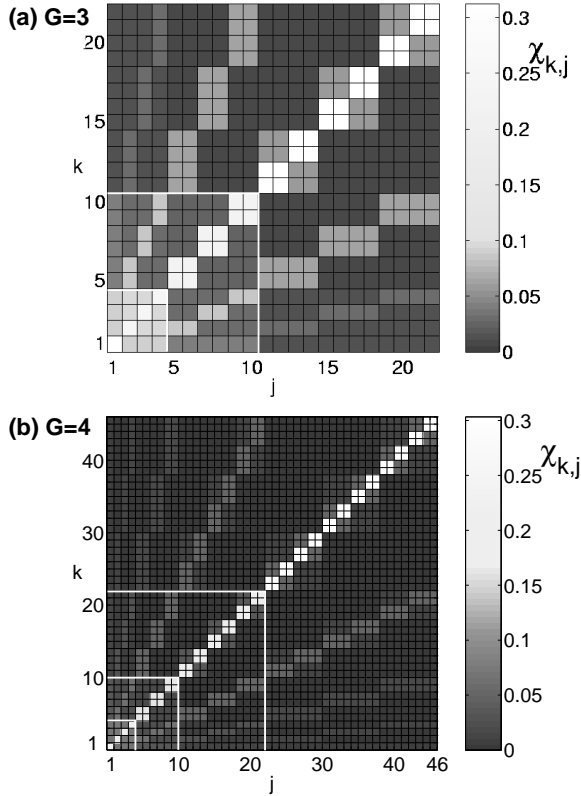}}
\caption{
Limiting probabilities for dendrimers with (a) $G=3$ and (b) $G=4$. The white
lines indicate the limiting distributions for the dendrimers of generations
smaller than $G$. From \cite{mbb2006a}.
}
\label{limiting}
\end{figure}

Classically, the LP is equipartitioned among all the nodes, i.e.\
$\lim_{t\to\infty} p_{k,j}(t) = 1/N$ for all nodes. Quantum mechanically
this is not the case. Figure \ref{limiting} shows the LPs $\chi_{k,j}$ as
a contour plot \cite{mbb2006a}.  Bright shadings correspond to high values
of the LP, whereas dark shadings correspond to low LPs.  The diagonal has
high values, meaning that an excitation starting at node $j$ has a high LP
to be found again at node $j$. The structures of the LP distributions of
dendrimers are self-similar, generation after generation.

Furthermore, different nodes $k$ and $l$ may have the same LP,
$\chi_{k,j}=\chi_{l,j}$. One hence combines all the nodes having (up to
our numerical precision, $10^{-10}$) the same LP into a cluster. Note,
however, that the separation of the nodes into clusters depends on the
initially excited node, namely on $j$. For an excitation starting at the
center (node $1$), the clusters correspond exactly to the different
generations of the dendrimer. In the general case, when starting from a
non-central node, one still finds from Fig.~\ref{limiting} that nodes
belonging to the same cluster also belong to the same generation (the
converse is not necessarily true).

\begin{figure}
\centerline{\includegraphics[clip=,width=6.5cm]{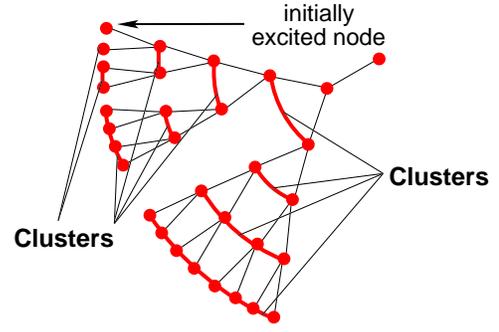}}
\caption{ (Color online). Clusters of the same limiting probability
$\chi_{k,j}$ in the branch with the initial excitation for a dendrimer of
generation $G$. Nodes connected by thick (red) lines belong to the same
cluster. From \cite{mbb2006a}.  }
\label{limiting_clust}
\end{figure}

For larger dendrimers, while the general cluster pattern is preserved,
some details change.  Figure~\ref{limiting_clust} shows the situation for
a dendrimer of dimension $G=5$, for an excitation starting at a peripheral
node.  Again one indicates clusters by connecting nodes by thick (red)
lines.  A change to be noticed is that for $G\ge5$ the initially excited
node does not form anymore a cluster with its next-nearest node of the
same generation. It appears as if such two nodes only belong to the same
cluster when the dendrimer has $G\le4$. Thus the total number of clusters
is $N_C \equiv (G^2+G+6)/2$ for $G\geq5$ and $N_C-1$ for $G\leq4$. 

\subsubsection{Husimi-cacti}

\begin{figure}
\centerline{
\includegraphics[clip=,width=0.95\columnwidth]{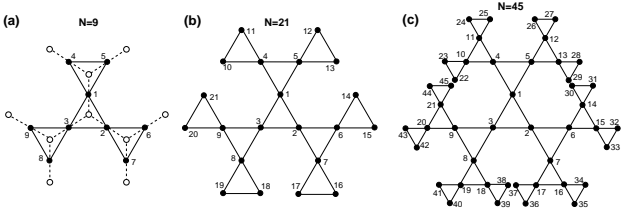}}
\caption{Finite Husimi cacti (filled circles) of size (a) $N=9$, (b)
$N=21$, and (c) $N=45$. (a) also shows with dashed lines and open circles
the corresponding dendrimer. From \cite{bbm2006a}.  }
\label{husimicactus}
\end{figure}

As shown in \cite{poliakov1999}, it may happen that the excitation
occupies preferentially the bonds between the branching points of the
dendrimer. Then, the essential underlying structure is different and is
given by sites localized at the mid-points of the bonds. The situation is
exemplified in Fig.~\ref{husimicactus}(a), starting from a dendrimer of
generation $2$ (open circles) and indicating the mid-points of the bonds
by filled circles. Connecting neighboring filled circles by new bonds, one
is led to the so-called Husimi cactus. Figure \ref{husimicactus} shows
three finite Husimi cacti of sizes $N=9,21$, and $45$.

In the same way as for the dendrimer, one finds numerically that for the
finite Husimi cactus consisting of $21$ nodes, as depicted in
Fig.~\ref{husimicactus}, the TPs are nearly periodic when the initial
excitation is placed on one of the (symmetrically equivalent) nodes $1$,
$2$, or $3$ of the inner triangle \cite{bbm2006a}.

\begin{figure}[t]
\centerline{
\includegraphics[clip=,width=0.9\columnwidth]{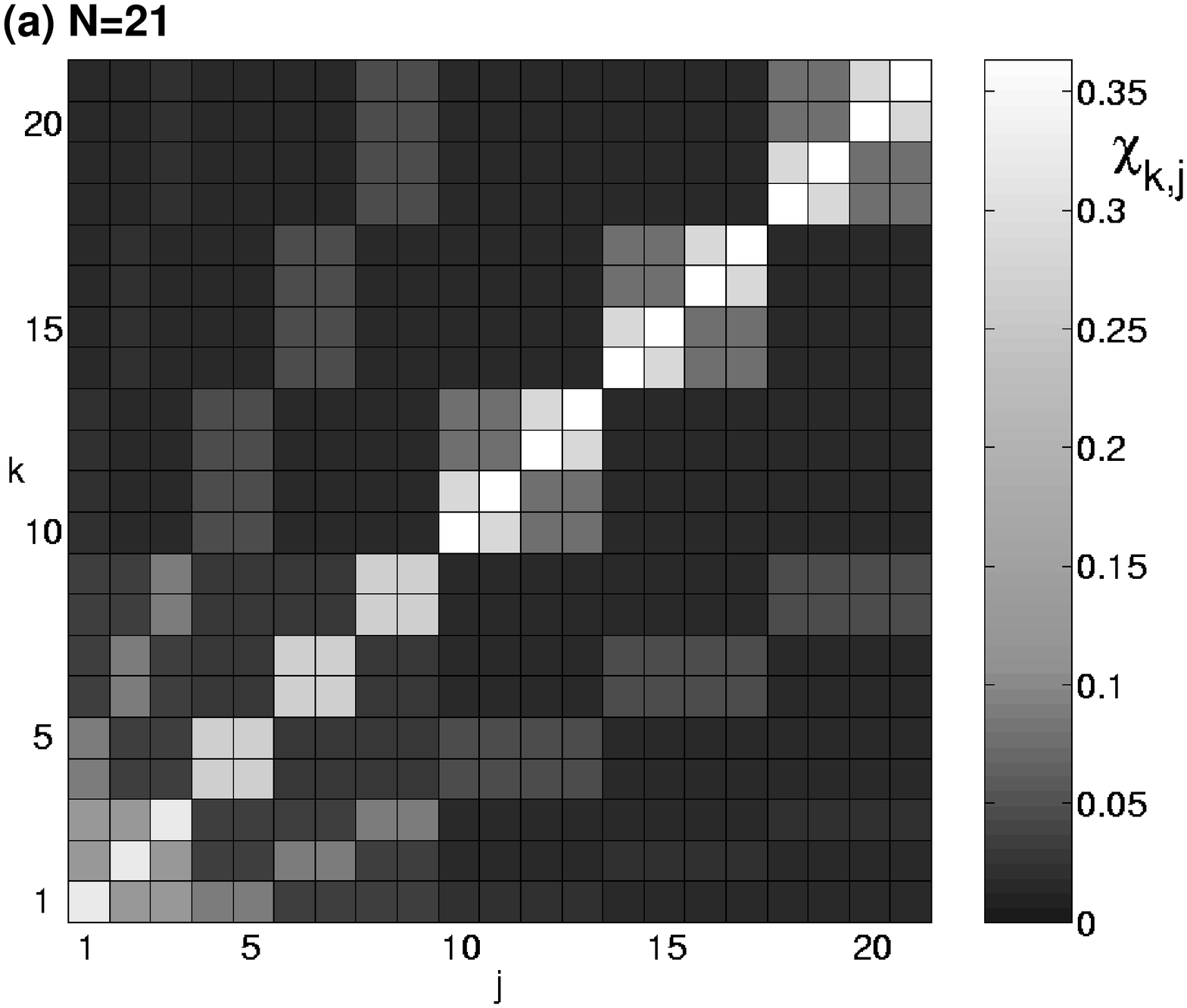}}
\centerline{
\includegraphics[clip=,width=0.9\columnwidth]{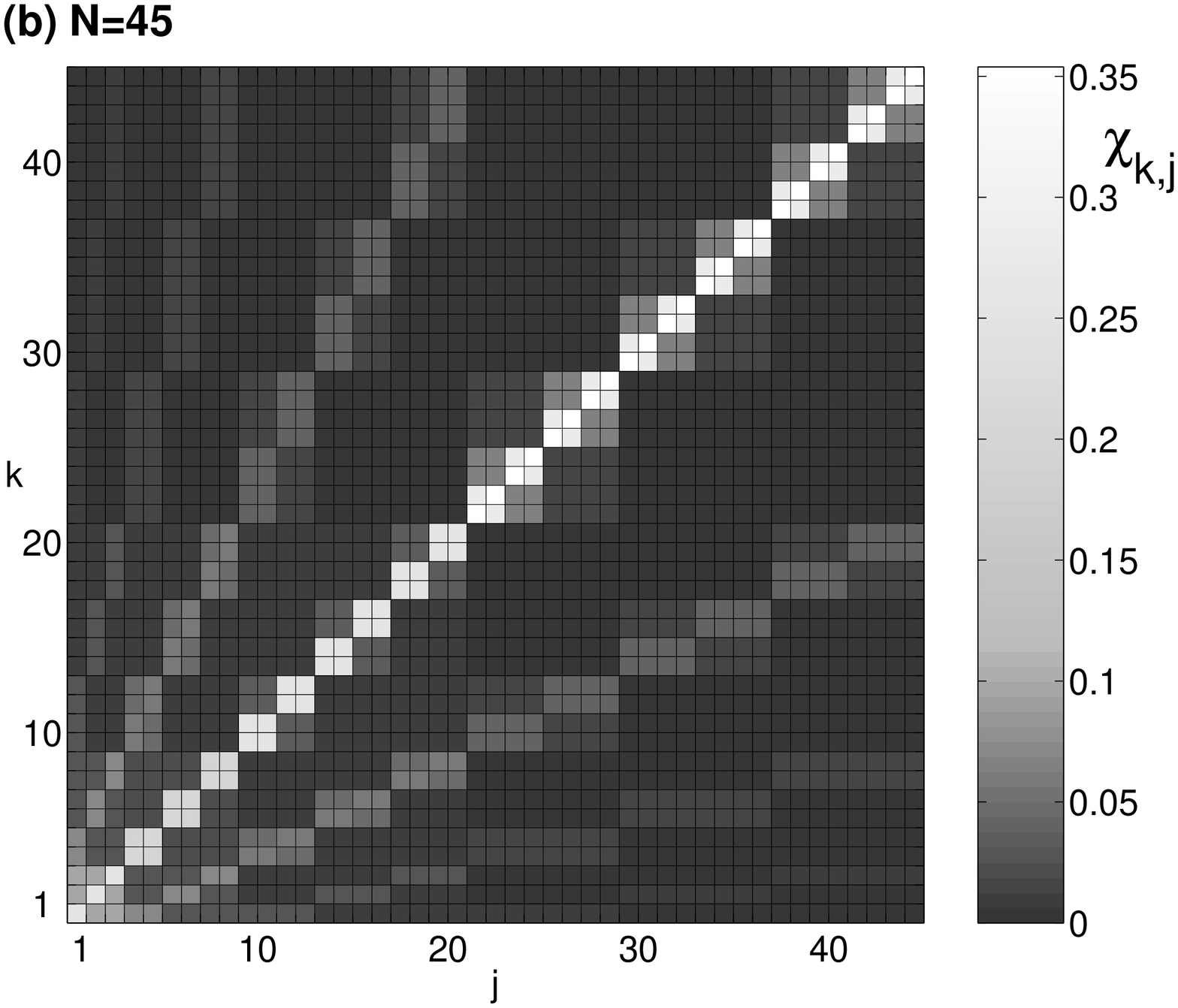}}
\caption{Limiting probabilites for finite Husimi cacti of sizes (a) $N=21$
and (b) $N=45$. From \cite{bbm2006a}.}
\label{limiting-husimi}
\end{figure}

Also in the long time limit, the similarities to the dendrimer are
obvious. Figure~\ref{limiting-husimi} presents the LPs for two sizes of
Husimi cacti, $N=21$ and $N=45$. The LP distributions are self-similar
generation after generation. Furthermore, for each size there are LPs
having the same value, i.e.\ $\chi_{k,j} = \chi_{l,j}$. One collects LPs
of the same value into clusters. Depending on where the excitation starts,
the clustering is different. This in analogous to the previous results for
dendrimers.  Since both dendrimers and Husimi cacti lead to similar
results, one can conclude that here the loops have no significant effect
on the transition probabilities.

\subsubsection{Glued Cayley trees}

\begin{figure}
\centerline{\includegraphics[width=0.9\columnwidth]{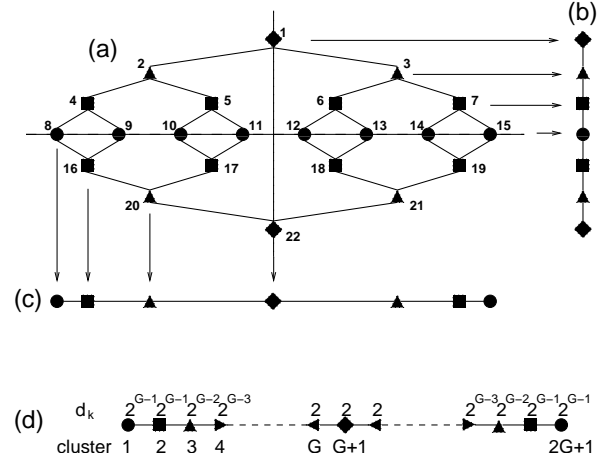}}
\caption{ (a) Graph consisting of two Cayley trees of generation $G=3$.
(b) horizontal projection of the graph following Ref.\ \cite{childs2002},
(c) vertical projection of the same graph. (d) Vertical projection of a
similar graph, obtained from two Cayley trees of general generation $G$,
indicating the new nodes (clusters) and the $d_k$, see text for details.
From \cite{mb2005a}.}
\label{cayley_comb}
\end{figure}

One particular example which shares the properties of both regular and
hyperbranched networks is a network which is composed of two Cayley-trees
with the same number of generations, where the nodes of the last
generation are shared by both trees, see Fig.~\ref{cayley_comb}.

Now, depending on the initial condition, the dynamics of the CTQW over
such a network can change dramatically. While the transport from the top
to the bottom node is very fast \cite{childs2002} and comparable to the
dynamics on a finite regular one-dimensional network, the transport from
the left-most node to the right-most node is very slow \cite{mb2005a}
compared to the transport in dendrimers or Husimi cacti when the
excitation starts at a peripheral node \cite{mbb2006a,bbm2006a}.

The authors of Refs.\ \cite{farhi1998} and \cite{childs2002} have analyzed
CTQW over such networks, focussing on walks which start at the top node,
and looking for the amplitude of being at the bottom node at time $t$. The
problem can then be simplified by considering only states which are
totally symmetric superpositions of states $|k\rangle$, involving all the
nodes $k$ in each row of Fig.~\ref{cayley_comb}(a), as indicated
schematically in Fig.~\ref{cayley_comb}(b). The transport gets then mapped
onto a one-dimensional CTQW \cite{childs2002}.

Other initial conditions for the CTQW are, indeed, possible, especially
when considering the high symmetry of the underlying graphs.  Note that,
using for instance the site enumeration of Fig.\ \ref{cayley_comb}(a), a
CTQW from node $8$ to node $15$ is equivalent to a CTQW from, say, node
$10$ to node $14$.  The graph's symmetry suggests to collect groups of
such nodes into clusters, while focussing on the transport from left to
right. It is then natural to view the nodes 8, 9, 10, and 11 as belonging
to the first cluster.  The second cluster consists then of the nodes 4, 5,
16, and 17, all of which are directly connected by one bond to the nodes
of the first cluster. The nodes 2 and 20 of the third cluster are all
nodes directly connected by one bond to the nodes of the second cluster,
while at the same time not belonging to the first cluster. In general, all
the nodes of the $(k+1)$st cluster are connected by one bond to nodes of
the $k$th cluster and at the same time do not belong to the $(k-1)$st
cluster.

Let us denote the number of nodes in cluster $k$ by $d_k$. The transport
occurs now from a cluster to the next, by which the original graph gets
mapped onto a line in which one {\sl new} node corresponds to a {\sl
group} of original nodes of the graph. For a new node at position $k \in
[2,G]$ one finds that $d_k = 2^{G-k+1}$, the same being true for the
mirror node value, i.e., $d_k=d_{2G+2-k}$.  Note that for the end nodes
$d_1 = d_{2G+1} = 2^{G-1}$, the same holds for the nodes next to them.
Moreover, for the middle node $d_{G+1} = 2$.

One now focuses on the transport via the states which are totally symmetric,
normalized, linear state-combinations for all the original nodes in each
cluster. Thus, for the $k$th cluster, whose sites are denoted by $n$, one
has as a new state \cite{mb2005a}
\be
| a_k \rangle = \frac{1}{\sqrt{d_k}}\sum_{n\in k} | n \rangle.
\label{line_states}
\ee

\begin{figure}
\centerline{\includegraphics[width=0.85\columnwidth]{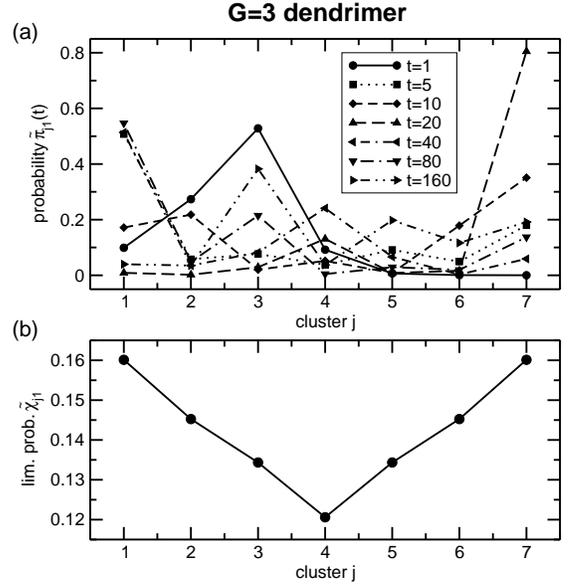}}
\caption{ (a) Transition probability $\tilde\pi_{j1}(t)$ for a CTQW
between different clusters $j$ of the $G=3$ graph. The CTQW  starts at the
first cluster, presented is the situation at times $t=1$, $5$, $10$, $20$,
$40$, $80$, and $160$.  (b) Limiting probability $\tilde\chi_{j1}$ for a
CTQW starting at the first cluster.  From \cite{mb2005a}.}
\label{cayley_22_map_time}
\end{figure}

The CTQW is now determined by the new Hamiltonian $\tilde{\bm H} = \gamma
\tilde{\bm A}$, where the matrix elements of $\tilde{\bm A}$ are obtained
from the new basis states $| a_k\rangle$ and from the matrix ${\bm A}$
through
\be
\tilde A_{jk} = \langle a_j | {\bm A} | a_k \rangle.
\ee
Given the properties of ${\bm A}$ and the construction of the
$|a_k\rangle$, Eq.(\ref{line_states}), $\tilde{\bm A}$ is a real and
symmetrical tridiagonal matrix, which implies a CTQW on a line. The
diagonal elements of $\tilde{\bm A}$ are given by
\bea
\tilde A_{kk} &=& \langle a_k | {\bm A} | a_k \rangle 
\nonumber \\
&=& 
\frac{1}{d_k}
\sum_{{n \in k}\atop{n'\in k}} \langle n' | {\bm A} | n \rangle = f_n
\equiv f_k,
\eea
where $f_k$ is the functionality of every node in the $k$th cluster.
For the sub- and super-diagonal elements of $\tilde{\bm A}$ one finds
\cite{mb2005a}
\bea
\tilde A_{k,k+1} &=& \tilde A_{k+1,k} = \langle a_k | {\bm A} | a_{k+1}
\rangle \nonumber \\
&=& \frac{1}{\sqrt{d_k d_{k+1}}} \sum_{{n \in k}\atop{n'\in k+1}}
\langle n' | {\bm A} | n \rangle 
\nonumber \\
&=& - \frac{b_k}{\sqrt{d_k d_{k+1}}},
\eea
where $b_k$ is the number of bonds between the clusters $k$ and $k+1$.

Now, except for the ends and the center of the graph, $b_k$ equals the
maximum of the pair $(d_k,d_{k+1})$. Between the central node
($d_{G+1}=2$) and its neighbors ($d_G=d_{G+2}=2$) the number of bonds is
$b_G = b_{G+2} = 2$. The number of bonds between the end node and its
neighbor is $b_1 = b_{2G+1} = 2 d_1 = 2^G$.

For the graph consisting of 22 {\sl original} nodes the new matrix
$\tilde{\bm A}$ is a tridiagonal $7\times7$ matrix, which can be readily
diagonalized.  The advantage of the procedure is clear: the new matrix
$\tilde{\bm A}$ depends on the number of clusters and grows with ($2G+1$),
whereas the full adjacency matrix, ${\bm A}$, grows with the total number
of nodes in the graph, namely with ($3\cdot2^G-2$).

From Eq.(\ref{line_states}) the transition amplitude between the state
$|a_k\rangle$ at time 0 and the state $|a_j\rangle$ at time $t$ is given
by \cite{mb2005a}
\be
\tilde \alpha_{jk}(t) = \langle a_j | e^{- i \tilde{\bm H} t} | a_k
\rangle
= \langle a_j | \tilde{\bm Q} e^{-i\gamma\tilde{\bm \Lambda} t} \tilde{\bm
Q}^{-1}| a_k \rangle, 
\ee
where $\tilde{\bm \Lambda}$ is the eigenvalue matrix and $\tilde{\bm Q}$
the matrix constructed from the orthonormalized eigenvectors of the new
matrix $\tilde{\bm A}$.

Now the quantum mechanical transition probabilities are given by
$\tilde\pi_{jk}(t) = |\tilde \alpha_{jk}(t)|^2$.  Figure
\ref{cayley_22_map_time}(a) shows the transition probabilities for CTQW
over clusters. Remarkably, now already during short periods of time, such
CTQW move from one end cluster to the other one. The limiting probability
distribution, $\tilde\chi_{jk}$, which is depicted in Fig.\
\ref{cayley_22_map_time}(b), also supports this finding. Note that Fig.\
\ref{cayley_22_map_time}(b) again reflects the symmetry of the original
graph.

\subsection{Fractals}

\subsubsection{Sierpinski Gaskets}

\begin{figure}
\begin{center}
\resizebox{0.85\columnwidth}{!}{\includegraphics{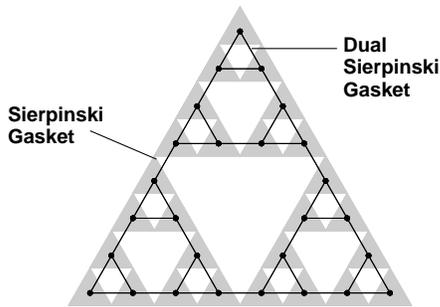}}
\end{center}
\caption{Dual transformation from Sierpinski Gasket to Dual Sierpinski
Gasket of generation $g=3$. From  \cite{abm2008}.} \label{fig:DualTransf}
\end{figure}

Regular networks have integer dimensions, while some tree-like stuctures
presented in the previous section, such as the dendrimers, can be of
``infinite fractal dimension'', when the (fractal) dimension is taken to
be given by $d_f=\lim_{R\to\infty} \ln N / \ln R$, $N$ being the number
of nodes within a sphere of radius $R$.. In contrast, (deterministic)
fractals have finite, in general, non-integer dimensions. One particular
example of a deterministic fractal is the dual Sierpinski gasket (DSG) for
which the exact spectrum of the eigenvalues of the connectivity matrix is
known \cite{abm2008}. A DSG is an exactly-decimable fractal which is
directly related, through a dual transformation, to the Sierpinski gasket
(SG). The DSG of generation $g$ can be constructed by replacing each small
triangle belonging to the SG with a node and by connecting such nodes
whenever the relevant triangles share a vertex in the original gasket (see
Fig.~\ref{fig:DualTransf}). It is straightforward to verify that the
number of nodes at any given generation $g$ is $N=3^g$.

The dual transformation does not conserve the coordination number (which
decreases from 4 to 3, while the coordination number of nodes
corresponding to the vertices of the gasket remains 2), but it does
conserve the fractal dimension $d_f$ and the spectral dimension
$\tilde{d}$, which are therefore the same as for the original Sierpinski
gasket, namely $d_f = \ln 3 / \ln 2 = 1.58496...$ and $\tilde{d}= 2 \ln 3
/ \ln 5=1.36521...$.  The eigenvalue spectrum of the DSG connectivity
matrix can be determined at any generation through the following iterative
procedure (for more details see \cite{cosenza1992,blumenjurjiu}): At any
given generation $g$ the spectrum includes the non-degenerate eigenvalue
$\lambda_{N}=0$, the eigenvalue $3$ with degeneracy $(3^{g-1} + 3)/2$ and
the eigenvalue $5$ with degeneracy $(3^{g-1} - 1)/2$. Moreover, given the
eigenvalue spectrum at generation $g-1$, then to each non-vanishing
eigenvalue $\lambda_{g-1}$ correspond two new eigenvalues
$\lambda_g^{\pm}$ according to
\be
\lambda_g^{\pm}=\frac{5 \pm \sqrt{25-4 \lambda_{g-1}}}{2};
\ee
both $\lambda_{g}^+$ and $\lambda_{g}^-$ inherit the degeneracy of
$\lambda_{g-1}$. The eigenvalue spectrum is therefore bounded in $[0,5]$.
As explained in \cite{cosenza1992}, at any generation $g$, one can
calculate the degeneracy of each distinct eigenvalue: apart from
$\lambda_{N}$ whose degeneracy is $1$, there are $2^r$ distinct
eigenvalues, each with degeneracy $(3^{g-r-1}+3)/2$, being
$r=0,1,...,g-1$, and $2^r$ distinct eigenvalues, each with degeneracy
$(3^{g-r-1}-1)/2$, being $r=0,1,...,g-2$. As can be easily verified, the
degeneracies sum up to $N=3^g$.

For the DSG the CTRW average return probability $\bar{p}(t)$ is readily
obtained without numerically diagonalizing the connectivity matrix, since
it only depends on the eigenvalues which can be calculated iteratively.
\begin{figure}
\centerline{\includegraphics[width=0.9\columnwidth]{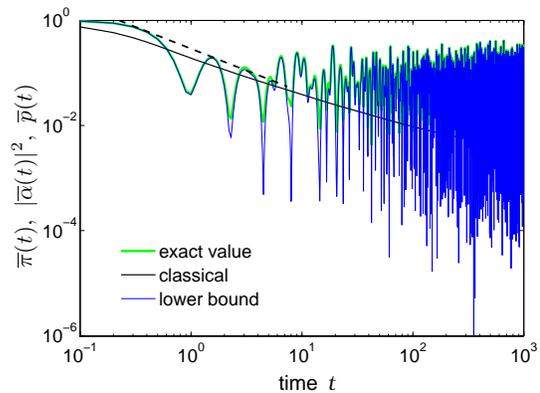}}
\caption{(Color online) Average return probability $\bar{\pi}(t)$ for the
DSG of generation $g=5$ on a log-log scale. The comparison with the
classical $\bar{p}(t)$ evidences that the classical random walk spreads
more efficiently than its quantum-mechanical counterpart. The dashed line
represents the envelope of $\bar{\pi}(t)$.  From  \cite{abm2008}.}
\label{fig:pi_compare}
\end{figure}
Figure \ref{fig:pi_compare} displays the averaged probabilities
$\bar{p}(t)$, $\bar{\pi}(t)$ and $|\bar{\alpha}(t)|^2$ as a function of
time, obtained for $g=5$. The classical $\bar{p}(t)$ decays monotonically
to the equipartition value $1 / N$ , while the quantum-mechanical
probabilities eventually oscillate around the value $0.7$, which is larger
than $3^{-g}$. Although the amplitude of fluctuations exhibited by the
lower bound is larger than that of the exact value, the agreement between
the two quantities is very good. In particular, the positions of the
extremal points practically coincide and the maxima of $\bar{\pi}(t)$ are
well reproduced by the lower bound.  This is analogous to the behavior of
walks on square lattices, Cayley trees, and stars, as described in the
previous sections. Notice, however, that for the square lattices the lower
bound turns out to be exact while for Cayley trees and for stars it is
only an approximation, which, moreover, turns out to be less accurate than
what is found for the DSG.

At short times ($t < 5 \gamma^{-1}$) it is possible to construct the
envelope of $\bar{\pi}(t)$, which depends algebraically on $t$
\cite{abm2008}. The exponent is $\approx -0.82$, to be possibly compared
with $\tilde{d}/2 \approx -0.68$ which is the exponent expected
classically for the infinite DSG. The decay of the average return
probability $\bar{\pi}(t)$ for the ST can be estimated as well: its
envelope goes like $t^{-2}$ (classically $\bar{p}(t) \sim t^{-1}$, see
above), implying a faster delocalization of the QW over the graph.

\begin{figure}
\centerline{\includegraphics[width=0.9\columnwidth]{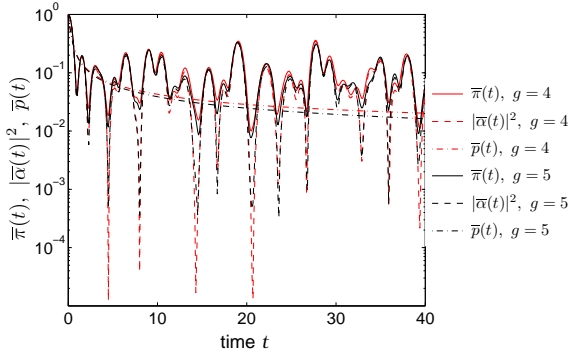}}
\caption{(Color online) Average return probability $\bar{\pi}(t)$ for the
DSG of generation $g=4$ (bright colour) and $g=5$ (dark colour). Its lower
bound $|\bar{\alpha}(t)|^2$ (dashed line) and the classical $\bar{p}(t)$
(dotted line) are also depicted, as shown by the legend. From
\cite{abm2008}.} \label{fig:pi_G4G5}
\end{figure}

Interestingly, for the DSG, the overall shape of $\bar{\pi}(t)$ does not
depend significantly on the size of the gasket (see Fig.~\ref{fig:pi_G4G5}
and \cite{abm2008}).  In fact, the behaviour of $\bar{\pi}(t)$ is mainly
controlled by the most highly degenerate eigenvalues. These do not change
when increasing the fractal size (i.e. its generation). These values are:
$3$ with degeneracy $m_g(3)=(3^{g-1}+3)/2$, $5$ with degeneracy
$m_g(5)=(3^{g-1}-1)/2$, $(5 \pm \sqrt{13})/2$ with degeneracy
$m_{g-1}(3)$.

The inhomogeneity of the pattern of the long-time averages $\chi_{k,j}$
mirrors the lack of translation invariance of the DSG itself. For
instance, being $v$ the label assigned to any vertex of the main triangle,
$\chi_{v,v}$ is a global maximum; off-diagonal local maxima correspond to
couples of connected nodes belonging to different minor triangles of
generation $g-1$. This allows to establish a mapping between the pattern
of $\chi_{k,j}$ and the structure of the relevant DSG.

\begin{figure}[t]
\centerline{\includegraphics[width=0.9\columnwidth]{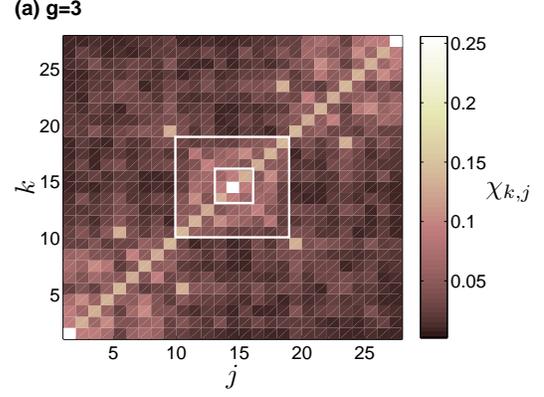}}
\centerline{\includegraphics[width=0.9\columnwidth]{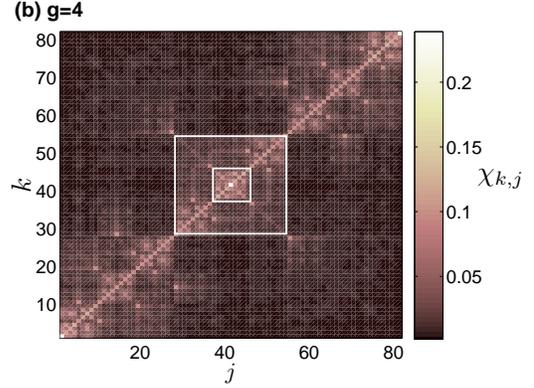}}
\caption{Limiting probabilities for the DSG of generation (a) $g=3$ and
(b) $g=4$, whose volumes are $N=27$ and $N=81$, respectively. The white
lines enclose the limiting distributions for gaskets of smaller
generations. Notice that the global maxima lay on the main diagonal and
correspond to $j=1,14,27$ and to $j=1,41,81$ for $g=3$ and for $g=4$,
respectively. From  \cite{abm2008}.} \label{fig:carpet}
\end{figure}

Since the spectrum of the DSG is known, one can calculate the lower bound
of the long-time average of the average return probability,
$\bar{\chi}_{lb}$, analytically.  At generation $g$ the spectrum of the
connectivity matrix displays $\tilde{N}$ distinct eigenvalues, where
\be
\tilde{N} = \sum_{r=0}^{g-1} 2^r + \sum_{r=0}^{g-2} 2^r + 1 =
3\times 2^{g-1}-1.
\ee
Let us denote the set of distinct eigenvalues by $\{ \tilde{\lambda}_i
\}_{i=1,...,\tilde{N}}$.  Being $m(\lambda_i)$ the degeneracy of the
eigenvalue $\lambda_i$, one can write
\be 
N^2 \bar{\chi}_{lb} = \sum_{n,m=1}^{\mathcal{N}}
\delta_{\lambda_n, \lambda_m} = \sum_{n=1}^{N}
m(\lambda_n) = \sum_{i=1}^{\tilde{N}} \left[
m(\tilde{\lambda}_i) \right]^2.
\ee
Now, one goes over to the space of distinct degeneracies, each
corresponding to a number $\rho$ of distinct eigenvalues and gets the
final, explicit formula \cite{abm2008}
\begin{eqnarray} \label{eq:exact_chi_bar_lb}
\bar{\chi} & \geq & \bar{\chi}_{lb} = \frac{1}{N^2}
\sum_{r=0}^{2g} [m(r)]^2
\rho(m(r))
\nonumber \\
& & = \frac{1}{N^2}  \Big\{  \sum_{r=0}^{g-1} \left[
\frac{3^{g-r-1}+3}{2} \right]^2 \times 2^r 
\nonumber \\
&& +  \sum_{r=0}^{g-2}
\left[ \frac{3^{g-r-1}-1}{2} \right]^2 \times 2^r  +1  \Big \}
\nonumber \\
& & = \frac{1}{3^{2g}} \left[ 3^g \left( 1 + \frac{3^g}{14}
\right) +\frac{10}{7}2^g - \frac{3}{2} \right],
\end{eqnarray}
such that $\bar{\chi} > 1/3^g$.  Interestingly, in the limit $g
\rightarrow \infty$, the LTA $\bar{\chi}$ is finite \cite{abm2008}:
\be
\bar{\chi} \geq \lim_{g \rightarrow \infty} \bar{\chi}_{lb}
= \frac{1}{14},
\ee
and $\bar{\chi}_{lb}$ reaches this asymptotic value from above.

\subsection{Statistical networks}

\subsubsection{Small-world networks}
\label{sec_swn}

\begin{figure}
\centerline{\includegraphics[clip=,width=0.6\columnwidth]{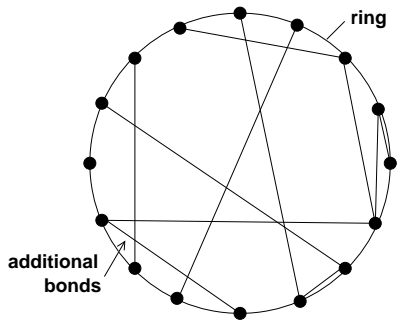}}
\caption{Sketch of a SWN of size $N=16$ containing $B=11$ additional
bonds. From \cite{mpb2007a}.}
\label{swn}
\end{figure}

The above examples dealt with deterministic networks. However, many real
systems have stochastic features; thus the connectivity can be random. To
model this, one can disrupt the periodicity of regular patterns by randomly
including $B$ additional bonds into the network \cite{mpb2007a}.  In such
a way one creates ``shortcuts'' and a walker can find shorter paths
between pairs of sites than on the regular network.  So-called
small-world-networks (SWN) are created by randomly adding bonds to a
regular one dimensional ring, see Fig.~\ref{swn}. Here we do not consider
self-connections, i.e., bonds connecting one node with itself.

The general behavior of CTQW on SWN can be analyzed by averaging over distinct
realizations $R$
\be
\langle \cdots \rangle_R \equiv \frac{1}{R} \sum_{r=1}^R [\cdots]_r,
\ee
where the index $r$ specifies the $r$th realization of the quantity in
question. In so doing one obtains statistical results which allow for a
comparison with the deterministic situation. In particular, we consider
here the realization-averaged transition probabilities
$\langle\pi_{kj}(t)\rangle_R$, the averaged probabilities
$\langle\overline{\pi}(t)\rangle_R$, their lower bound
$\langle\overline{\alpha}(t)\rangle_R$, and their classical analog
$\langle\overline{p}(t)\rangle_R$. Furthermore, we also discuss the long
time average (LTA) of each of these quantities:
\be
\left\langle \lim_{T\to\infty} \frac{1}{T} \int_0^T dt \
\cdots \right\rangle_R.
\ee

In the absence of any additional bond, the excitations travel along the
ring and interfere in a very regular manner, producing discrete quantum
carpets \cite{mb2005b}. Typical for these carpets is that they show,
depending on $N$, full or partial revivals at specific times
\cite{mb2005b}.

\begin{figure}
\centerline{\includegraphics[clip=,width=0.9\columnwidth]{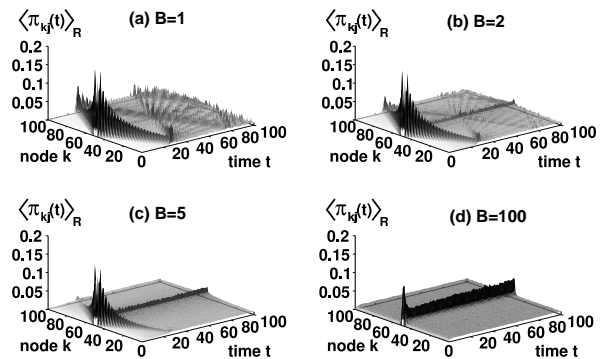}}
\caption{Time dependence of the averaged transition probabilities
$\langle\pi_{kj}(t)\rangle_R$ for SWN of size $N=100$ with (a) $B=1$, (b)
$B=2$, (c) $B=5$, and (d) $B=100$. The initial node is $j=50$ and the
number of realizations is $R=500$. From \cite{mpb2007a}.}
\label{pikj}
\end{figure}

For SWN the situation is quite different. Already a few additional bonds
obliterate the quantum carpets; the patterns fade away \cite{mpb2007a}. By
adding more bonds, only the initial node retains a significant value for
$\langle\pi_{jj}(t)\rangle_R$ at all times $t$. Furthermore, already for
SWN with as little as $B=5$ the pattern of $\langle\pi_{jj}(t)\rangle_R$
becomes quite regular after a short time, see Fig.~\ref{pikj}(c).  This
almost regular shape is reached very quickly when $B$ gets to be
comparable to $N$ [Fig.~\ref{pikj}(d)]. One notes, however, that
particular realizations may still show (depending on their actual
additional bonds) strong interference patterns. These features are washed
out by the ensemble average, so that only the dependence on the initial
node stands out.

Since CTQW on SWN always carry the information of their initial node $j$,
the averaged probabilities to return to $j$ are a good measure to quantify
the efficiency of the transport on such networks, see \cite{mb2006b}.

\begin{figure}[t]
\centerline{\includegraphics[clip=,width=0.9\columnwidth]{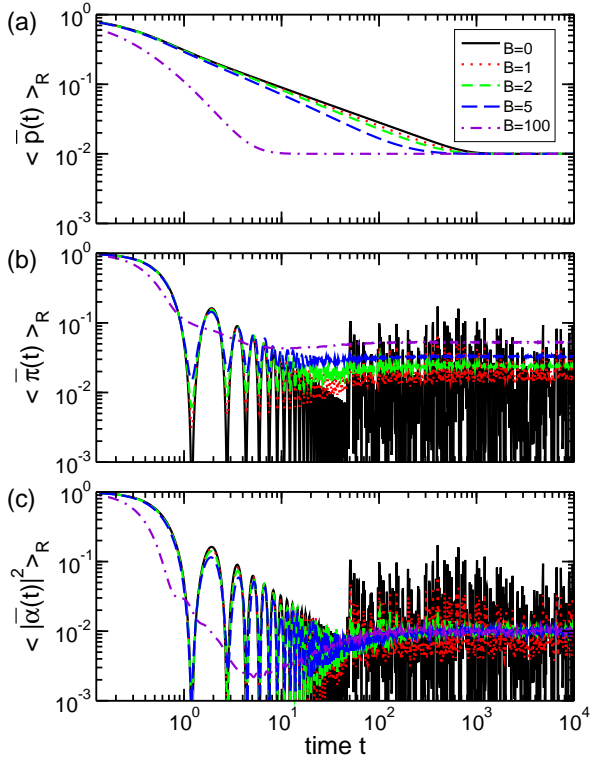}}
\caption{(Color online) Time dependence of the averaged probabilities (a)
$\langle\overline{p}(t)\rangle_R$, (b)
$\langle\overline{\pi}(t)\rangle_R$, and (c)
$\langle|\overline{\alpha}(t)|^2\rangle_R$ for SWN of size $N=100$ with
$B=1$, $2$, $5$, and $100$. The number of realizations is $R=500$. From
\cite{mpb2007a}}
\label{prob_kk}
\end{figure}

Figure \ref{prob_kk} shows in double-logarithmic scales the ensemble
averages $\langle\overline{p}(t)\rangle_R$,
$\langle\overline{\pi}(t)\rangle_R$, and
$\langle\overline{\alpha}(t)\rangle_R$ for SWN with $N=100$ nodes and
$B=1$, $2$, $5$, and $100$. For classical transport
[Fig.~\ref{prob_kk}(a)] the initial decay of
$\langle\overline{p}(t)\rangle_R$ occurs faster for larger $B$.  The decay
at intermediate times follows a power-law ($t^{-1/2}$) for the ring (as is
clear from the linear behavior in the scales of the figure) and changes to
a stretched exponential-type when $B$ is large \cite{jespersen2000}. Thus,
a classical excitation will quickly explore the whole SWN, so that it will
occupy each site with equal probability of $1/N$ already after a
relatively short time, see the final plateau in Fig.~\ref{prob_kk}(a).

Quantum mechanically, however, the situation is more complex.
Fig.~\ref{prob_kk}(b) shows the ensemble average
$\langle\overline{\pi}(t)\rangle_R$. For a ring of $N$ nodes and for times
shorter than roughly $N/2$ $\langle\overline{\pi}(t)\rangle_R$ displays a
quasiperiodic pattern (black curve), the maxima of which decay as
$t^{-1}$. At longer times interference sets in and leads to an irregular
behavior at times longer than $N/2$ \cite{mb2006b}. Now, for SWN, as long
as $B$ is considerably less than $N$, the periodic pattern still remains
visible; in Fig.~\ref{prob_kk}(b) one can follow how an increase in $B$
(red, green, and blue curves) is smoothing out the curves, so that both
the heights of the first maxima and the depths of the minima decrease.  At
longer times the SWN patterns are flattened out and
$\langle\overline{\pi}(t)\rangle_R$ tends towards a limiting value.  With
increasing $B$ this asymptotic domain is reached more quickly, such that
for larger $B$ the crossover from the quasiperiodic behavior at short
times to a smoothed out pattern at longer times is shifted to smaller $t$.

Figure~\ref{prob_kk}(c) shows the lower bound of $\overline{\pi}(t)$,
namely $\langle|\overline{\alpha}(t)|^2\rangle_R$ averaged over the
realizations. One notices that the overall behavior of
Figs.~\ref{prob_kk}(b) and \ref{prob_kk}(c) is quite similar. However, the
limiting values at long times differ. For the LTA of
$\langle\overline{\pi}(t)\rangle_R$ one has (see also Eq.~(17) of
Ref.~\cite{mvb2005a})
\bea
\langle\overline{\chi}\rangle_R
&\equiv& 
\Big\langle
\lim_{T\to\infty}\frac{1}{T} \int_0^T dt \ \overline{\pi}(t) \Big\rangle_R
\nonumber \\
&=& 
\frac{1}{RN} \sum_{r,j,n,n'}
\delta(E_{n,r}-E_{n',r}) 
\nonumber \\
&& \times 
\big|\langle j | \Phi_{n,r} \rangle \langle j |
\Phi_{n',r} \rangle \big|^2,
\label{pikk_avg}
\eea
where $\delta(E_{n,r}-E_{n',r})=1$ for $E_{n,r}=E_{n',r}$ and
$\delta(E_{n,r}-E_{n',r})=0$ otherwise.  For
$\langle|\overline{\alpha}(t)|^2\rangle_R$ the long-time values for
different $B$ collapse to one value. In fact, the LTA of
$\langle|\overline{\alpha}(t)|^2\rangle_R$ obeys
\bea
&& 
\Big\langle
\lim_{T\to\infty}\frac{1}{T} \int_0^T dt \ |\overline{\alpha}(t)|^2
\Big\rangle_R
\nonumber \\
&& 
= \frac{1}{RN^2} \sum_{r,n,n'} \delta(E_{n,r}-E_{n',r}),
\label{alphakk_avg}
\eea
as can be immediately inferred from Eq.~(\ref{piavg}). Thus this quantity
is only a function of the eigenvalues $E_{n,r}$ and does not depend on the
eigenstates $|\Phi_{n,r}\rangle$.  In order to quantify the differences
between Eqs.~(\ref{pikk_avg}) and (\ref{alphakk_avg}) for SWN, one assumes
that all the eigenvalues are nondegenerate (this assumption is, of course,
not valid for the ring, see below). In Eq.~(\ref{alphakk_avg}) the triple
sum adds then to $RN$, so that the rhs equals $1/N$. On the other hand,
Eq.~(\ref{pikk_avg}) leads to \cite{mpb2007a}
\be
\langle\overline{\chi}\rangle_R = \frac{1}{RN}  \sum_{r,j,n}
\big|\langle j |
\Phi_{n,r} \rangle \big|^4.
\label{chi_avg}
\ee
This expression depends on the eigenstates; in fact the rhs of
Eq.~(\ref{chi_avg}) is the ensemble average of the average participation
ratio of the eigenstates $| \Phi_{n,r} \rangle$.  Equation (\ref{chi_avg})
is well known in the theory of quantum localization, see, e.g.,
Sec.~V.~A.\ in \cite{heller1987}. For the ring the eigenstates are Bloch
states,
\be
|\Phi_n \rangle = \frac{1}{\sqrt{N}} \sum_{j=1}^N e^{ iE_n j} |j\rangle,
\label{bloch}
\ee
from which $\big|\langle k | \Phi_{n} \rangle \big|^4 = 1/N^2$ follows for
all $|\Phi_{n} \rangle$.

Now, increasing $B$ results in an increase of
$\langle\overline{\chi}\rangle_R$, starting from the corresponding value
for the ring ($B=0$, only one realization, and $N$ even)
\be
\langle\overline{\chi}_{\rm ring}\rangle_R \equiv \overline{\chi} =
\frac{1}{N} \sum_{j} \chi_{jj} = \frac{2N-2}{N^2},
\label{pikk_avg_ring}
\ee
where $\chi_{jj} = (2N-2)/N^2$. Equation~(\ref{alphakk_avg}) yields a
$1/N$ dependence for the LTA of
$\langle|\overline{\alpha}(t)|^2\rangle_R$, which by rescaling with
$\langle\overline{\chi}_{\rm ring}\rangle_R\sim 1/N$ would result in a
constant value for large $N$ \cite{mpb2007a}.  However, rescaling
$\langle\overline{\chi}\rangle_R$ with $\langle\overline{\chi}_{\rm
ring}\rangle_R$ shows an increase with $N$ of
$\langle\overline{\chi}\rangle_R/\langle\overline{\chi}_{\rm
ring}\rangle_R$ which is less than linear, thus,
$\langle\overline{\chi}\rangle_R$ depends on $N$ as $1/N^\nu$, with
$\nu\in[1,2]$.

The fact that $\langle\overline{\chi}\rangle_R$ for SWN increases with
increasing $B$ points towards a change of $\big|\langle k | \Phi_{n}
\rangle \big|^4$ from the value $1/N^2$.  The situation may be visualized
as follows: For the ring all eigenstates are Bloch states and hence are
completely delocalized. Going over to SWN and increasing the number of
additional bonds $B$ leads to localized states at the band edges and to
fairly delocalized states well inside the band. The increase of
$\langle\overline{\chi}\rangle_R$ is thus mainly due to the localized band
edge states.

\subsubsection{Erd\"os-R{\'e}nyi networks}

Somewhat similar to the SWN is the Erd\"os-R{\'e}nyi network (ERN). One
starts with $N$ disconnected nodes, every pair of nodes is then connected
with the probability $p$, where only single connections between two nodes
are allowed. One can in turn associate to $p$ an average degree $\overline
k$ of the nodes, which is related to $p$ by $\overline k=p(N-1)$. For
large network sizes $N$, the degree distribution $P(k)$ of the ERN is
Poissonian peaked at $\overline k$. 

Xu and Liu showed in Ref.~\cite{xu2008e} that in the ensemble average, the
average return probabilities display a behavior very similar to the SWN, see
Sec.~\ref{sec_swn} and \cite{mpb2007a}. Although the probabilities decay,
they do so only until they reach a plateau (for $N=100$), at a level
considerably higher than the equipartition value of $1/N$. The behavior is
only weakly affected by the value of the average degree $\overline k$.

The height of the plateau is determined by the long-time behavior of the
transition probabilities. Due to the ensemble average, all structure of
the non-diagonal elements of $\langle \Xi_{k,j}\rangle$ disappears and
only the main diagonal $\langle \Xi_{j,j}\rangle$ remains.

\subsubsection{Scale-free networks}

\begin{figure}
\centerline{\includegraphics[clip=,width=0.95\columnwidth]{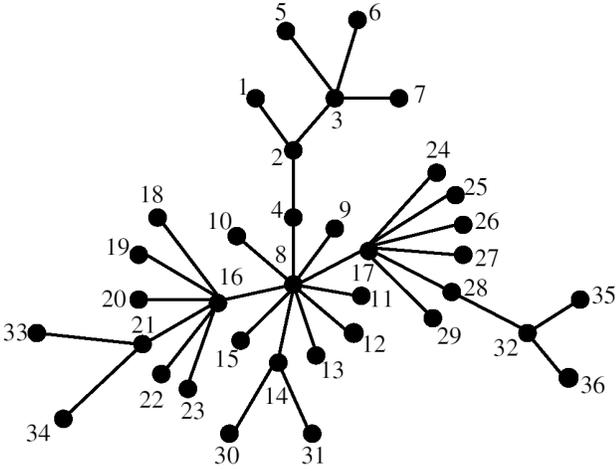}}
\caption{Example of a scale-free network with $N=36$ nodes and $\phi=5/2$.}
\label{sfn_example}
\end{figure}

The distribution $P(k)$ for the number $k$ of bonds emanating from a node
does not need to be Poissonian in general. Networks for which the
distribution $P(k)$ follows a power-law, i.e., $P(k) \sim k^{-\phi}$, are
called scale-free networks (SFN), see Fig.~\ref{sfn_example}. These have
been proven useful in various fields of research from biology to social
sciences. Xu and Liu have considered CTQW over such structures
\cite{xu2008a}. They distinguish between deterministic scale-free
networks (DSFN) and random scale-free networks (RSFN): the latter may obey
distinct building procedures which can lead to either tree-like
structures or structures containing loops.

For the DSFN, Xu and Liu determined the return probablities
$\pi_{j,j}(t)$ and long-time averages $\chi_{k,j}$. They find that there
is a strong dependence of $\pi_{j,j}(t)$ on the initial node $j$, which
can even result in (almost) complete revivals of the initial condition.
The transition probabilities translate directly into the long-time
averages $\chi_{k,j}$, where one can identify clusters of nodes having the
same $\chi_{k,j}$. However, the patterns obtained are quite distinct from
the previously found patterns for the dendrimers \cite{mbb2006a} or for
the Husimi cacti \cite{bbm2006a}. As mentioned earlier, this is a direct
consequence of the fact that the $\chi_{k,j}$ mirror the topology of the
network.

For RSFN, Xu and Liu calculated the ensemble average over many
realizations of individual RSFN \cite{xu2008a}. In the ensemble average
the return probabilities do not oscillate but rather reach a stationary
value, which differs for different initial nodes. In all cases, however,
this stationary value is roughly one order of magnitude larger than the
classical equipartition value $1/N$ obtained for CTRW. Moreover, the
higher the symmetry of the initial node (where on average the central
nodes have the largest symmetry), the larger is the stationary value.
Large average return probablities are also found in the long-time averages
$\chi_{k,j}$, where in the ensemble average the values on the diagonal
(for $k=j$) are much larger than the values for $k\neq j$.

\subsubsection{Apollonian networks}

\begin{figure}
\centerline{\includegraphics[clip=,width=0.75\columnwidth]{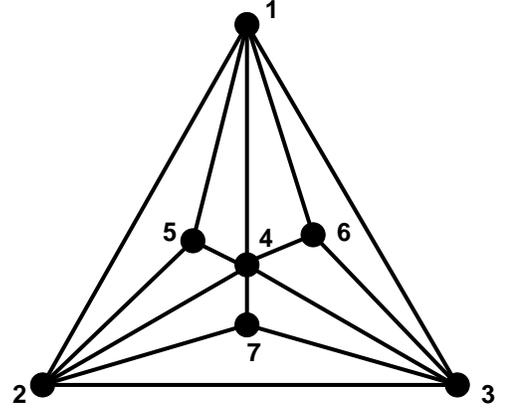}}
\caption{Example of an apollonian network of generation $G=2$.}
\label{apollonian}
\end{figure}

So-called Apollonian networks are models for networks which have
small-world as well as scale-free properties. CTQW on such structures have
been investigated by Xu {\sl et al.} \cite{xu2008b}. The network
is generated from a triangle ($N=3$ at generation $g=0$). In generation
$g=1$ a single node is added which is connected with all three nodes of
$g=0$, this divides the triangle of $g=0$ into three distinct triangles.
In $g=2$ three new noded are added which are placed at the center of each
of the three triangles of $g=1$, see Fig.~\ref{apollonian}. These three
nodes divide each triangle of $g=1$ into three new triangles, such that
there are now nine new triangles in total. The iteration proceeds in the
same manner, such that in generation $G$ the total number of nodes  is
$N=3 + (3^G-1)/2$. 

For $G=1$ the Apollonian network is identical to the complete graph of
size $N=4$, therefore the transition probabilities for CTQW are fully
periodic. It also turns out that for $G=2$ ($N=7$), see
Fig.~\ref{apollonian}, and when choosing as the initial node the central,
most symmetric node $4$, the transition probabilities are also fully
periodic, being namely \cite{xu2008a}
\be
\pi_{k,4}(t) = 
\begin{cases}
\big[37 + 12 \cos (7t) \big]/49 & \mbox{for } k=4 \\
\big[2-2\cos(7t)\big]/49 & \mbox{for } k\neq 4.
\end{cases}
\ee

However, also here Xu {\sl et al.} observe a strong dependence on the initial
condition. If the initial node is not a central node being, say, node $1$,
the transition probabilities $\pi_{1,1}(t)$ are still periodic, although
there is no perfect revival. Enlarging the networks lets the strong
dependence on the initial condition be more pronounced. But, similar to
the dendrimers \cite{mbb2006a}, Xu {\sl et al.} also identify clusters
of nodes which, in the long-time average, have the same value for their
probabilities $\chi_{k,j}$.

\section{Extensions}

\subsection{Systems with long-range interactions}
\label{sec-longrange}

So far the Hamiltonian for CTQW and the transfer matrix for CTRW have been
directly related to the connectivity matrix. However, this matrix is
purely topological (indicating whether or not two nodes are connected) and
does not take metric (i.e., distance dependent) aspects into account.
Some adjustment is thus necessary if one considers distance-dependent
interactions. As an example, consider the network being embedded in the
$d$-dimensional space; then to every node $j$ one associates a vector $\bm
j$ whose coordinates are $\{x_{j1},\dots,x_{jd}\}$. The distance between
two nodes $j$ and $k$ is then given by the euclidean norm $R\equiv |\bm k
- \bm j|$.  Consider now systems in which the interaction between the two
nodes $j$ and $k$ decreases with increasing distance. An example for such
a system could be a collection of dipoles, interacting via dipole-dipole
forces, whose potential decreases, to a good approximation, as $R^{-3}$.
Now, in terms of the connectivity matrix, a node $j$ is not only connected
to its nearest neighbors but also to other nodes. Thus, due to the
decaying interaction potential the transition rates are not the same for
all bonds but they depend on the distance.

Take as a first example a one-dimensional network with periodic boundary
conditions (i.e., a discrete ring) \cite{mpb2008a}. Here, when the
interactions go as $R^{-\gamma}$, the Hamiltonian has the following
structure:
\be
{\bm H}_\gamma = \sum_{n=1}^{N} \sum_{R=1}^{R_{\rm max}}R^{-\gamma}\Big(2 | n
\rangle \langle n| - | n-R \rangle
\langle n | - | n+R \rangle \langle n |\Big) ,
\ee
where $R_{\rm max}$ is a cut-off for finite systems. Note, that in the
infinite system limit one first takes $N\to\infty$ before taking also
$R_{\rm max}\to\infty$.  For the cases considered here, namely
$\gamma\geq2$ and $N$ of the order of a few hundred nodes, a resonable
cut-off is $R_{\rm max}=N/2$, which is also the largest distance between
two nodes on the discrete ring. In this way, to each pair of sites a
single (minimal) distance and a unique interaction is assigned.

For all $\gamma$, the eigenstates are again the Bloch states
$|\Psi_\theta\rangle$ given above. The fact that the eigenstates are not
affected by the long-range interactions is  due to the translational
invariance along the ring. For other systems without such an invariance
the eigenstates will also change depending on the type of the interaction.
From the eigenstates one obtains the eigenvalues which now do depend on
the interaction range \cite{mpb2008a}:
\be
E_\gamma(\theta) = \sum_{R=1}^{R_{\rm max}} R^{-\gamma} \big[ 2 -
2\cos(\theta
R)\big].
\label{evals}
\ee

The DOS $\rho_\gamma(E)$ is obtained by inverting Eq.~(\ref{evals}) and
taking the derivative with respect to $E_\gamma$.  In the NN-case
($\gamma=\infty$) one gets the known DOS 
\be
\rho_\infty(E) =
\big(\pi\sqrt{4E-E^2}\big)^{-1}. 
\ee
For $\gamma=2$ one can approximate the sum by letting $R_{\rm max} \to
\infty$, which yields $E_2(\theta) =
\pi\theta - \theta^2/2$, and one obtains
\be
\rho_2(E) = \big( \pi\sqrt{2}\sqrt{\pi^2/2 - E}\big)^{-1}.  
\ee
In the intermediate range there is an analytic solution for $\gamma=4$,
namely $E_4(\theta) = \theta^4/24 - \pi\theta^3/6 + \pi^2\theta^2/6$ (see
Eq.~1.443.6 of \cite{gradshteyn}), which yields \cite{abramowitz}
\be
\rho_4(E) = \Big[2\pi (2/3)^{1/4}\sqrt{E (\pi^2/\sqrt{24}) - E^{3/2}}
\Big]^{-1}.
\ee
One assumes the following general form for the DOS in order to interpolate
between $\rho_2(E)$ and $\rho_\infty(E)$ \cite{mpb2008a}:
\be
\rho_\gamma(E)
\sim \Big[ \sqrt{c_\gamma E^\alpha - E^\beta} \Big]^{-1}
\label{dos_gen}
\ee
with $\alpha\in[0,1]$ and $\beta\in[1,2]$; $c_\gamma$ is a constant
related to the maximal energy, $c_\gamma \equiv (E_{\gamma,{\rm
max}})^{\beta-\alpha}$. 

Having the DOS at hand, the integrals in Eqs.~(\ref{pclavginf}) and
(\ref{pqmavginf}) can be calculated - at least asymptotically - for large
$t$.  In the classical case Eq.~(\ref{pclavginf}) will be dominated by
small values of $E$ when $t$ becomes large.  The DOS yields
\be
\overline{p}_\gamma(t) \sim \begin{cases} t^{-1/2} & \mbox{for } \alpha=1 \\
t^{\alpha/2-1} & \mbox{for } \alpha<1. \end{cases}
\ee
Quantum mechanically one knows that for the NN-case
$\overline{\pi}_\infty(t) \sim t^{-1}$, see for instance \cite{mb2006b}.
Considering now the other limiting case, $\gamma=2$, one has
\cite{mpb2008a}
\be
\overline{\pi}_2(t) = \Bigg| \int_0^{\pi^2/2} dE \ \frac{\exp(-i E
t)}{\pi\sqrt{2}\sqrt{\pi^2/2 - E}} \Bigg|^2 \sim t^{-1}.
\label{pi2}
\ee
Thus, the behavior for long times is the same for $\overline{\pi}_2(t)$
and $\overline{\pi}_\infty(t)$, which suggests that for all one-dimensional
lattices with extensive ($\gamma\geq2$) interactions the long time
dynamics of the excitations is similar, no matter how long- or short-range
the step lengths are. This is in contrast to the classical case, where
only CTRW with $\gamma>3$ belong to the same universality class.

These results are corroborated by analytically evaluating
$\overline{\pi}_\gamma(t)$ using the stationary phase approximation (SPA)
\cite{Bender}. For large $N$, $\overline{\alpha}_\gamma(t)$ can be written
in integral form
\be
\overline{\alpha}_\gamma(t) = \frac{1}{2\pi}\int_0^{2\pi} d\theta \
\exp(iE_\gamma(\theta) t).
\ee
The SPA asserts now that the main contribution to this integral comes from
those points where $E_\gamma(\theta)$ is stationary
[$dE_\gamma(\theta)/d\theta\equiv E_\gamma'(\theta)=0$]. For $\gamma=2$,
$E_2(\theta)$ has only one stationary point in $\theta\in[0,2\pi[$, namely
$\theta_0=\pi$, leading to 
\be
\overline{\pi}_2(t) = |\overline{\alpha}_2(t)|^2
\approx \frac{1}{2\pi t |E_2''(\pi)|} \sim t^{-1}, 
\ee
which does not show any oscillations and coincides with the long time
limit of Eq.~(\ref{pi2}). For $\gamma>2$, $E_\gamma(\theta)$ has two
stationary points in the interval $\theta\in[0,2\pi[$, namely $\theta_0=0$
and $\theta_0=\pi$. Then $\overline{\alpha}_\gamma(t)$ is approximately
given by the sum of the contributions of the two stationary points.
Consequently \cite{mpb2008a},
\bea
&& 
\overline{\pi}_\gamma(t) \approx
\frac{1}{2\pi t} \Bigg(
\frac{1}{|E_\gamma''(0)|} + \frac{1}{|E_\gamma''(\pi)|}
\nonumber \\
&&
+
\frac{2\cos\{t[E_\gamma(0)-E_\gamma(\pi)]+\pi/2\}}{
\sqrt{|E_\gamma''(0)E_\gamma''(\pi)|}} \Bigg) \sim t^{-1}. 
\label{pi_spa}
\eea

\begin{figure}
\centerline{\includegraphics[clip=,width=0.95\columnwidth]{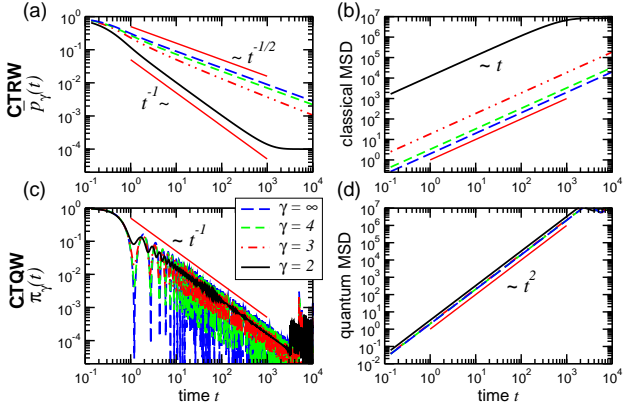}}
\caption{(Color online) (a) Classical $\overline{p}_\gamma(t)$ and (b)
quantum mechanical $\overline{\pi}_\gamma(t)$ for a discrete ring with
$N=10000$ nodes with $\gamma=2$, $3$, $4$, and $\infty$. From
\cite{mpb2008a}.  }
\label{returnprob}
\end{figure}

The classical and quantum mean square displacements (MSD) are in line with
these findings. Now, the MSD for CTRW/CTQW on the discrete ring with
initial site $j$ are given by \cite{mpb2008a}
\be
\langle R_\gamma^2 (t) \rangle_{\rm cl;~qm} = \frac{1}{N}\sum_{k=1}^{N}
|k-j|^2 {\cal P}_{k,j}^{(\gamma)} (t),
\ee
where ${\cal P}_{k,j}^{(\gamma)} (t) = p_{k,j}^{(\gamma)} (t)$ for CTRW
and ${\cal P}_{k,j}^{(\gamma)} (t) = \pi_{k,j}^{(\gamma)} (t)$ for CTQW.
Figure~\ref{returnprob} shows numerical calculations of
$\overline{p}_\gamma(t)$ and $\overline{\pi}_\gamma(t)$ for different
$\gamma$ and a discrete ring of $N=10000$ nodes.  Clearly,
$\overline{p}_\gamma(t)$ changes when increasing the step width from NN
steps to long-range steps, see Fig.~\ref{returnprob}(a).  While
$\overline{p}_\gamma(t)$ for $\gamma>3$ decays as $t^{-1/2}$, the power
law changes to $t^{-1}$ for $\gamma=2$. In contrast, the decay of the
maxima of the quantum return probability $\overline{\pi}_\gamma(t)$
follows $t^{-1}$ for all $\gamma$, Fig.~\ref{returnprob}(c).  Long-range
steps lead only to a damping of the oscillations and to an earlier
interference once the excitation has propagated around half of the ring.

As an extension of the ring topology just discussed \cite{mpb2008a}, Xu
considered networks where not only the nearest neighbors are connected but
in which each node is connected to its $2m$ nearest neighbors
\cite{xu2008d}. This model differs from the one with long-range
interactions, where the interaction decreases with the distance $|k-j|$
between the nodes $k$ and $j$. 

Now the action of the Hamiltonian on state $|j\rangle$ reads
\be
H|j\rangle = (2m+1) |j\rangle - \sum_{z=-m}^m |j+z\rangle.
\ee
It turns out that the Bloch states are also the eigenstates of this
Hamiltonian: their eigenvalues read \cite{xu2008d}
\be
E_n = 2m - 2\sum_{j=1}^m \cos (j\theta_n). 
\ee
Inserting these values into Eqs.~(\ref{pi_allg}) and (\ref{meq-solution})
allows to study the dependence of CTQW and of CTRW on $m$. As has been
shown by Xu \cite{xu2008d}, increasing $m$ results (as intuitively
expected) in a faster transport, both for CTQW and CTRW. Morevoer, when
considering the long-time average $\chi_{kj}$, Xu finds characteristic
peaks which depend on $m$. Especially for even $N$, the probability to be
at the initial node and the probability to be at the exactly opposite
node, i.e., the node $j\pm N/2$, are not necessary equal \cite{xu2008d},
as it is the case for only nearest-neighbor couplings. Depending on $m$,
these two values may differ \cite{xu2008d}: an explanation of this effect
is still lacking.

\subsection{Systems with disorder and localization}
\label{ctqw-disorder}

In real physical systems, under the influence of the surroundings, the
couplings between the nodes may differ. In a static picture, one can
introduce disorder by adding to the unperturbed Hamiltonian ${\bm H}_{0}$
a disorder operator $\bm \Delta$, i.e., by setting ${\bm H} = {\bm H}_{0}
+ {\bm\Delta}$ \cite{mbb2007a}. The disorder matrix $\bm \Delta =
(\Delta_{l,j})$ is taken to have non-zero entries only at the positions
for which $H_{l,j}\neq0$. For different strengths of disorder, the
elements $\Delta_{l,j} = \Delta_{j,l}$ are chosen randomly (drawn from a
normal distribution with the zero mean and unit variance, and then
multiplied by a factor of $\Delta$ which takes values from the interval
$[0,1/2]$).  Note that under these assumptions for the (static) disorder
the connectivity of the graph is essentially unchanged, i.e., there are no
new connections created nor are existing connections destroyed. Therefore,
the only non-zero matrix elements of ${\bm H}$ are those of the initial
${\bm A}$.  The action of the new Hamiltonian ${\bm H}$ on a state
$|j\rangle$ reads then
\bea
&&{\bm H} |j\rangle  = 
\Big({\bm H}_{0} + {\bm \Delta}\Big) |j\rangle 
\nonumber\\
&&=
2 |j\rangle - |j-1\rangle - |j+1\rangle 
\nonumber \\
&&+ 
2\Delta_{j,j} |j\rangle - \Delta_{j,j-1} |j-1\rangle - \Delta_{j,j+1}
|j+1\rangle. \nonumber \\
\eea

In the following two cases of disorder are considered:

(A)
Diagonal disorder (DD), where $\Delta_{j,j}\ne0$ and $\Delta_{l,j}=0$ for
$l\ne j$.  Here, a random number is assigned to each $\Delta_{j,j}$, a
procedure which leads to $N$ random numbers.

(B)
Diagonal and off-diagonal disorder (DOD), where a random number is chosen
for each $\Delta_{j,j}$ and for each $\Delta_{j,j-1}$. For this, $2N$
random numbers are needed.

Introducing disorder into the system in this way has consequences for the
relation between the CTQW and the CTRW. In CTRW the transition rates,
given by the entries of the the transfer matrix ${\bm T}$, are correlated,
i.e., for each site the sum of the non-diagonal rates for transmission
from it and the diagonal rate of leaving it are the same.  In the cases
considered here, a direct identification of the Hamiltonian ${\bm H}$ with
a classical transfer matrix is not possible anymore. However, the DOD and
DD Hamiltonians are widely used in quantum mechanical nearest-neighbor
hopping models, to which also the CTQW belong.  Furthermore, we still
consider transport processes on graphs which have the connectivity matrix
${\bm A}$, but the direct connection between ${\bm H}$ and ${\bm T}$ is
lost.

\begin{figure}
\centerline{\includegraphics[clip=,width=0.95\columnwidth]{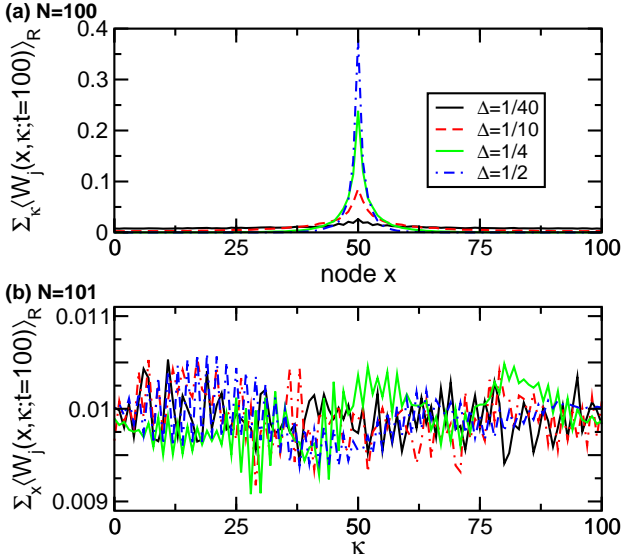}}
\caption{(Color online) Long-time average $\chi_{k,j}$ for different
$\Delta$ and (a) $N=100$ and (b) $N=101$. From \cite{mbb2007a}.}
\label{marg_lim_wigner}
\end{figure}

Now, consider again rings of $N$ nodes, where at time $t=0$ the excitation
is assumed to be localized at node $j$.  The above system is similar to
the Anderson model \cite{anderson1958}, which has been found to show
(strong) localization around the initial condition. The same happens here.
In the ensemble average as well as for single realizations, an excitation
starting at $j$ remains localized in the vicinity of $j$ \cite{mbb2007a}.
This effect is best seen in the long-time average. Figure
\ref{marg_lim_wigner} shows $\chi_{k,j}$ for two different ring sizes,
$N=100$ and $N=101$, and varying $\Delta$. Clearly, the larger is $\Delta$
the more is the long time average localized around the initial node
$j=50$.

Quantum walks in random linear environments were also studied by Yin {\sl
et al.} \cite{yin2008}. They calculated numerically the quantum carpet
structures for DD. It turns out that in the course of time the excitation
stays localized around its initial node; also depending on the strength of
the disorder, some structure of the original quantum carpets may still
remain visible.

Yin {\sl et al.} also studied dynamic diagonal disorder, where the
diagonal elements of $\bm H$ are rapidly varying with time and the update
of $(H)_{jj}$ is done at times comparable or much smaller than the time
step of the numerically determined dynamical changes  \cite{yin2008}. For
dynamic disorder, the interference patterns making up the quantum carpets
are washed out. However, distinct from the case of static disorder, no
localization can be seen. Moreover, after the temporal range in which
interference is lost, the dynamics becomes classical. This crossover
behavior is also observed in the mean square displacement, which changes
in a certain time range - depending on the strength of disorder - from the
ballistic behavior $\sim t^2$ to the diffusive behavior $\sim t$
\cite{yin2008}.

\section{Systems with absorption}

In general, an excitation does not stay forever in the system in which it
was created; the excitation either decays (radiatively or by exciton
recombination) or, e.g., in the case of biological light-harvesting
systems, it gets absorbed at the reaction center, where it is transformed
into chemical energy.  In such cases, the total probability to find the
excitation within the network is not conserved. Such loss processes can be
modelled phenomenologically by changing the transfer matrix or the
Hamiltonian \cite{mbagrw2007,mpb2008b,amb2010,mb2010a}.  To fix the ideas,
we consider networks in which the excitation can only vanish at certain
nodes. These nodes will be called trap-nodes or traps. In the absence of
traps, let the transfer matrix and the Hamiltonian of the corresponding
network be $\bm T_0$ and $\bm H_0$, respectively.  Take now $M$ out to the
$N$ total nodes to be traps and denote them by $m$, so that $m\in{\cal
M}$, with ${\cal M}\subset\{1,\dots,N\}$.  The trapping process is now
modelled by introducing a trapping matrix $\bm \Gamma$ which is given by a
sum over all trap nodes; $\bm\Gamma$ has only diagonal elements, i.e., 
\be
\bm \Gamma \equiv\sum_m \Gamma_m 
|m\rangle\langle m|
\ee 
(in the following one assumes that $\Gamma_m = \Gamma >0 $ for all $m$).

CTRW with decreasing exciton probabilities due to trapping are well
described through the following transfer matrix:
\be
\bm T \equiv \bm T_0 - \bm \Gamma.
\ee
The total Hamiltonian $\bm H$ corresponding to trapping is then:
\be
\bm H \equiv \bm H_0 - i \bm
\Gamma.
\ee
Note that the connection between CTRW and CTQW is now less direct than
before.  For CTRW the term corresponding to trapping has only real
elements and the total transfer matrix stays real. For CTQW, however, the
trapping term has purely imaginary elements. As a result, ${\bm H}$ is
non-hermitian and has $N$ complex eigenvalues, $E_l = \epsilon_l -
i\gamma_l$ ($l=1,\dots,N$).  In general, ${\bm H}$ has $N$ left and $N$
right eigenstates $|\Phi_l\rangle$ and $\langle\tilde\Phi_l|$,
respectively.  It turns out that in most cases the eigenstates of $\bm H$
form a complete and biorthonormal set, see, e.g.,
Ref.~\cite{sternheim1972}, 
\be
\sum_{l=1}^N
|\Phi_l\rangle\langle\tilde\Phi_l|={\bm 1} 
\qquad \mbox{and} \qquad 
\langle
\tilde\Phi_l | \Phi_{l'} \rangle = \delta_{ll'}
\ee
Both, eigenvalues and eigenstates, will be different for CTRW and CTQW,
because the incorporation of the trapping process is different.

If the trapping strength $\Gamma$ is small  compared to the couplings
between neighboring nodes, perturbation theory allows to relate the real
part of the eigenvalues to the eigenvalues of the unperturbed Hamiltonian
$\bm H_0$. Let $|\Psi^{(0)}_l\rangle$ be the $l$th eigenstate and
$E^{(0)}_l\in \mathbb{R}$ be the $l$th eigenvalue of the unperturbed
system with Hamiltonian ${\bm H}_0$. Up to first-order, the eigenvalues of
the perturbed system are given by \cite{mpb2008b}
\be
E_l = E^{(0)}_l + E^{(1)}_l = E^{(0)}_l - i \Gamma \sum_{m\in{\cal M}}
\Big| \langle m |\Psi^{(0)}_l\rangle \Big|^2.
\label{evals_perturb}
\ee
Therefore, the correction term determines the imaginary parts $\gamma_l$,
while the unperturbed eigenvalues are the real parts $\epsilon_l =
E^{(0)}_l$. Moreover, the imaginary parts are solely determined by the
contribution of the eigenstates of the network without traps at the trap
nodes $m$. 

\subsection{Average survival probability}

In an ideal experiment one would excite exactly one node, say
$j\not\in{\cal M}$, and read out the outcome $\pi_{kj}(t)$, i.e., the
probability to be at node $k\not\in{\cal M}$ at time $t$.  However, it is
easier to keep track of the total outcome at all nodes $k\not\in{\cal M}$,
namely of $\sum_{k\not\in{\cal M}} \pi_{kj}(t)$. Since the states
$|k\rangle$ form a complete, orthonormal basis set one has
$\sum_{k\not\in{\cal M}} | k \rangle\langle k | = {\bm 1} -
\sum_{m\in{\cal M}} | m \rangle\langle m|$, which leads to
\cite{mbagrw2007}:
\bea
&& \sum_{k\not\in{\cal M}} \pi_{kj}(t)
= \sum_{k\not\in{\cal M}} |\alpha_{kj}(t)|^2
\nonumber \\
&&= \sum_{l=1}^N  e^{-2\gamma_lt} \langle j | \Phi_l
\rangle\langle \tilde\Phi_l | j \rangle
-\sum_{l,l'=1}^N e^{-i(E_l-E_{l'}^*)t}
\nonumber \\
&& \times 
\sum_{m\in{\cal M}} \langle j |
\Phi_{l'} \rangle \langle \tilde\Phi_{l'} | m \rangle \langle m| \Phi_l
\rangle \langle \tilde\Phi_l | j \rangle.
\label{pi_avg_k}
\eea

By averaging over all $j\not\in{\cal M}$, the mean survival probability is
given by \cite{mbagrw2007}
\bea
&&\pavg \equiv \frac{1}{N-M} \sum_{j\not\in{\cal M}} \sum_{k\not\in{\cal
M}} \pi_{kj}(t) \\
&&=\frac{1}{N-M} \Bigg\{\sum_{l=1}^N e^{-2\gamma_lt} \Big[ 1 - 2 \sum_{m\in{\cal
M}} \langle \tilde\Phi_l | m \rangle \langle m | \Phi_l \rangle \Big]
\nonumber \\
&&
+ \sum_{l,l'=1}^N e^{-i(E_l-E_{l'}^*)t} 
\Big[
\sum_{m\in{\cal M}} \langle \tilde\Phi_{l'} | m \rangle \langle m| \Phi_l
\rangle \Big]^2 \Bigg\}.
\label{pi_avg}
\eea

For long $t$ and small $M/N$, Eq.~(\ref{pi_avg}) simplifies considerably:
At long times the oscillating term on the right hand side drops out and
for small $M/N$ one has $2 \sum_{m\in{\cal M}} \langle \tilde\Phi_l | m
\rangle\langle m | \Phi_l \rangle \ll~1$. Thus, $\pavg$ is mainly a sum of
exponentially decaying terms \cite{mbagrw2007}:
\be
\pavg \approx \frac{1}{N-M} \sum_{l=1}^N \exp[-2\gamma_lt].
\label{pi_avg2}
\ee

Asymptotically, Eq.~(\ref{pi_avg2}) is dominated by the $\gamma_l$ values
closest to zero. If the smallest one, $\gamma_{\rm min}$, is well
separated from the other values, one is led for $t \gg 1/\gamma_{\rm min}$
to the exponential decay found in earlier works, $\pavg = \exp( - 2
\gamma_{\rm min} t)$ \cite{parris1989b}.

Such long times are not of much experimental relevance (see also below),
since most measurements highlight shorter times, at which many $\gamma_l$
contribute. In the corresponding energy range the $\gamma_l$ often scale,
so that in a large $l$ range one finds $\gamma_l \sim a l^\mu$. The
prefactor $a$ depends only on $\Gamma$ and $N$ \cite{parris1989b}. For
densely distributed $\gamma_l$ and at intermediate times one has, from
Eq.~(\ref{pi_avg2}), \cite{mbagrw2007}
\bea
\pavg &\approx& \int dx \ e^{- 2atx^\mu}
\nonumber \\
&=& \int dy \ \frac{e^{-y^\mu}}{(2at)^{-1/\mu}} \sim t^{-1/\mu}.
\label{pi_avg_pl}
\eea

Analogously, the mean survival probability for CTRW is given by
\be
P_M(t) \equiv \frac{1}{N-M} \sum_{j\not\in{\cal
M}} \sum_{k\not\in{\cal M}} p_{kj}(t).
\ee
If the smallest eigenvalue, $\lambda_1$, is well separated from the rest,
$P_M(t)$ turns very quickly into a simple exponential decay. Then, for not
too small times, it can be shown that \cite{mpb2008b}
\be
P_M(t) \approx \frac{1}{N-M} e^{-\lambda_1 t} \Big|\sum_{k\not\in{\cal
M}} \langle k | q_1\rangle\Big|^2.
\ee

\subsection{Regular networks}

\subsubsection{Ring with traps}

Some of the nodes of a ring of $N$ nodes are now taken to be traps.
Depending on the particular choice of trap arrangements, the average
survival probability $\pavg$ shows different features \cite{amb2010}.

As already discussed, the eigenstates of the ring without traps are
Bloch states.  Furthermore, if the trapping strength is small compared to
the interaction strength between the nodes, the imaginary parts of the
Hamiltonian of a network with traps are given by the contributions of the
eigenstates at the trap positions, see Eq.~(\ref{evals_perturb}). However,
since all eigenvalues except $E_1=0$ (and for even $N$ also except
$E_{N/2}=4$) are two-fold degenerate, some care is in order when applying
perturbation theory. 

Take now $N$ to be even. Then, for $l= 1$ and for $l=N/2$ one has
\begin{equation} \label{eq:nondeg}
E_{l}^{(1)} = - i \Gamma \sum_{m \in \mathcal{M}} \left| \langle m |
\Phi_l^{(0)} \rangle \right|^2.
\end{equation}
Furthermore,
\begin{equation} \label{eq:correction_nondeg}
 E_{1} = 4 - i \Gamma \frac{M}{N}\; \; \;  \mathrm{and} \; \; \;E_{N/2} =
- i \Gamma \frac{M}{N} .
\end{equation}
For $l$ different from $1$ and from $N/2$ one sets
\be
V_{i,j} \equiv \langle \Phi_i^{(0)} | -i \bm \Gamma | \Phi_j^{(0)}
\rangle
\ee
and applies the expression valid for two-fold degenerate solutions of $\bm
H_0$ \cite{amb2010}:
\bea
\label{eq:deg}
&& 
E_{l}^{(1)} = \frac{1}{2} \left ( V_{l,l} + V_{N-l,N-l} \right ) 
\nonumber \\
&& 
\pm \frac{1}{2} \left[ \left( V_{l,l} - V_{N-l,N-l} \right)^2 + 4
|V_{l,N-l}|^2  \right ]^{1/2},
\eea
where one takes the positive sign for $l \in [1,N/2-1]$ and the negative
sign for $l \in [N/2+1,N-1]$. Now one has
\begin{equation} \label{eq:V11}
V_{l,l} \equiv V_{N-l,N-l} = - i \Gamma \frac{M}{N},
\end{equation}
independently of the trap arrangement and
\bea
\label{eq:V12}
&& 
V_{l,N-l} = -i \frac{\Gamma}{N} \sum_{j=1}^{M}  \exp \{ 2 i \pi m_j [l -
(N-l)]/N \}
\nonumber \\
&& 
= -i \frac{\Gamma}{N} \sum_{j=1}^{M} \exp (4 i \pi l m_j /N).
\eea
Inserting the last results into Eq.~(\ref{eq:deg}) yields \cite{amb2010}
\begin{equation} \label{eq:correction_deg}
E_{l}^{(1)} = \frac{-i \Gamma}{N} \left ( M \pm  \left| \sum_{j=1}^{M}
e^{2 i \pi 2 l m_j /N} \right| \right ).
\end{equation}
Notice that for special trap arrangements the $E_l^{(1)}$ can be
calculated exactly: The most striking results are obtained when the
exponential in the sum in Eq.~(\ref{eq:correction_deg}) equals one of the
values from the set $\{ 1, i, -1, -i \}$. Then the absolute value of the
sum reduces to $|\sum_{j=1}^{M} \exp(i 4 \pi l m_j /N)| = M$.

Now, for a single trap, which without loss of generality is placed at
position $m_j=j=1$, one has $E_l^{(1)} = -i\Gamma /N [1 \pm \exp(i4\pi
l/N)]$. Therefore, one has $E_l^{(1)} = 0$ for $l=N$ and for $l=N/2$.  As
a consequence, $\pavg$ will not decay to zero but to a constant value
given by $1/(N-1)$.

\begin{figure}
\includegraphics[width =0.9\columnwidth]{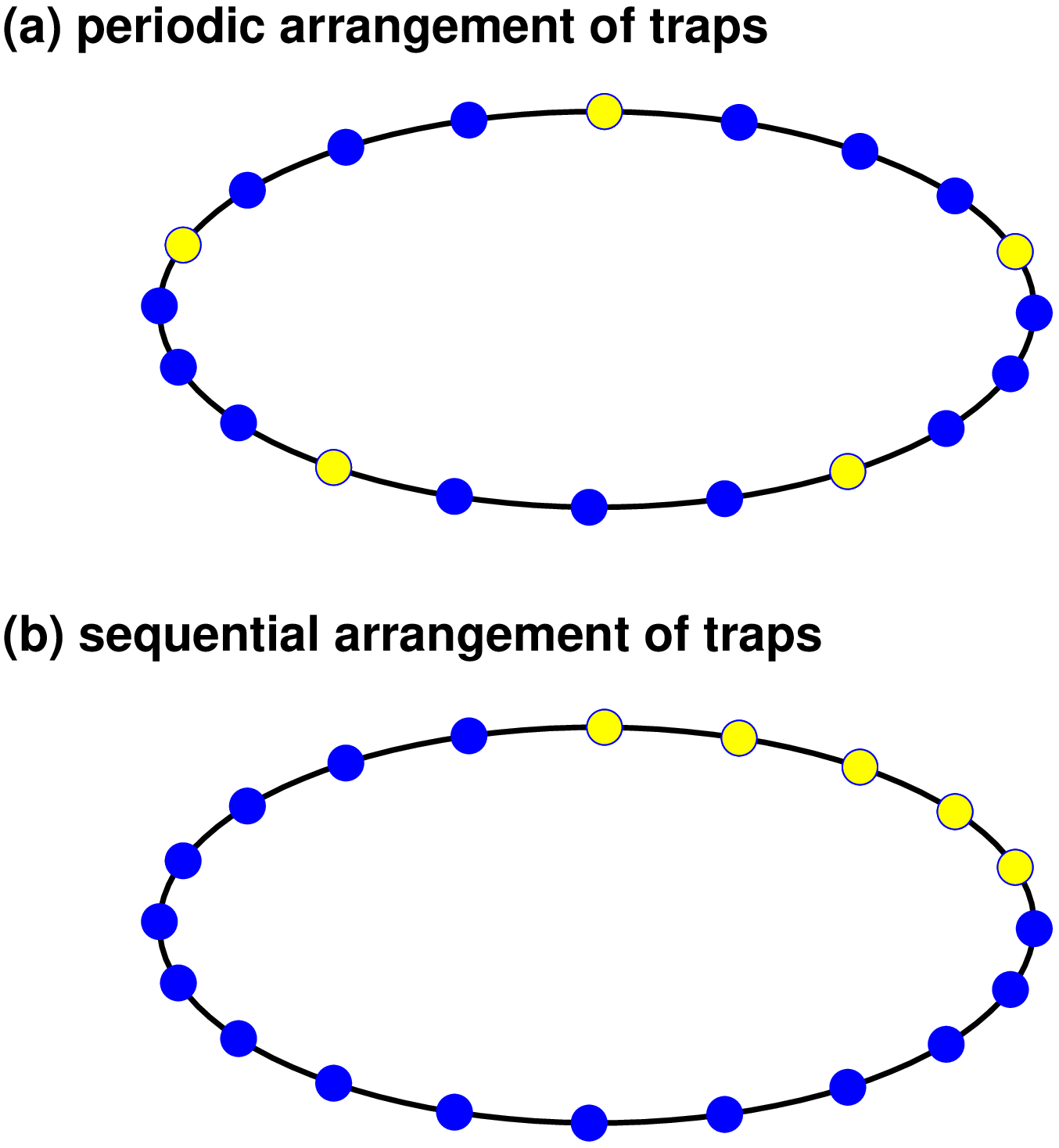}
\caption{Periodic (a) and sequential (b) arrangements of traps on a ring.}
\label{ringswithtraps}
\end{figure}

\begin{figure}
\includegraphics[width =0.9\columnwidth]{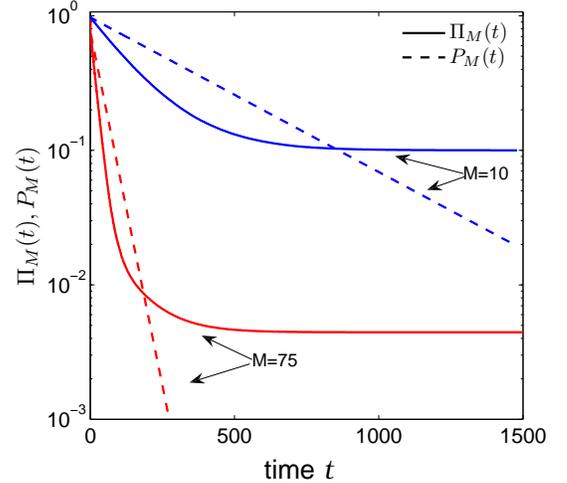}
\caption{Survival probabilities $\Pi_M(t)$ (continuous line) and $P_M(t)$
(dotted lines) on a ring of size $N=300$ and $\Gamma=0.01$ in the presence
of $M=10$ and of $M=75$ traps arranged periodically, i.e. $m_j = j N/M$.
Note the semilogarithmic scales. From \cite{amb2010}.}  \label{fig:per}
\end{figure}

Another example is a periodic distribution of traps with $m_j = j N/M$,
while $N/M \in \mathbb{N}$, see Fig.~\ref{ringswithtraps}(a). Then it is
straightforward to show that if $2l/M \in \mathbb{N}$ the sum in
Eq.~(\ref{eq:correction_deg}) equals one of the values from the set $\{ 1,
i, -1, -i \}$. The total number of such values is given by \cite{amb2010}
\begin{equation} \label{eq:upsi}
|\Upsilon| =
\left\{
\begin{array}{cr}
\lfloor (N-2) /M \rfloor & \mathrm{\; \; for \; even} \; M, \\
\lfloor (N-2) / 2M \rfloor & \mathrm{\; \; for \; odd} \; M,
\end{array}
\right.
\end{equation}
where $\lfloor x \rfloor$ denotes the largest integer less than or equal
to $x$.  In particular, for both $M=1$ and $M=2$, $| \Upsilon| =N/2-1$.
Hence, for large structures with $M \ll N$, $\Pi_M(t)$ decays
asymptotically to $1/M$ (even case) and to $1/(2M)$ (odd case). Figure
\ref{fig:per} shows results obtained for a ring of size $N=300$ with a
periodic arrangement of $M=10$ ($|\Upsilon| = 29$) and $M=75$ ($|\Upsilon|
= 1$) traps. Consequently, the survival probability $\Pi_M(t)$ decays to
the constant values $1/10$ and $1/225$, respectively.  From a physical
point of view, the finite limit for the survival probability stems from
the existence of stationary states to which the nodes in $\mathcal{M}$ do
not contribute, so that they never ``see'' the traps. This genuine
quantum-mechanical effect has no counterpart in the classical world where,
for finite structures, the survival probability always decays to zero in
the presence of traps. In particular, as shown in Fig.~\ref{fig:per},
$P_M(t)$ decays exponentially, as expected.

\begin{figure}
\includegraphics[width =0.9\columnwidth]{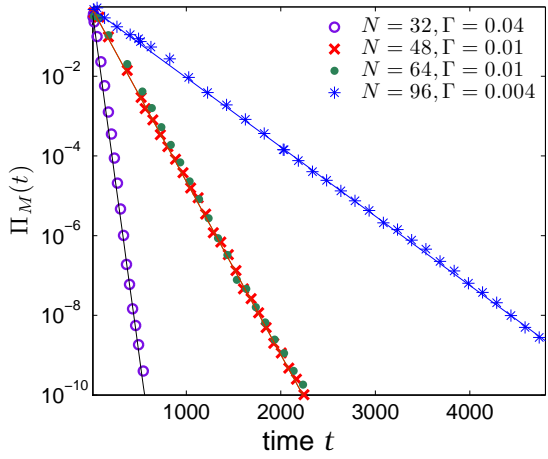}
\caption{Survival probability $\Pi_M(t)$ on rings of size $N=32, 48, 64$
and $96$ with a sequential arrangement of $M=N/2$ traps for $\Gamma=0.04,
0.01, 0.004$, as indicated.The straight lines represent
Eq.~(\ref{eq:pi_seq_spec}). From \cite{amb2010}.} \label{fig:seq_q}
\end{figure}

For a sequential arrangement of traps, such that $m_j=j$ and $j=1,....,M$,
see Fig.~\ref{ringswithtraps}(b), Eq.~\ref{eq:V12} can be written as
\begin{eqnarray} \label{eq:V12_seq}
\lefteqn{V_{l,N-l} = -i \Gamma /N  \sum_{j=1}^{M} \exp (4 i \pi l j /N)
}\\
\nonumber
& & = \frac{-i \Gamma}{N} \; \frac{  \exp (4 i \pi l M/N ) -1 } { \exp (4
\pi i l /N ) -1 } \; \exp (4 \pi i l /N )\\
\nonumber
& & = \frac{-i \Gamma}{N} \; \frac{\sin(2 \pi M l /N)}{\sin(2 \pi l /N)}
\; \exp [2 i \pi l (M+1)/N],
\end{eqnarray}
which yields
\begin{equation} \label{eq:correction_deg_seq}
E_l^{(1)} = \frac{-i \Gamma}{N} \left( M  \pm \frac{\sin(2 \pi M l
/N)}{\sin(2 \pi l /N)} \right).
\end{equation}
Notice that since $l \neq N/2$ and $l \neq N$ then $2l/N \notin
\mathbb{N}$, while for $2lM/N \in \mathbb{N}$ then $E_l^{(1)} =
E_{N-l}^{(1)} = - i \Gamma M/N$. In particular, when $M=N/2$, $\gamma_l =
M/N$ for each value of $l \in \left[ 1,N \right]$.  As a result, and by
neglecting oscillations, one has \cite{amb2010}
\begin{equation} \label{eq:pi_seq_spec}
\Pi_M (t) \approx \frac{M}{N-M} e^{-2 \Gamma t M/N} \sim e^{-\Gamma t},
\end{equation}
which is independent of $N$. As shown in Fig.~\ref{fig:seq_q}, the
exponential behaviour predicted by Eq.~(\ref{eq:pi_seq_spec}) holds also
for intermediate times.

\subsubsection{Line with traps}
\label{sec-linewithtraps}

An example of a network with traps which allows to study different time
scales is a finite line of nodes with traps at each end. The Hamiltonian
is thus \cite{mbagrw2007}
\bea
{\bm H} &=& \sum_{n=2}^{N-1} \Big(2 | n \rangle \langle n| - | n-1 \rangle
\langle n | - | n+1 \rangle \langle n |\Big)  \nonumber \\
&& +|1\rangle\langle 1| - |2\rangle\langle 1| + |N\rangle\langle N| -
|N-1\rangle\langle N| 
\nonumber \\
&& 
+ i\Gamma \Big( | 1 \rangle \langle 1 | +  |N\rangle\langle N| \Big).
\eea
Without loss of generality, an eigenstate of the finite chain without
traps can be written as ($l=1,\dots,N$) \cite{mpb2008b}
\be
|\Psi^{(0)}_l\rangle =
\begin{cases}
\displaystyle
\sqrt{\frac{1}{N}} \sum_{j=1}^N | j\rangle & \mbox{for } l=N  \\
\displaystyle
\sqrt{\frac{2}{N}} \sum_{j=1}^N \cos\big[(2j-1)\theta_l/2\big] |j\rangle &
\mbox{else},
\end{cases}
\label{statesNN}
\ee
where for convenience one takes $\theta_l \equiv \pi(N-l)/N \in [0,\pi[$;
the corresponding eigenvalues are $E^{(0)}_l = 2 - 2\cos\theta_l$ (note
that the smallest eigenvalue is $E^{(0)}_N=0$). Thus, first order
perturbation theory yields from Eqs.~(\ref{evals_perturb}) and
(\ref{statesNN}) as imaginary parts $\gamma_N=2\Gamma/N$ and $\gamma_l =
(4\Gamma/N) \cos^2\big(\theta_l/2\big)$ for $l=1,\dots,N-1$.  Indeed, for
$l \ll N$ this means that $\gamma_l \sim l^2$, so that the average
survival probability scales in the corresponding time interval as $\pavg
\sim t^{-1/2}$, see Eq.~(\ref{pi_avg_pl}).

\begin{figure}
\centerline{\includegraphics[clip=,width=0.95\columnwidth]{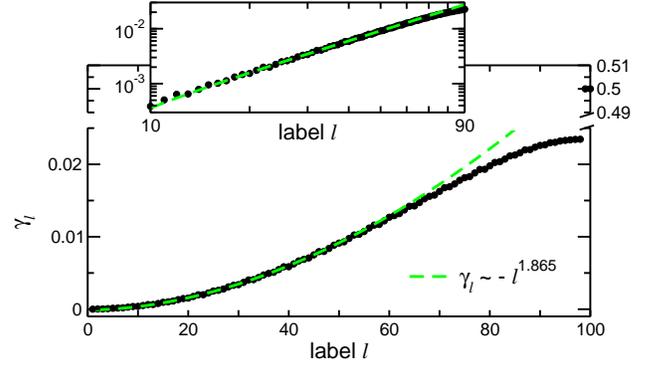}}
\caption{(Color online) Imaginary parts $\gamma_l$ (dots) in ascending
order for $N=100$ and $\Gamma=1$. Note the shortened $y$ axis.
The inset shows $\gamma_l$ in log-log scale for $l=10,\dots,90$. From
\cite{mbagrw2007}.}
\label{evals_imag_n100}
\end{figure}

If the trapping stength increases, one cannot employ perturbation theory
anymore. Nevertheless, one can always calculate the eigenvalues of $\bm H$
numerically. As it will turn out, the scaling $\gamma_l \sim l^\mu$ still
holds in this case and extends even over a wider range of $l$-values.
Figure~\ref{evals_imag_n100} shows the spectrum of $\gamma_l$ for $N=100$
and $\Gamma=1$; the double logarithmic plot (see inset) demonstrates that
scaling holds for $10\leq l \leq 60$, where the exponent $\mu$ is about
$\mu=1.865$. 

\begin{figure}
\centerline{\includegraphics[clip=,width=0.95\columnwidth]{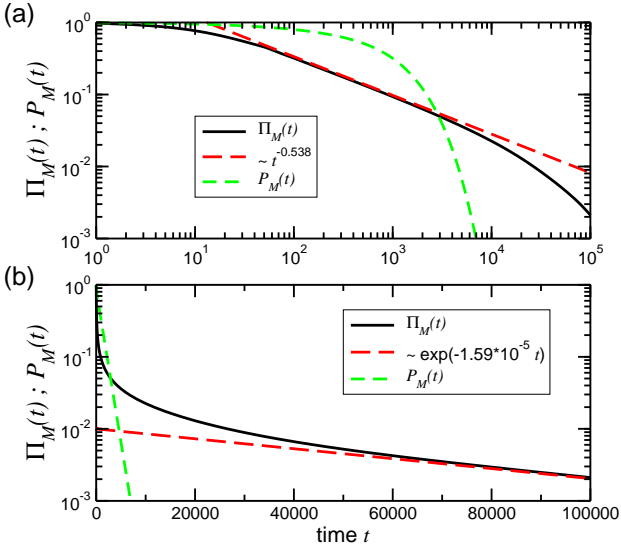}}
\caption{(Color online) Temporal decay of $\pavg$ (solid black lines) and
$P_M(t)$ (short dashed green lines) for $N=100$ and $\Gamma=1$ in (a)
double logarithmic scales and in (b) logarithmic scales. Indicated are the
fits to $\pavg$ (long dashed lines) in the intermediate (upper red) and
the long (lower blue) time regime. From \cite{mbagrw2007}.}
\label{decay_n100}
\end{figure}

Figure~\ref{decay_n100} compares, for a linear system with $N=100$ and
$\Gamma=1$, the classical $P_M(t)$ to the quantum mechanical survival
probability $\pavg$ \cite{mbagrw2007}. Evidently, $P_M(t)$ and $\pavg$
differ strongly: the $P_M(t)$ decay established for CTRW is practically
exponential. $\pavg$, on the other hand, shows two regimes: a power-law
decay at intermediate times (panel (a)) and an exponential decay (panel
(b)) at very long times.

\begin{figure}
\centerline{\includegraphics[clip=,width=0.95\columnwidth]{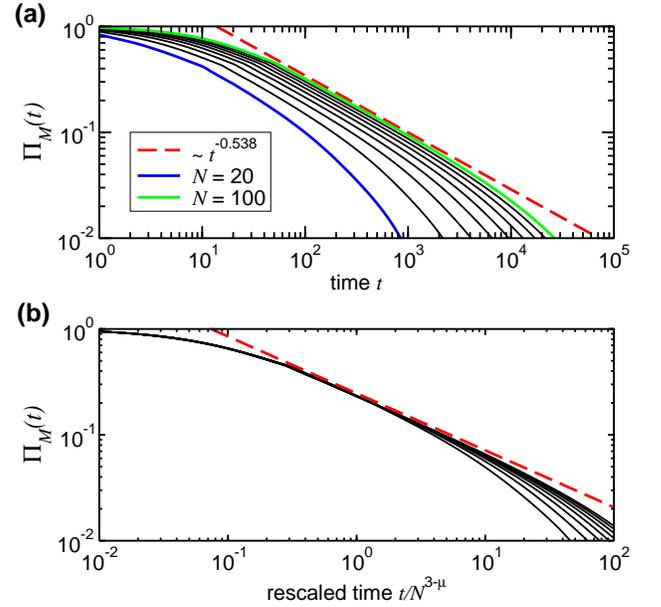}}
\caption{(Color online) Panel (a) shows the $N$-dependence of $\pavg$ for $\Gamma=1$; $N$
increases in steps of $10$ from $20$ (blue line) to $100$ (green line).
Panel (b) shows $\pavg$ versus the rescaled time $t/N^{3-\mu}$. From
\cite{mbagrw2007}.}
\label{avg_all_n}
\end{figure}

Turning now to the parameter dependences of $\pavg$, Fig.~\ref{avg_all_n}
shows the dependence of $\pavg$ on $N$ \cite{mbagrw2007}. Note that the
scaling regime, where $\pavg\sim t^{-1/\mu}$ holds, gets larger with
increasing $N$. The cross-over to this scaling region from the domain of
short times occurs around $t\approx N/2$.  For larger $N$ and in the
intermediate time domain, $\pavg$ scales nicely with $N$. In this case,
the power-law approximation [Eq.~(\ref{pi_avg_pl})] holds and by rescaling
$l$ to $l/N$ one has from Eq.~(\ref{pi_avg2}) that \cite{mbagrw2007}
\bea
\pavg &\sim& \sum_l e^{-2N^{-3}l^\mu t}
\nonumber \\
&=& \sum_l \exp\Big[-2(l/N)^\mu
N^{-(3-\mu)}t\Big], \nonumber \\
\label{tau_approx}
\eea
where it was assumed that $a\sim N^{-3}$ for a linear system
\cite{parris1989b}.  Thus, when rescaling $l$ to $l/N$, the time has to be
rescaled by the factor $N^{-(3-\mu)}$. Indeed, all curves for which a
power-law behavior is visible fall on a master curve; see the inset in
Fig.~\ref{avg_all_n}.

\subsubsection{Line with traps and long-range interactions}

As mentioned above, the interaction range does not need to be restricted
to nearest neighbor interactions (NNI). When the interactions between two
nodes go as $R^{-\nu}$, the Hamiltonian of the networks without traps has
the following structure \cite{mpb2008b}:
\bea
&&
{\bm H}_0(\nu) = \sum_{n=1}^N \Bigg[ \sum_{R=1}^{n-1} R^{-\nu} \Big( | n
\rangle \langle n| - | n-R \rangle \langle n | \Big) 
\nonumber \\ 
&& 
+ \sum_{R=1}^{N-n} R^{-\nu} \Big( | n \rangle \langle n| - | n+R \rangle
\langle n | \Big) \Bigg].
\label{hamil_long-ring}
\eea
Note that in the case of a line, the states $|k\rangle$ with
$k\notin\{1,\dots,N\}$ are implicitly excluded from the summation.

For large exponents $\nu$ the LRI can be regarded as a small perturbation
to the NNI, i.e., having ${\bm H}_0(\nu) = {\bm H}_0 + {\bm H}_{\nu}$,
where ${\bm H}_\nu$ contains only the correction terms to the NNI case
${\bm H}_0$. This allows one to calculate from the unperturbed states
$|\Psi^{(0)}_l\rangle$ the perturbed eigenstates $|\Psi_l\rangle$ up to
first order. Taking the states $|\Psi_l\rangle$ to be the eigenstates of
the LRI system without traps, one readily obtains the imaginary parts
$\gamma_l$ for small trapping strength from Eq.~(\ref{evals_perturb}) as
$\gamma_l = 2\Gamma  \big| \langle 1 |\Psi_l\rangle \big|^2$, where
\cite{mpb2008b}
\be
\langle 1 | \Psi_l\rangle = \langle 1 | \Psi^{(0)}_l\rangle + \sum_{r\neq
l} \frac{ \langle \Psi^{(0)}_r | {\bm H}_\nu | \Psi^{(0)}_l
\rangle}{E^{(0)}_l - E^{(0)}_r} \langle 1 | \Psi^{(0)}_r\rangle.
\label{states_lr}
\ee

For large $\nu$ the coupling to the next-next-nearest neighbor is by a
factor of $(3/2)^\nu$ smaller, for $\nu=10$ this is about one and a half
orders of magnitude. Taking, for fixed $\nu$, only nearest and
next-nearest neighbor couplings into account allows one to obtain simple
analytic expressions.  Thus, Eq.~(\ref{states_lr}) yields
\bea
\langle 1 | \Psi_l\rangle &=&
\sqrt{\frac{2}{N}}\cos\Big(\frac{\theta_l}{2}\Big) 
\nonumber \\
&& + 2^{-\nu}
\sqrt{\frac{2}{N}} \sin\big(2\theta_l\big)
\sin\Big(\frac{\theta_l}{2}\Big),
\label{states_lr_approx}
\eea
($\theta_l \equiv \pi(N-l)/N \in [0,\pi[$) which results in
\cite{mpb2008b}
\be
\gamma_l \approx \gamma_l^{(0)} + 2^{-\nu} \gamma_l^{(1)} + {\cal
O}(2^{-2\nu}),
\label{gamma_approx}
\ee
where $\gamma_l^{(0)}$ is the NNI expression, discussed in
Sec.~\ref{sec-linewithtraps}, and $\gamma_l^{(1)} = (8\Gamma/N)
\cos\big(\theta_l/2\big) \sin\big(2\theta_l\big) \sin\big(\theta_l/2\big)$
is the correction due to the LRI.  The smallest $\gamma_l$-values are
those for which $l \ll N$, which leads to a decrease of the imaginary
parts $\gamma_l$ because $\gamma_l^{(1)} < 0$ for $l \ll N$. Here, one can
approximate the imaginary parts by a power-law, i.e., $\gamma_{ l} \sim
l^\mu$. A rough estimate of the scaling exponent $\mu$, assuming
$\nu\gg1$, can be readily given \cite{mpb2008b}:
\be
\mu \approx
\frac{\ln\gamma_{l+1} - \ln\gamma_{l}}{\ln
(l+1) - \ln l} \approx \mu^{(0)} + 2^{-\nu} \mu^{(1)}.
\ee
Since $\mu^{(1)}$ is strictly positive for small $l$, the inclusion of LRI
leads to a decrease of $\gamma_{l}$ when compared to the NNI case. In
turn, this results in a slower decay of $\pavg$.

\begin{figure}
\centerline{
\includegraphics[clip=,width=0.9\columnwidth]{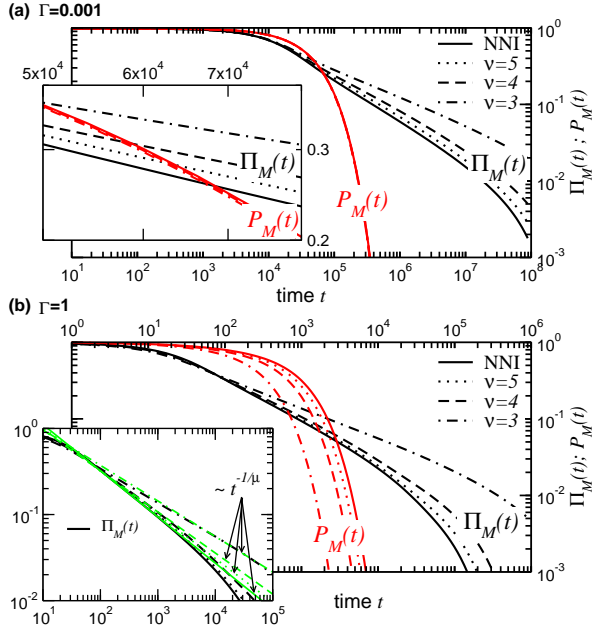}
}
\caption{(Color online) $\nu$-dependence of the quantum mechanical $\pavg$
and the classical $P_M(t)$ decay behaviors for a chain of $N=100$ sites;
here (a) $\Gamma=0.001$ and (b) $\Gamma=1$. The inset in (a) shows a
close-up picture of the region where $\pavg$ and $P_M(t)$ cross. The inset
in (b) shows power-law fits to $\pavg$ in the intermediate time regime
with exponents $1/\mu$, where the $\mu$ are taken from Fig.~3(b) of
Ref.~\cite{mpb2008b}. From \cite{mpb2008b}.}
\label{decay_n100lr}
\end{figure}

Figure~\ref{decay_n100lr} displays for comparison the quantum mechanical
$\pavg$ and the classical $P_M(t)$ behaviors for different $\nu$ and
$\Gamma$; $\pavg$ and $P_M(t)$ were obtained by numerically diagonalizing
the corresponding Hamiltonian ${\bm H}(\nu)$ and transfer matrix ${\bm
T}(\nu)$, respectively \cite{mpb2008b}. Clearly, for both $\Gamma$-values
the LRI lead to a slower decay of $\pavg$, i.e., to a slower trapping of
the excitation, which is somewhat counterintuitive since the opposite
effect is observable for classical systems, where the decay of $P_M(t)$
becomes faster for decreasing $\nu$, see below. By increasing the trapping
strength $\Gamma$, the difference between the quantum and the classical
behaviors become even more pronounced, compare Figs.~\ref{decay_n100lr}(a)
and \ref{decay_n100lr}(b).  Generally, for $\pavg$ a change in $\Gamma$
leads mainly to in a rescaled time axis, since the imaginary parts
$\gamma_l$ turn out to be of the same order of magnitude when rescaled by
$\Gamma$.

Xu has studied a ring where the interaction strength does not decay with
distance, as in \cite{mpb2008b}, but ranges  from a given node to the $m$
nearest neighbors \cite{xu2009b}.  By choosing for $N=100$ five nodes
randomly as trap nodes, Xu has shown numerically that both the CTQW and
the CTRW survival probabilities decay with increasing $m$ \cite{xu2009b},
thus showing an opposite effect to that mentioned above. However, an
explanation was not given. 

\subsection{Fractals}

\subsubsection{Hyperbranched fractals}

\begin{figure}
\centerline{
\includegraphics[clip=,width=0.9\columnwidth]{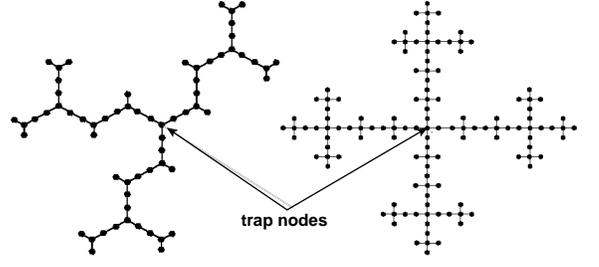}
}
\caption{Examples of two regular hyperbranched fractals of generation
$g=3$ with functionalities $f=3$ (left) and $f=4$ (right).}
\label{rhf-volta}
\end{figure}

Unlike the regular network discussed in the previous sections,
hyperbranched fractals allow one to study the survival probability in the
presence of highly degenerate eigenvalues. Volta investigated so-called
regular hyperbranched fractals, see Fig.~\ref{rhf-volta}, for which the
eigenvalue spectra of the connectivity matrices can be calculated
recursively \cite{volta2009}. As shown previously \cite{blumen2004}, the
eigenvalues of the $(g+1)$st generation can be obtained from the
eigenvalues of the $g$th generation by solving
\be
P(\lambda^{g+1}) = \lambda^g,
\ee
where $P(\lambda) = \lambda(\lambda-3)(\lambda-f-1)$, $f$ being the
functionality of the fractal. The three roots of the polynomial are given
by the Cardano solutions \cite{blumen2004}
\be
\lambda_j = \frac{f+4}{7} +\frac{2}{3} |f(f-1)+7|^{1/2} \cos\big[(\phi +
2\pi j)/3\big],
\ee
with $j=1,2,3$. Thus, each eigenvalue of the $g$th generation gives rise
to three new eigenvalues, not all of which are different from the previous
ones. Therefore, there appear highly degenerate eigenvalues, e.g., the
eigenvalue $\lambda=1$ has in the $g$th generation the degeneracy
$\Delta_g = (f-2)(f+1)^{g-1}+1$.

Placing now a trap node at the center of the fractal, Volta is able to
calculate the (complex) spectrum of the Hamiltonian including the trap
\cite{volta2009}. It turns out that only the nondegenerate eigenvalues
depend on the trapping strength $\Gamma$. Thus, the degenerate eigenvalues
can be again calculated based on the Cardano solution. In particular,
these eigenvalues are real and, therefore, do not contribute to the decay
of the survival probability. At generation $g$ there are
$(3^g-1)/2+(f+1)^g-3^g$ real degenerate eigenvalues, such that
Eq.~(\ref{pi_avg2}) yields
\bea
\Pi_M(t) &\approx& \frac{1}{N-1} \Big[ (3^g-1)/2+(f+1)^g-3^g 
\nonumber \\
&& +
\sum_{\gamma_l\neq 0} \exp(-2\gamma_l t)\Big].
\eea
Clearly, in the limit of $t\to\infty$ this leads to a constant value.
As Volta also shows (extending the results for the line), increasing the
trapping strength above the value of $\Gamma=V=1$ does not lead to a
faster $\Pi_M(t)$ decay.

\subsection{Random networks}

\subsubsection{Disordered system with one trap}

\begin{figure}
\centerline{
\includegraphics[clip=,width=0.9\columnwidth]{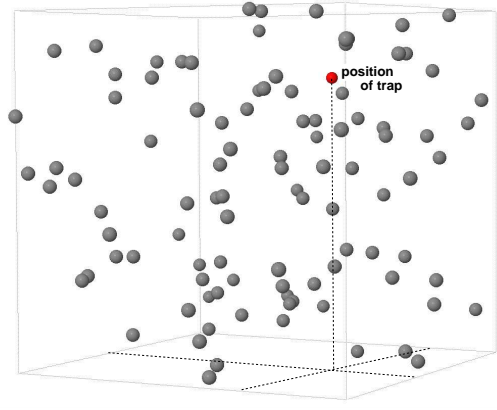}
}
\caption{Random configuration of $99$ nodes and a single trap. From
\cite{mb2010a}.}
\label{randomnetwork}
\end{figure}

Introducing long-range interactions also allows one to study topologically
disordered systems. To illustrate this, take a random configuration of
$(N-1)$ identical nodes and {\sl one} trap node \cite{mb2010a}, see
Fig.~\ref{randomnetwork}. All $N$ nodes are placed at random in a
$3$-dimensional box with Cartesian coordinates $\{x^{(i)}\}$, with
$i=1,2,3$. Then the distance between two nodes $j$ and $k$ is given by
\be
R_{j,k} = \Bigg[\sum_{i=1}^3
\big(x_j^{(i)}-x_k^{(i)}\big)^2\Bigg]^{1/2},
\ee
where the coordinates $x_j^{(i)}$ and $x_k^{(i)}$ are homogeneously
distributed random numbers in the interval $[0,N]$. To relate this to the
energy transfer dynamics within Rydberg gases one considers interactions
decaying as $R_{j,k}^{-3}$. In the absence of traps the corresponding
Hamiltonian ${\bm H}_0$ has the following matrix elements
\be
\langle k | {\bm H}_0 | j\rangle =
\begin{cases}
\displaystyle -R_{j,k}^{-3} & \mbox{for} \ k\neq j \\
\displaystyle \sum_{k\neq j} R_{j,k}^{-3} & \mbox{for} \ k=j.
\end{cases}
\label{hamil_long}
\ee
Now, choose one of the $N$ nodes to be a trap, i.e., for this node the
full Hamiltonian ${\bm H}$ has an additional purely imaginary matrix
element $-i\Gamma$. Since the configuration of nodes is random, one can
(without any loss of generality) assume in the following that the node
labeled $1$ is the trap.

For disordered systems, one calculates averages over $\cal R$ different
realizations following again, as for the SWN,
\be
\big\langle \cdot \big\rangle_{\cal R} \equiv \frac{1}{\cal R}
\sum_{r=1}^{\cal R} \big[\cdot \big]_r,
\ee
where $\big[ \cdot \big]_r$ denotes the realization $r$. In the
calculations one also assumes the trapping strength $\big[\Gamma\big]_r$
to be realization dependent, because it is required that it be
proportional to the diagonal element $\big[\langle 1 | {\bm H}_0 | 1
\rangle\big]_r$ in that particular realization, namely $\big[\Gamma\big]_r
\equiv \Gamma \big[\langle 1 | {\bm H}_0 | 1 \rangle\big]_r$. As it turns
out, the dependence of the decay on the value of $\Gamma$ is quite weak -
different $\Gamma$ mainly rescale the time axis. In Ref.~\cite{mb2010a}
only two extreme cases were considered: (a) $\Gamma=10^{-6}$, for which a
perturbation theoretical treatment can be justified, and (b) $\Gamma=1$,
such that the average trapping strength is of the same order as the
diagonal elements of ${\bm H}_0$ at the node of the trap.

While each realization $\big[\Pi(t)\big]_r$ leads to a specific spectrum
of the $\big[\gamma_l\big]_r$, the relation between $\langle \Pi(t)
\rangle_{\cal R}$ and the average $\langle\gamma_l\rangle_{\cal R}$ is not
that straightforward.  However, for all $t$, the function $\exp(-2\gamma_l
t)$ is convex, therefore, Jensen's inequality applies, see paragraph 12.41
of \cite{gradshteyn}, such that one obtains for a given $l$ \cite{mb2010a}
\be
\big\langle \exp(-2\gamma_l t) \big\rangle_{\cal R} \geq \exp\big(-2 t
\langle\gamma_l\rangle_R \big).
\ee
From this one gets a lower bound for $\big\langle \Pi(t) \big\rangle_{\cal
R}$:
\be
\big\langle \Pi(t) \big\rangle_{\cal R} \geq \frac{1}{N} \sum_{l=1}^N \exp\big(-2
t \langle\gamma_l\rangle_{\cal R} \big).
\label{pi_lb}
\ee

\begin{figure}
\centerline{
\includegraphics[clip=,width=0.95\columnwidth]{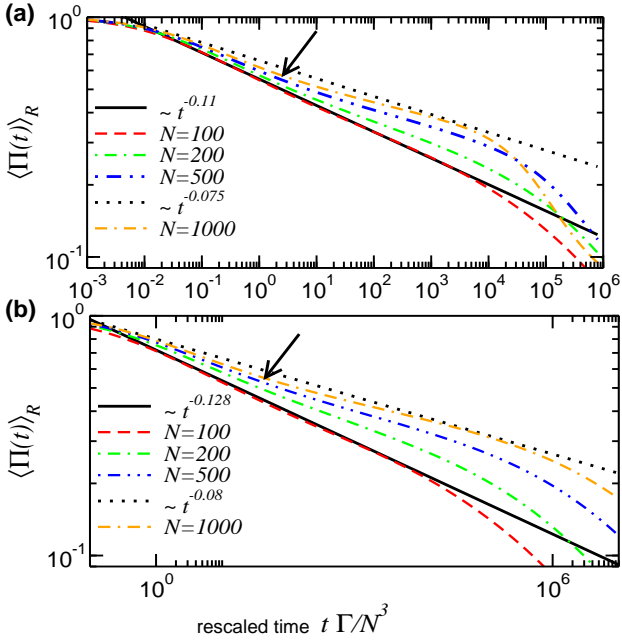}
}
\caption{(Color online) Ensemble averages $\langle \Pi(t) \rangle_R$ for
different $N$: (a) for $\Gamma=10^{-6}$ and (b) for $\Gamma=1$. The
scalings in the intermediate time regions for $N=100$ and $N=1000$ are
shown as solid black lines along with the appropriate scaling law. The
arrows are guides to the eye pointing at the bend of $\langle \Pi(t)
\rangle_R$ for $N=1000$. From \cite{mb2010a}.  }
\label{probs_avg_n}
\end{figure}

In Fig.~\ref{probs_avg_n} $\langle \Pi(t) \rangle_{\cal R}$ is displayed
for the two values $\Gamma=10^{-6}$ and $\Gamma=1$, for different $N$. In
all cases the intermediate time decay can be fitted by a power-law
\cite{mb2010a}
\be
\langle \Pi(t)
\rangle_{\cal R} \sim t^{-\eta(N)},
\ee
where, different from the regular linear case \cite{mbagrw2007,mpb2008b},
the exponent $\eta$ is now $N$-dependent. It turns out that approximating
the exponent by
\be
\eta(N) = \eta_0 N^\mu,
\label{etaN}
\ee
while keeping $\Gamma$ fixed, reproduces the curves well.  Note, however,
that $\eta_0$ and also $\mu$ can still depend on $\Gamma$.

Now, one can estimate $\mu$ based on the results of $\eta$ for $N_1=100$
and $N_2=1000$. Based on $\eta_0 = \eta(N_1)/N_1^\mu = \eta(N_2)/N_2^\mu$
one has \cite{mb2010a}
\be
\mu = \frac{\ln\eta(N_2) - \ln\eta(N_1)}{\ln N_2 - \ln N_1}.
\ee
From the numerical values given in Fig.~\ref{probs_avg_n} one obtains
approximately $\mu\approx-0.166$ and $\eta_0=0.0349$ for $\Gamma=10^{-6}$
and $\mu\approx-0.204$ and $\eta_0=0.0313$ for $\Gamma=1$.

From Fig.~\ref{probs_avg_n} as well as from Eq.~(\ref{etaN}) one sees that
the exponent $\eta(N)$ decreases with increasing $N$.  Certainly, if $N$
becomes very large it becomes quite improbable (in the ensemble average)
for an exciton to encounter the single trap. Therefore, the decay of
$\langle \Pi(t) \rangle_R$ can only be observed at very long times.

\section{Relations between CTRW/CTQW and other approaches}

\subsection{Phase space approaches}

In order to obtain a unifying framework for describing both quantum
mechanical and classical transport one introduces a quantum analog for the
classical dynamics, see, e.g.~\cite{Schleich}. One particular approach is
attributed to Wigner and uses the so-called Wigner-function (WF) in the
(quantum mechanical) $2d$ phase space.  If the phase space is spanned by
the continuous variables $X$ and $K$, the WF is given by
\cite{Wigner1932,Hillery1984}
\be
W(X,K;t) = \frac{1}{\pi}\int \ dY \ e^{iKY} \ \langle X - Y/2 | \hat
\rho(t) | X + Y/2
\rangle ,
\label{wigner}
\ee
where $\hat \rho(t)$ is the density operator and thus for a pure state,
$\hat \rho (t) = | \psi(t) \rangle \langle \psi(t) |$. Here, $\psi(X;t) =
\langle X | \psi(t) \rangle$ is the wave function of the particle. The WF
is a quasi-probability (in the sense that it can become negative).
Integrating $W(X,K;t)$ along lines in phase space gives marginal
distributions, e.g., when integrating along the $K$-axis one has,
\cite{Wigner1932,Hillery1984},
\be
\int d K \ W(X,K;t) = |\psi(X;t)|^2.
\label{marg_x}
\ee

In the case of a discrete system, given, for instance, by $N$ discrete
positions on a network (enumerated as $0,1,\dots,N-1$) the functions
$\psi(x)$ are only defined for integer values of $x=0,1,\dots,N-1$, and
the form of Eq.(\ref{wigner}) has to be changed from an integral to a sum.
There have been several attempts in doing so, see, for instance,
\cite{Wootters1987,Cohendet1988}. However, the definition of the discrete
WFs might depend on whether the length $N$ of the system is even or odd
\cite{Leonhardt1995,Leonhardt1996}.

For a one-dimensional system of length $N$ with periodic boundary
conditions (exemplified by a ring) one has $\psi(x)\equiv \psi(x\pm rN)$
for all $r\in\mathbb{N}$. It follows that each and every one of the
products $\psi^*(x-y';t) \psi(x+y';t)$ is identical to (at least) one of
the $N$ forms $\psi^*(x-y;t) \psi(x+y;t)$, where $y=0,1,\dots,N-1$.

Now the WF has the form of a Fourier transform; a unique transformation of
these $N$ products requires $N$ different $k$-values. These $k$-values may
evidently be chosen as $k=2\pi\kappa/N$, again having
$\kappa=0,1,\dots,N-1$. One is thus led to propose for integer $x$ and $y$
the following discrete WF \cite{mb2006a}
\be
W(x,k;t) = \frac{1}{N} \sum_{y=0}^{N-1} e^{iky} \ \psi^*(x-y;t)
\psi(x+y;t).
\label{wigner_discrete}
\ee

\subsubsection{WF for a ring}

Since the eigenstates of the ring are Bloch states, the WF for a CTQW on a
ring of $N$ nodes reads \cite{mb2006a}
\bea
W_j(x,\kappa;t) &=& \frac{1}{N^2} \sum_{n=0}^{N-1}
\exp\big[
-i2\pi(2n+\kappa)(x-j)/N
\big]
\nonumber \\
&& \times
\exp\big\{
-i2t[\cos(2\pi (\kappa + n)/N) 
\nonumber \\
&& - \cos(2\pi n/N)]
\big\}
,
\label{wigner_bloch_3}
\eea
where it has been used that $\theta=2\pi n/N$ and $k=2\pi \kappa/N$.
Furthermore, one may note from Eq.(\ref{wigner_discrete}) that
\bea
&& 
\sum_\kappa
W_j(x,\kappa;t) 
=
|\psi(x;t)|^2
\nonumber \\
&&
= \frac{1}{N} \sum_\kappa \sum_y e^{-i2\pi\kappa y/N}
\psi^*(x-y;t) \psi(x+y;t) \nonumber \\
\label{wigner_disc_marg}
\eea

\begin{figure}
\centerline{\includegraphics[clip=,width=0.9\columnwidth]{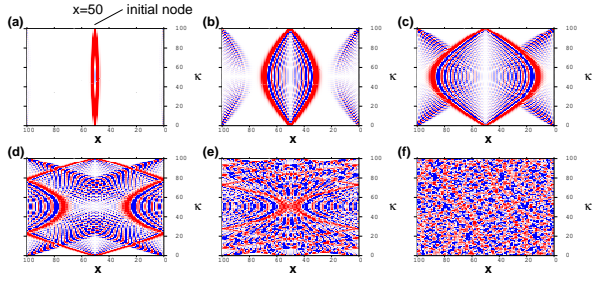}}
\caption{(Color online) WFs of a CTQW on a cycle of length $N=101$ at
times $t=1,10,20,40$ [(a)-(d)] as well as $t=100,500$ [(e),(f)]. The
initial node is at $j=50$.  Red regions denote positive values of
$W_j(x,\kappa;t)$, blue regions negative values and white regions values
close to $0$. From \cite{mb2006a}.  }
\label{wigner_bloch_101_time}
\end{figure}

In general, the WFs have a very complex structure. Figure
\ref{wigner_bloch_101_time} shows a contour plot of the WF of a CTQW on a
cycle of length $N=101$ at different times. Note that at $t=0$ the WF is
localized on the strip at the initial point $j$. At $t=1$, the WF is still
mostly localized about $j$. As time increases other sites get populated.
On short time scales, the WF develops a very regular structure in phase
space, with ``wavefronts'' originating from the initial point $j$.
Additionally, one also notes ``wavefronts'' starting from the region
opposite to the initial point, which are much weaker in amplitude.  As
time progresses, these two types of waves start interfering with each
other.

\begin{figure}
\centerline{\includegraphics[clip=,width=0.9\columnwidth]{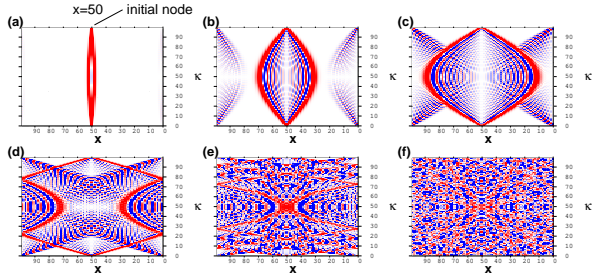}}
\caption{(Color online) Same as Fig.\ref{wigner_bloch_101_time}, for
$N=100$ and $j=50$. From \cite{mb2006a}.}
\label{wigner_bloch_100_time}
\end{figure}

Figure~\ref{wigner_bloch_100_time} shows the WF of a CTQW for $N=100$.  At
short times, the structure of the WF is quite similar to the situation for
$N=101$.  Nonetheless, there are differences at larger times, visible by
comparing Figs.\ref{wigner_bloch_101_time}(e) and
\ref{wigner_bloch_101_time}(f) to Figs.\ref{wigner_bloch_100_time}(e) and
\ref{wigner_bloch_100_time}(f).

However, although at long times the interference effects are quite
intricate, typical patterns are still visible.  At $t=500$, one finds less
regularities in phase space for $N=101$ than for $N=100$, reflecting the
higher symmetry of CTQW for even $N$. One notes, moreover, that the phase
space patterns give a much richer picture of the underlying dynamics than
the transition probabilities $|\psi_j(x;t)|^2$ alone.

Furthermore, one should remark that in an infinite system the WFs have a
much simpler structure, because there one does not face the problem of
distinct wave fronts running into opposite directions and interfering with
each other because of the closure of the ring.

In general, the marginal distribution $\sum_\kappa W_j(x,\kappa;t)$ is
obtained from Eq.(\ref{wigner_bloch_3}) as \cite{mb2006a}
\bea
&&\sum_\kappa
W_j(x,\kappa;t)
= \frac{1}{N} \sum_{n=0}^{N-1}\exp[i2\pi n(x-j)/N]
\nonumber \\
&&\times \exp[-i2t\cos(2\pi
n/N)]
\nonumber \\
&&\times
\frac{1}{N} \sum_{\kappa=0}^{N-1}\exp[i2\pi(\kappa+n)(x-j)/N]
\nonumber \\
&&\times \exp[i2t\cos(2\pi (\kappa
+ n)/N)].
\eea
Since the system is periodic, the arguments $(\kappa+n)$ in the
exponentials can be written as $(\kappa+n)\equiv~N-\nu$, where
$\nu=0,1,\dots,N-1$. This yields
\bea
&&\sum_\kappa
W_j(x,\kappa;t)
\nonumber \\
&&=\left| \frac{1}{N} \sum_{n=0}^{N-1} e^{i2\pi n(x-j)/N} e^{-i2t\cos(2\pi
n/N)}
\right|^2, \nonumber \\
\label{marg_x_o}
\eea
which is exactly what also follows, see Eq.(\ref{wigner_disc_marg}), from
calculating $|\psi(x;t)|^2$ directly from the Bloch ansatz, see
Eq.(\ref{ampl_bloch2}) and \cite{mb2005b}.

For the marginal distribution $\sum_x W_j(x,\kappa;t)$ one also gets from
Eq.(\ref{wigner_bloch_3})
\bea
&&\sum_{x} W_j(x,\kappa;t)
= \frac{1}{N^2} \sum_{n=0}^{N-1} N\delta_{2n,-\kappa}
\ e^{i2\pi(2n+\kappa)j/N}
\nonumber \\
&& \times \exp\big\{
-i2t[\cos(2\pi (\kappa + n)/N) - \cos(2\pi n/N)]
\big\}
\nonumber \\
&&= \begin{cases} 1/N & \mbox{for } N \ \mbox{odd and all} \ \kappa \\
2/N & \mbox{for } N \ \mbox{even and} \ \kappa \ \mbox{even} \\
0 & \mbox{for }  N \ \mbox{even and} \ \kappa \ \mbox{odd},
\end{cases}
\label{marg_k_o}
\eea
all of these expressions are independent of $t$. Eq.(\ref{marg_k_o}) can
be confirmed directly by taking the Fourier transform of $|\psi(x;t)|^2$.
The whole phase space volume is normalized to unity, as can be seen by
summing Eq.(\ref{marg_k_o}) over all $\kappa$, with
$\kappa=0,1,\dots,N-1$.

In the long-time average and for odd $N$ (superscript $^o$) most points in
the quantum mechanical phase space have a weight of $1/N^2$, namely one
has has \cite{mbb2007a}
\be
{\overline W}^o_j(x,\kappa) = \begin{cases} 1/N^2 & \mbox{for } \kappa\neq0 \
\mbox{and any} \ x \\ 
1/N &\mbox{for }  \kappa=0 \ \mbox{and} \ x=j \\ 0 &
\mbox{else.} \end{cases}
\label{lim_wf_0}
\ee
For even $N$ (superscript $^e$), the limiting WF reads
\be
{\overline W}^e_j(x,\kappa) = \begin{cases} 2/N^2 & \kappa\neq0, \ \kappa
\ \mbox{even}\ \mbox{and any} \ x \\ 1/N & \kappa=0 \ \mbox{and} \
x=j,j+ N/2 \\ 0 & \mbox{else.} \end{cases}
\label{lim_wf_0e}
\ee
The long time averages of the WFs for even $N$ are somewhat peculiar,
since values different from zero appear only for even $\kappa$, whereas
the WFs themselves have values different from zero at arbitrary times for
all $\kappa$. These stripes in the long-time average are due to the
periodicity of the ring. For even $N$ one finds constructive interference
patterns in the transition probabilities, since the number of steps in
both directions is the same, see also Ref.~\cite{mb2005b}. For a finite
line with even $N$, there are no stripes in the long time average.

\subsubsection{Rings with energetic disorder}

By introducing (static) disorder into the system, as in
Sec.~\ref{ctqw-disorder}, the Bloch property is lost \cite{mbb2007a}. In
order to have a global picture of the effect of the disorder on the
dynamics, one considers ensemble averages of the WFs. For this one
calculates the WF for different realizations of ${\bm H}$ and averages
over all realizations, i.e., for $R$ realizations: 
\be
\langle W_j(x,\kappa;t) \rangle_R \equiv \frac{1}{R} \sum_{r=1}^R
\big[W_j(x,\kappa;t)\big]_r,
\ee
where $\big[W_j(x,\kappa;t)\big]_r$ is the WF of the $r$th realization of
${\bm H}$.

\begin{figure}[t]
\centerline{\includegraphics[clip=,width=0.99\columnwidth]{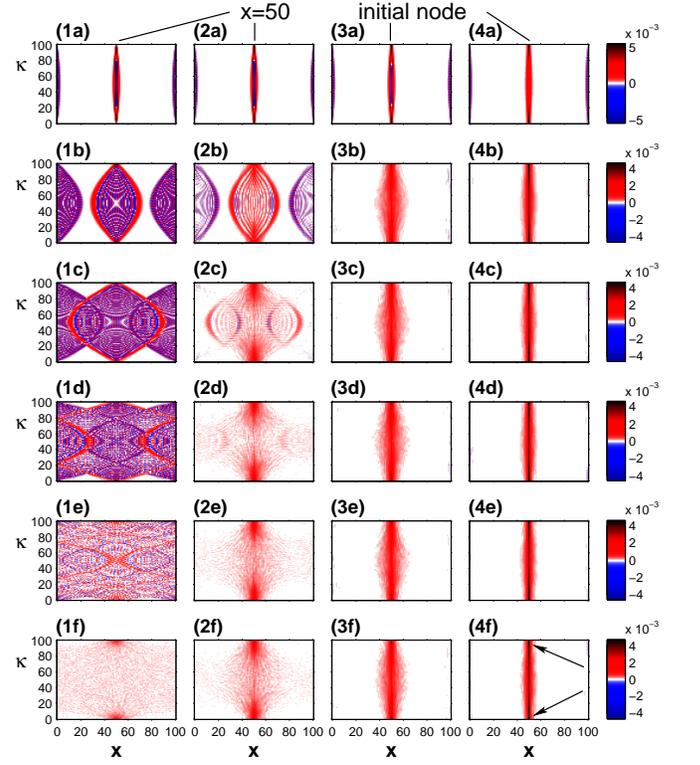}}
\caption{(Color online) Ensemble average of WFs of the quantum dynamics on
a ring of length $N=101$ with diagonal and off-diagonal disorder for
$\Delta=1/40,1/10,1/4$, and $1/2$ [columns (1)-(4)], each at times
$t=1,10,20,40,100$, and $500$ [rows (a)-(f)]. The initial node is always
$j=50$ and the average is over $R=1000$ realizations. Red regions denote
positive values of the averaged WFs, blue regions negative values and
white regions values close to $0$. The colormaps are always chosen to be
the same for each row but might differ in different rows.  The maximal
values of $\langle W_j(x,\kappa;t) \rangle_R$ are denoted by small black
regions; these are highlighted and   exemplified by the arrows in panel
(4f). From \cite{mbb2007a}.  }
\label{wigner_101_diag}
\end{figure}

It turns out that the particular type of disorder, i.e., diagonal or
diagonal and off-diagonal, does not change the picture significantly. For
diagonal and for off-diagonal disorder Figure~\ref{wigner_101_diag} shows
snapshots of $\langle W_j(x,\kappa;t) \rangle_R$ at different times for
$N=101$ and for different values of $\Delta$. 

The first column shows $\langle W_j(x,\kappa;t) \rangle_R$ for
$\Delta=1/40$ at times $t=1,10,20,40,100$, and $500$
[Fig.~\ref{wigner_101_diag}(1a)-(1f)].  For this quite weak disorder, the
patterns in phase space are similar to the unperturbed case, where
``waves'' in phase space emanate from the initial site $x=j=50$ and start
to interfere after having reached the opposite site of the ring (see
Fig.~3 of \cite{mb2006a}). However, at longer times differences become
visible, Fig~\ref{wigner_101_diag}(1f).  The pattern for the unperturbed
case is quite irregular but with alternating positive and negative regions
of the WF of approximately the same magnitude. For $x$ close to the
initial site $j=50$, the disorder causes a decrease of $\langle
W_j(x,\kappa;t) \rangle_R$ for $\kappa$-values in the middle of the
interval $[0,N-1]$ when compared to what is found for values of $\kappa$
close to $0$ or to $N-1$.

Increasing the disorder parameter $\Delta$, the patterns change
profoundly. The wave structure gets suppressed and for all $\kappa$ a
localized region forms about the initial site $j=50$, already for small
disorder ($\Delta=1/10$) and short times ($t=20$), see
Fig.~\ref{wigner_101_diag}(2c).

For even larger values of $\Delta$, for all $\kappa$ the formation of a
localized region about $j=50$ becomes even more pronounced. Already for
$\Delta=1/4$ this localized region forms for times as short as $t=10$, see
Fig.~\ref{wigner_101_diag}(3b).  At $\Delta=1/2$, $\langle W_j(x,\kappa;t)
\rangle_R$ stays localized at all times
[Fig.~\ref{wigner_101_diag}(4a)-(4f)].  Also here, the values of the WF at
about $\kappa\approx N/2$ are rather low, whereas the values of the WF for
$\kappa$ close to the interval borders $0$ and $N-1$ remain rather large
for $x\approx j$, as indicated by the thin black region (see also the
arrows), e.g., in Fig.~\ref{wigner_101_diag}(4f).  One further notes that
at high disorder the localized averaged WF is always positive, i.e., all
fluctuations, present for small disorder, have vanished. One recalls that
the WF is normalized to unity when integrated over the whole phase space.
Having an even number of nodes in the graph does not alter the picture
significantly.

\begin{figure}
\centerline{\includegraphics[clip=,width=0.99\columnwidth]{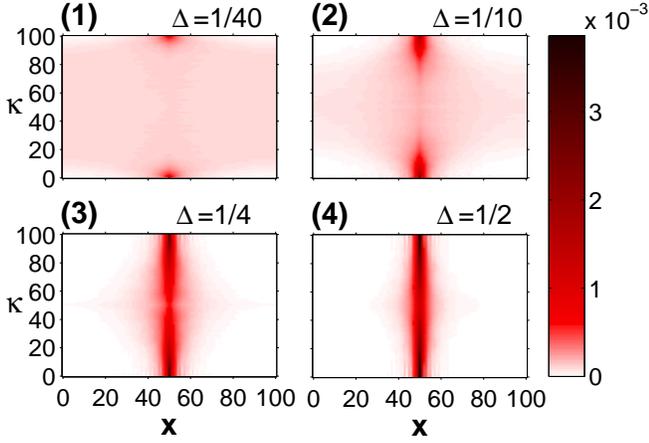}}
\caption{(Color online) DOD: Limiting averaged WF, $\langle{\overline
W}_j(x,\kappa)\rangle_R$, for $N=101$ and $\Delta=1/40$, $1/10$, $1/4$,
and $1/2$ [panels (1) to (4), respectively], according to
Eq.(\ref{wf_avg_tavg}). From \cite{mbb2007a}.}
\label{lim_wigner}
\end{figure}

The ensemble average of the long-time averaged WF follows as
\cite{mbb2007a}
\bea
&&\langle {\overline W}_j(x,\kappa)\rangle_R
\equiv \Big\langle\lim_{T\to\infty}\frac{1}{T} \int\limits_0^T dt \
W_j(x,\kappa;t) \Big\rangle_R \nonumber \\
&&= \lim_{T\to\infty}\frac{1}{T} \int\limits_0^T dt \
\langle W_j(x,\kappa;t) \rangle_R.
\label{wf_avg_tavg}
\eea
Now, the disorder changes also the limiting WF quite drastically.
Starting from high disorder of $\Delta=1/2$, one expects from
Figs.~\ref{wigner_101_diag}(4a)-(4f) that the long time average of the
{\sl averaged} WF will look basically the same. Figure \ref{lim_wigner}
shows the limiting averaged WF for $N=101$ and DOD according to
Eq.~(\ref{wf_avg_tavg}). Indeed, for large $\Delta$, the limiting averaged
WF is comparable to the corresponding averaged WF, compare
Figs.~\ref{lim_wigner}(4). Close to the initial node $x=j=50$ and there
along the $\kappa$-direction, $\langle{\overline W}_j(x,\kappa)\rangle_R$
has large (positive) values for $\kappa$ about $0$ and $N-1$ which
decrease by going toward $\kappa=N/2$. Here, the decrease depends on the
particular type of disorder involved.

Also for small $\Delta$ there are significant differences to the case
without disorder. From Fig.~\ref{lim_wigner}(1) one sees that the disorder
``smears out'' the localized value $\overline{W}_j(j,0)=1/N$ in the
absence of disorder, see Eq.(\ref{lim_wf_0}). Specifically, the
$(x=j,\kappa=0)$-value decreases, while the neighboring ones increase.
For $\Delta=1/10$, the onset of localization about $x=j=50$ can already be
seen, e.g., Fig.~\ref{lim_wigner}(2), and becomes more and more pronounced
as $\Delta$ increases, Fig.~\ref{lim_wigner}(3). Furthermore, all other
values of $\langle{\overline W}_j(x,\kappa)\rangle_R$ for $x\neq j$
decrease with increasing disorder, as can be seen by the decreasing size
of the light pink region, which corresponds to values close to $1/N^2$.
In fact, the values of the limiting averaged WF for $x$ distant from
$x=j=50$ drop to zero, shown as white regions in Fig.~\ref{lim_wigner}.

\begin{figure}
\centerline{\includegraphics[clip=,width=0.99\columnwidth]{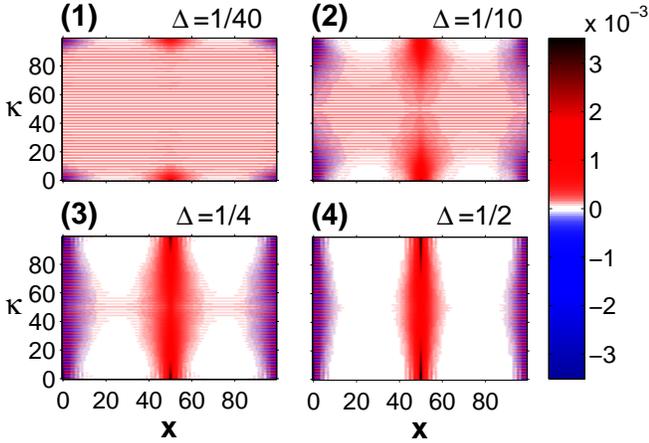}}
\caption{(Color online) Same as Fig.~\ref{lim_wigner} but for $N=100$.
From \cite{mbb2007a}.}
\label{lim_wigner_even}
\end{figure}

For even $N$ one finds that without disorder the limiting WF shows a
peculiar ``striped'' distribution, caused by the PBCs, see
Eq.(\ref{lim_wf_0e}). By switching on the disorder, these peculiarities
vanish. Figure \ref{lim_wigner_even} shows the limiting averaged WF for
DOD and $N=100$. Although for small $\Delta$ there are some remainders of
stripes left, these disappear completely for higher values of $\Delta$,
compare Figs.~\ref{lim_wigner_even}(1)-(4). Note further that the second
peak of ${\overline W}^e_j(x,\kappa)$ at $x=j+N/2$, see
Eq.(\ref{lim_wf_0e}), transforms for increasing disorder to an oscillatory
line in the $\kappa$-direction at $x=j+N/2$.

\subsubsection{Long-range interaction cycles}

Xu and Liu have considered the WF, defined in Eq.~(\ref{wigner_discrete})
and \cite{mb2006a}, for rings, where the $m$ next nearest neighbors of a
given node are connected to this node by bonds \cite{xu2008c}, see also
Sec.~\ref{sec-longrange}. The eigenstates of such a translationally
invariant system are still Bloch states. However, the eigenvalues read
\cite{xu2008c}
\be
E_n = 2m - 2\sum_{j=1}^m \cos{2nj\pi/N}.
\ee
As in Eq.~(\ref{wigner_bloch_3}), the WF follows as
\bea
W_j(x,\kappa:t) &=& \frac{1}{N^2} \sum_{n=0}^{N-1}
e^{-it(E_n-E_{N+\kappa-n})} 
\nonumber \\
&& \times e^{i2\pi(2n-\kappa)(x-j)/N}.
\eea
Thus, inserting the results for the eigenvalues one obtains the WF for
this type of long-range interacting cycles.

Now, while for $m=1$ one recovers the results of the previous section,
increasing $m$ leads to more complex structures of the WF. The fronts
emanating from the initial value $x=j$ have, in the case of $m=1$, only a
single maximum value at about $\kappa=N/2$, see
Figs.~\ref{wigner_bloch_101_time} and \ref{wigner_bloch_100_time}. For
$m>1$ there appear $m$ front maxima at about $\kappa=(2l-1) N/2m$ with
$l=1,\dots,m$ \cite{xu2008c}. With increasing number $m$ the structure of
the WFs become more complex, but the resulting patterns can still be
attributed to the interfering fronts emanating from $x=j$ \cite{xu2008c}.

\subsubsection{Disordered networks}

Xu and Liu have also studied the WF for disordered systems, namely for the
Watts-Strogatz (WS) small-world model \cite{xu2008c}. The WS model starts
from an ordered ring with only nearest neighbor interactions. Now, every
bond is rewired with probability $p$, implying that for $p=0$ one retains
the ordered ring and for $p=1$ one has a completely disordered network of
$N$ nodes with $N$ bonds. The WF for such systems shows essentially the
same features as for rings with energetic disorder, see above. In the
ensemble average there is stronger and stronger localization with
increasing $p$. Moreover, here also the distinction between even and odd
$N$ remains, namely even $N$ lead to a stripe-patterned WF caused by
constructive interference while odd $N$ do not show this behavior
\cite{xu2008c}.

\subsection{Quantum master equations} 
\label{sec-qme}

CTQW and CTRW are the two extreme cases of possible transport processes,
i.e., a purely coherent transport for CTQW and a purely incoherent
transport for CTRW. Now, if a quantum mechanical system cannot be regarded
as being isolated from its environment, this environment will influence
the dynamics of the system. Depending on the strength of the coupling
(where also the temperature is assumed to be controlled by the
environment), the dynamics may change from quantal to classical. Since the
total system (i.e., the system $S$ and the environment/reservoir $R$)
resides in a Hilbert space of huge dimension (this due to the many degrees
of freedom (DOF) of the environment), one often considers the system only,
where the environmental DOF have been traced out
\cite{Breuer-Petruccione}. 

Let $\bm W(t)$ be the density operator of the total system, then $\bm
\rho(t) \equiv {\rm tr}_R\{ \bm W(t) \}$ is called the reduced density
operator. Then the dynamics of $\bm W(t)$ is still governed by
Schr\"odinger's equation, which can be recast into the Liouville-von~Neumann
(LvN) equation
\be
\dot{\bm W}(t) = -i [\bm H_{\rm tot}, \bm W(t)],
\ee
where $\bm H_{\rm tot}$ is the Hamiltonian of the total system. $\bm
H_{\rm tot}$ usually comprises three parts: the Hamiltonians $\bm H_S$ for
the system, $\bm H_R$ for the environemnt, and $\bm H_{SR}$
for the coupling between system and environment: 
\be
\bm H_{\rm tot} = \bm H_S + \bm H_R + \bm H_{SR}.
\ee

Inserting this into the LvN equation for $\bm W(t)$ and integrating out
the environmental DOF one arrives at an equation for the reduced density
operator (for details see \cite{Breuer-Petruccione}):
\be
\dot{\bm\rho}(t) = -i [\bm H_S , \bm\rho(t)] + {\cal D}[\bm\rho(t)].
\label{qme}
\ee
Here, ${\cal D}[\bm\rho(t)]$ is called the {\it dissipator}. Note that in
order to arrive at Eq.~(\ref{qme}) several assumptions, such as bilinear
couplings between system and environment or the Markov approximation, have
been used \cite{Breuer-Petruccione}. 

Under certain conditons, such as a weak coupling between the system and
the environment, the dissipator can also be expressed by operators acting
on the Hilbert space of the system $S$. In particular,
Eq.~(\ref{qme}) can be brought into the so-called Lindblad form
\cite{Breuer-Petruccione}
\bea
\dot{\bm\rho}(t) &=& -i [\bm H_S , \bm\rho(t)] - \sum_{j=1}^{N^2} \Big(2 \bm L_j
\bm\rho(t) \bm L_j^\dagger \nonumber \\
&& - \bm L_j^\dagger \bm L_j \bm\rho(t) -
\bm\rho(t)  \bm L_j^\dagger \bm L_j\Big),
\label{lqme}
\eea
where the operators $\bm L_j$, the Lindblad operators, mimick the
influence of the environemnt on the dynamics.  Equation~(\ref{lqme}) is
also called Lindblad quantum master equation (LQME).

Obviously, if the coupling to the environment vanishes, one arrives again
at the LvN equation for the density operator $\bm\rho(t)$ of the system
$S$. Since the resulting LvN equation is equivalent to
the Schr\"odinger equation of a closed system with Hamiltonian $\bm H$, it
is also equivalent to the CTQW formulation of the previous sections.  Now,
when increasing the coupling to the environment, Eq.~(\ref{lqme}) allows
to study the onset of the quantum to classical cross-over.

\subsubsection{Decoherence on rings}

Take now a ring of $N$ nodes and $N$ Lindblad operators each of which
projects onto a single node, i.e., $\bm L_j = \sqrt{\lambda_j}
|j\rangle\langle j|$ for $j=1,\dots,N$. The parameters $\lambda_j$ specify
the strength of the coupling to the environment. If $\lambda_j = \lambda$
for all $j$, Eq.~(\ref{lqme}) simplifies:
\bea
\dot{\bm\rho}(t) &=& -i [\bm H_S , \bm\rho(t)] 
\nonumber \\
&& - 2\lambda \sum_j \big(\bm\rho(t)
- \rho_{jj}(t)\big) |j\rangle\langle j|,
\label{lvne-ring}
\eea
where $\rho_{kj}(t) \equiv  \langle k| \bm\rho(t) | j \rangle$.  The rate
$\lambda$ can be estimated  from the spectral density $J(\omega)$
describing the environment within the Caldeira-Leggett
model~\cite{caldeira_leggett__Ann.Phys._(1983-4)a,caldeira_leggett__Ann.Phys._(1983-4)b}
at a given temperature $T$. Taking $J(\omega) = 2\pi \alpha \omega
\exp(-\omega/\omega_c)$ and using the Markov approximation one arrives at
$\lambda = \pi \alpha k_B T$ \cite{Breuer-Petruccione}. One has to bear in
mind that Eq.~(\ref{lvne-ring}) is an approximation with a limited range
of validity: For a very large coupling strength $\lambda$,
Eq.~(\ref{lvne}) leads to the quantum Zeno limit rather than to a
classical master/rate equation.  In matrix form Eq.~(\ref{lvne-ring}) is
equivalent to the so-called Gurvitz model, see for instance
\cite{gurvitz2003}, considered in \cite{solenov2006a, solenov2006b,
fedichkin2006}
\bea
\dot{\rho}_{kj}(t) &=& -i \langle k| [\bm H_S , \bm\rho(t)] |j\rangle 
\nonumber \\
&& 
- 2
\lambda (1-\delta_{kj}) \rho_{kj}(t).
\eea

Solenov and Fedichkin showed analytically that the probability
distribution to be at node $k$, where for a ring one assumes without loss
of generality that the initial node is $j=0$, follows from the diagonal
elements of the reduced density matrix \cite{solenov2006b, fedichkin2006}
\bea
&&\rho_{k,k}(t) = \frac{1}{N} + \sum_{n,m=0}^{N-1}
\frac{1-\delta_{n+m,0}-\delta_{n+m,N}}{N^2} 
\nonumber \\
&& \times \Big[ \delta_{n,m}
e^{-2\lambda[(N-1)/N]t} 
+ (1-\delta_{n,m}) e^{-2\lambda[(N-2)/N]t}\Big]
\nonumber \\
&& \times \exp\Big[it \sin\frac{\pi(n+m)}{N}
\cos\frac{\pi(n-m)}{N} 
\nonumber \\
&& + \frac{2\pi i}{N}(n+m) k\Big]. 
\nonumber \\
&& ~
\label{lqme-cycle}
\eea
Clearly, Eq.~(\ref{lqme-cycle}) reduces to the results of
Sec.~\ref{sec-detnet} in the limit of $\lambda\to0$. For $\lambda\neq0$,
due to the exponentially decaying factors, the probability distribution
will eventually decay to the classical equilibrium values, in this case
the equipartition value $1/N$.

The so-called mixing time $t_m$ is used to compare the spreading of CTQW
to that of CTRW. It is implicitly defined by looking at the deviation from
the classical uniform distribution $1/N$ \cite{solenov2006a}. One gets
thus
\be
\sum_{k=0}^{N-1} \left|\rho_{k,k}(t_m) - \frac{1}{N}\right| \leq \varepsilon,
\ee
where $\varepsilon$ is a small dimensionless constant representing the
degree of mixing. By using Eq.~(\ref{lqme-cycle}), Solenov and Fedichkin
showed that the mixing time is bound from above by
\be
t_{\rm mix} \equiv \min t_m < \frac{N}{2\lambda(N-2)} \ln
\left(\frac{N+1}{\varepsilon}\right).
\ee

The same authors have generalized this approach to so-called hypercycles.
A hypercycle of dimension $d$ and size $N$ is build by considering $N$
copies of the $d-1$ dimensional hypercycle and connecting the
corresponding nodes, see \cite{solenov2006a} for details. Thereby, the
total number of nodes is $N^d$. The corrsponding upper bound for the
mixing time $t_m$ now follows as
\be
t_{\rm mix}^{(d)} \leq \frac{d}{2\lambda} \frac{N}{N-1} \ln
\left[\frac{d(N+1)(1+\varepsilon N^d)}{\varepsilon}\right].
\ee

Salimi and Radgohar have considered the long-range case of the single
cycle, where a given node interacts with its $2m$ nearest neighbors ($m$
to each side) \cite{salimi2009b}, see also Sec.~\ref{sec-longrange}. The
probability distribution to be at node $k$ is then obtained for even $m$
as 
\bea
&&\rho_{k,k}(t) = \frac{1}{N} + \sum_{n,l=0}^{N-1}
\frac{1-\delta_{n+l,0}-\delta_{n+l,N}}{N^2}
\nonumber \\
&& \times
\exp\Big[-2\lambda[(N-1)/N]t\Big]
\nonumber \\
&& \times \exp\Big[it \sin\frac{\pi(n+l)}{N}
\cos\frac{\pi(n-l)}{N} + \frac{2\pi i}{N}(n+l) k\Big]
\nonumber \\
&& \times \exp\Big[i^{m+1}t \sin\frac{\pi m(n+l)}{N}\sin\frac{\pi
m(n-l)}{N}\Big].
\eea
and for odd $m$ as 
\bea
&&\rho_{k,k}(t) = \frac{1}{N} + \sum_{n,l=0}^{N-1}
\frac{1-\delta_{n+l,0}-\delta_{n+l,N}}{N^2}
\nonumber \\
&& \times \Big[ \delta_{n,l}
e^{-2\lambda[(N-1)/N]t}
+ (1-\delta_{n,l}) e^{-2\lambda[(N-2)/N]t}\Big]
\nonumber \\
&& \times \exp\Big[it \sin\frac{\pi(n+l)}{N}
\cos\frac{\pi(n-l)}{N} + \frac{2\pi i}{N}(n+l) k\Big]
\nonumber \\
&& \times \exp\Big[i^{m+1}t \sin\frac{\pi m(n+l)}{N}\sin\frac{\pi
m(n-l)}{N}\Big].
\eea
While the upper bounds of the mixing time turn out to be independent of
$m$, they do differ for even and for odd $m$, one has namely
\be
t_{\rm mix} \leq \frac{N}{N-1} \ln \frac{N}{\varepsilon} \qquad \mbox{for
even }m 
\ee
and
\be
t_{\rm mix} \leq \frac{N}{N-2} \ln \frac{N}{\varepsilon} \qquad \mbox{for
odd }m.
\ee
Therefore, mixing turns out to be faster for even $m$ than for odd $m$.

\subsubsection{Dimer with traps}

Within the phenonmenological approach presented above, introducing traps
into the system leads to non-Hermitian Hamiltonians. The LvN equation
changes, therefore, to
\be
\dot{\bm W}(t) = -i \big[ {\bm H_0}, {\bm W}(t) \big] -
\big\{{\bm\Gamma},{\bm W}(t) \big\},
\ee
where $\{\cdot,\cdot\}$ is the anti-commutator and $\bm H_0$ the
Hamiltonian without traps. Writing the Lindblad operators in the form
$\sqrt{\lambda}\bm L_j$, the LQME reads \cite{mmsb2010}
\bea
\dot{\bm \rho}(t) &=& -i \big[ {\bm H_0}, {\bm \rho}(t) \big] - 
\big\{{\bm\Gamma},{\bm\rho}(t) \big\} 
\nonumber \\
&& 
- 2\lambda 
\sum_j \Big(\bm\rho(t) - \rho_{jj}(t)\Big){\bm L}_j.
\label{lvne}
\eea

Now, consider a dimer which is coupled to an external bath. This situation
allows to solve Eq.~(\ref{lvne}) analytically and to compare the LQME
results to the numerically exact Path Integral Monte Carlo (PIMC)
calculations. For details on PIMC techniques see, e.g.,
\cite{LM_JCP_121a,LM_JCP_121b} and references therein. The Hamiltonian of
the dimer without any coupling to the surroundings can be expressed
through the Pauli matrices ${\bm\sigma_z}$ and ${\bm\sigma_x}$,
\begin{equation}
{\bm H} = E \ {\bm 1} - V{\bm\sigma_x} - i\frac{\Gamma}{2}({\bm 1} -
{\bm\sigma_z}) \,, 
\end{equation}
where $E$ is the on-site energy, which is choosen to be the same for both
nodes, and $V$ is the coupling between the two nodes. It is easily
verified that the eigenvalues are \cite{mmsb2010}
\begin{equation}
E_\pm= E \pm V e^{\pm i\phi} = E \pm \sqrt{V^2-\Gamma^2/4} - i \Gamma/2,
\end{equation}
where $\phi = \arcsin(\Gamma/2V)$. For $\Gamma\to0$ ($\phi\to0$) this
yields the correct eigenvalues $E\pm V$ of ${\bm H}_0$. Note that for
$\Gamma\leq2V$ the negative imaginary part of $E_\pm$ is identical for
both eigenvalues, i.e., $\gamma_+=\gamma_-=\Gamma/2$. The
bi-orthonormalized eigenstates of ${\bm H}$ are of the form
\cite{mmsb2010}
\begin{equation}
|\Phi_\pm\rangle \equiv \frac{1}{\sqrt{2\cos\phi}} \left( \begin{matrix}
e^{\pm i\phi/2} \\
\pm e^{\mp i\phi/2} \end{matrix} \right) 
\end{equation}
and
\begin{equation}
|\tilde\Phi_\pm\rangle \equiv \frac{1}{\sqrt{2\cos\phi}} \left(
\begin{matrix} e^{\mp i\phi/2} \\
\pm e^{\pm i\phi/2} \end{matrix} \right),
\end{equation}
where the phases $\phi$ depend on $\Gamma$, such that in the limit
$\Gamma\to0$ one recovers the eigenstates of ${\bm H}_0$.

When the coupling to the environment vanishes ($ \lambda\to 0$), one
obtains the survival probability directly from the eigenstates and
eigenvalues of $\bm H$. For $\Gamma\leq2V$ one has \cite{mmsb2010}
\begin{equation}
\Pi(t) = e^{-\Gamma t}\frac{\cos^2(\phi +
tV\cos\phi)}{\cos^2\phi} \quad (\mbox{for} \ \lambda=0).
\label{pi_l0}
\end{equation}
Note that for values $\Gamma> 2V$ the dimer is overdamped.

When considering the dimer without traps ($\Gamma=0$) but coupled to the
environment, Eq.~(\ref{lvne}) simplifies and, from the solution for ${\bm
\rho}$, one obtains the transition probabilities \cite{mmsb2010}
\begin{eqnarray} 
\pi_{1,1}^{(0)}(t) &=& \frac{1}{2} + \frac{e^{-\lambda t}}{2} \Bigg[
\frac{\lambda\sin\left(t\sqrt{4V^2-\lambda^2}\right)}{\sqrt{4V^2-\lambda^2}}
\nonumber \\
&& +
\cos\left(t\sqrt{4V^2-\lambda^2}\right)
\Bigg] \quad (\mbox{for } \Gamma=0). \nonumber \\  
\label{pi_g0}
\end{eqnarray}
Moreover, $\pi_{2,1}^{(0)}(t) = 1 - \pi_{1,1}^{(0)}(t)$. From
Eq.~(\ref{pi_g0}) one recovers for $\lambda\to 0$ the simple oscillatory
behavior of the transition probabilities [namely, $\lim_{\lambda\to
0}\pi_{1,1}^{(0)}(t)=\cos^2(Vt)$]. When the coupling to the surroundings
does not vanish, for $\lambda>0$, the transition probabilities still show
oscillations, which are superimposed on an exponential decay which tends
for long times to the classical equipartition value of $1/2$.

\begin{figure}[t]
\centerline{\includegraphics[clip=,width=0.9\columnwidth]{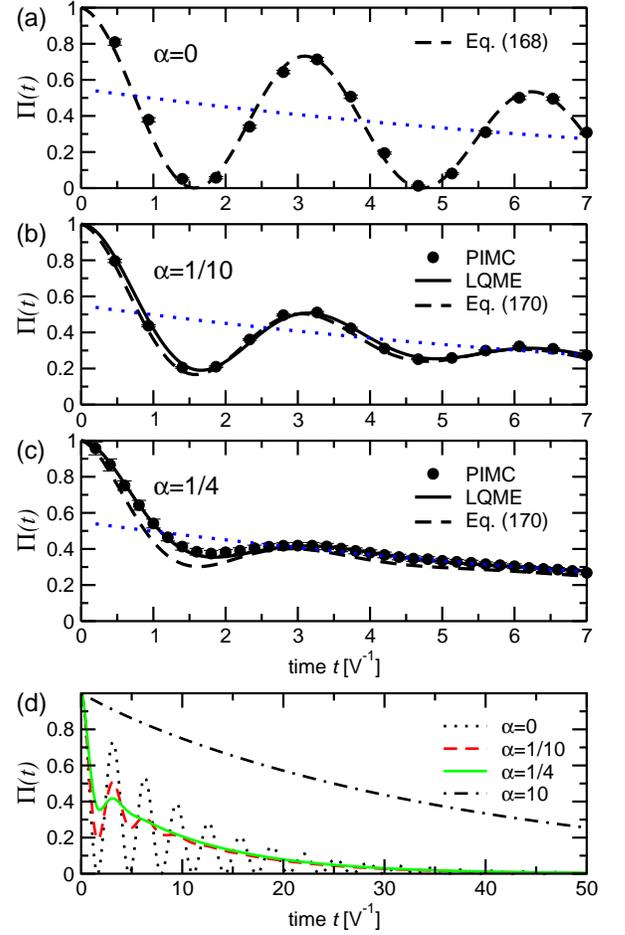}}
\caption{(Color online) PIMC results (circles) for a dimer with
$\Gamma=0.1$ and different system-bath couplings $\alpha=\lambda/\pi$: (a)
$\alpha=0$, (b) $\alpha=1/10$, and (c) $\alpha=1/4$. The solid lines
represent the numerical solution of the LQME equation, the dashed lines
show the corresponding analytical results obtained from Eq.~(\ref{pi_l0})
for $\alpha=\lambda=0$ and from Eq.~(\ref{pi_lr_approx}) for $\alpha=1/10$
and $\alpha=1/4$. The dotted blue line shows the long time limit $\Pi(t)
\sim \exp(-\Gamma t)$. Panel (d) shows the corresponding long-time
behavior of the numerical LQME solution for the three different values of
$\alpha$ and additionally the behavior for large couplings $\alpha=10$.
From \cite{mmsb2010}.}
\label{pimc_dimer}
\end{figure}

In order to combine the results of Eq.~(\ref{pi_l0}) -- taking values
$\Gamma <2V$ -- and Eq.~(\ref{pi_g0}), one expands in both equations all
terms except the exponentials to first order in $\Gamma$ and $\lambda$,
respectively. Note that for Eq.~(\ref{pi_l0}) one obtains a product of
$\exp(-\Gamma t)$ and $\cos^2Vt$, which is the simple oscillatory behavior
of the dimer in the absence of the trap. Now, the coupling to the
environment affects all transitions but still conserves probabilities.
Therefore, one replaces the term $\cos^2Vt$ by the expansion of
Eq.~(\ref{pi_g0}), such that one obtains \cite{mmsb2010}

\begin{eqnarray}
\Pi(t) &\approx& e^{-\Gamma t} \pi_{1,1}^{(0)}(t)  \nonumber \\
&\approx&  e^{-\Gamma t} \Big[ \frac{1}{2} +\frac{e^{-\lambda t}}{2}
\Big(\cos 2Vt +\frac{\lambda}{2V}\sin 2Vt \Big) \Big]. \nonumber \\~
\label{pi_lr_approx}
\end{eqnarray}

Figure \ref{pimc_dimer} compares the survival probabilities of a
dissipative dimer obtained from the approximative LQME approach to the
numerically exact PIMC calculations for a bath with ohmic spectral density
with exponential cutoff, $J(\omega) = 2 \pi \alpha \omega
e^{-\omega/\omega_c}$.  The initial condition is $\pi_{1,1}(0)=1$, i.e.,
at $t=0$ the system is localized in the non-trap node $1$ of the dimer.
Here, the on-site energies $E$ and the coupling elements $V$ have been
taken to be equal, $E=V=1$, one sets $\omega_c=5V$ and the temperature is
fixed to $k_B T =V$.

For small trapping strength ($\Gamma=0.1$) and vanishing coupling to the
environment ($\alpha=0$), Fig.~\ref{pimc_dimer}(a), the PIMC calculations
coincide with the result of Eq.~(\ref{pi_l0}). A moderate increase of the
coupling ($\alpha=1/10$), Fig.~\ref{pimc_dimer}(b), still leads for
Eqs.~(\ref{lvne}) (solid lines) and (\ref{pi_lr_approx}) (dashed lines) to
results which are in excellent agreement with the findings of the PIMC
calculations (symbols).  When increasing the coupling further to
$\alpha=1/4$, Fig.~\ref{pimc_dimer}(c), however, the approximate solution,
Eq.~(\ref{pi_lr_approx}), begins to deviate from the LQME and the PIMC
calculations, which are still in very good agreement \cite{mmsb2010}.

As the numerical effort of real-time PIMC simulations grows exponentially
with time, they can cover only short-to-intermediate time scales.
However, the agreement between the LQME and the PIMC calculations in the
weak coupling regime permits to compensate for this shortcoming; it allows
namely to use for longer times the LQME procedure, see
Fig.~\ref{pimc_dimer}(d).

\section{Outlook}

The application of CTQW to transport for large classes of (physical,
chemical, and biological) phenomena involving different types of networks
has turned out to be very successful in recent years. However, only little
is known about the detailed relations between topology, interaction
ranges, dimensions and the transport efficiencies. While CTRW fall into
different universality classes depending on the (fractal) dimensions and
the interaction ranges, there are only few indications on the analogous
emergence of ``quantum universality classes''. Take as an example the
dynamics on a chain with long-range interactions decaying as
$|k-j|^{-\gamma}$, see Sec.~\ref{sec-longrange}; in this case the mean
square displacement of CTQW increases as $t^2$ also for $\gamma=2$. This
is not the case for CTRW, where only systems with $\gamma>3$ lead to the
same power-law behavior of the mean square displacement. 

Therefore, a thorough investigation of the influence of different
topological aspects on the dynamics is clearly necessary. Moreover, once a
systematic classification of CTQW gets established, one needs to examine
the transition from the quantum to the classical picture, in order to
understand the changes in the universality patterns. For this, the
environment of each quantum system has to be properly taken into account.
In a phenomenological ansatz one can incorporate this either by using the
quantum master equation approach, sketched in Sec.~\ref{sec-qme}, or by
using so-called generalized master equations (GME) \cite{Kenkre}. Here,
the Markovian master equation (as used in CTRW) is extended by a
memory-kernel (MK) to account for non-Markovian effects. In the limit of a
MK of delta-type one recovers the CTRW, while in the limit of a constant
MK one arrives for CTQW at a wave equation similar to the Schr\"odinger
equation. Both approaches allow to study the influence of the environment
on the dynamics. Clearly, the validity of both methods has to be checked.
One option to do so is to compare the results - for not too large systems
- to those of numerically exact PIMC simulations. 

Such a comparison has several advantages. First, it sets the range of
validity of the phenomenological approaches. Second, once this range is
established, it allows to extend the short-time PIMC results to (in
principle) arbitrary long times. Thus, joining different methods together
may lead to a deeper understanding of the complex transport processes
found in nature. 

\section*{Acknowledgements}
Support from the Deutsche Forschungsgemeinschaft (DFG), the Fonds der
Chemischen Industrie and the Ministry of Science, Research and the Arts of
Baden-W\"{u}rttemberg (AZ: 24-7532.23-11-11/1) is gratefully acknowledged.
We thank Elena Agliari, Veronika Bierbaum, Lothar M{\"u}hlbacher, Volker
Pernice, Tobias Schmid, and Antonio Volta for many fruitful discussions.

% References

\bibliographystyle{elsarticle-num}

\end{document}